\documentclass{JHEP3}

\usepackage{latexsym}
\usepackage{graphics}

\def\comment #1{}
\def\cf {{\it cf. }}

\def\refer #1{{(\ref{#1})}}
\def\fullref #1{\ref{#1} (p.\pageref{#1})}

\def\bra #1{\left\langle{#1}\right|}
\def\ket #1{\left|{#1}\right\rangle}
\def\bracket #1#2{\left\langle{#1}|{#2}\right\rangle}

\def\of #1{\!\left({#1}\right)}

\def\Z {\theIntegers}
\def\N {\mathrm{I\!N}}
\def\R {\mathrm{I\!R}}
\def\C {\mathrm{\,I\!\!\!C}}

\def\gop {{\hat g}}

\def\gopg {{\gop_{\bf G}}}

\def\set #1{\left\lbrace{#1}\right\rbrace}

\def\brackets #1{\left[{#1}\right]}
\def\braces #1{\left\lbrace{#1}\right\rbrace}

\def\order #1{\mathcal{O}\of{#1}}

\def\commutator #1#2{\brackets{{#1},{#2}}}
\def\antiCommutator #1#2{\braces{{#1},{#2}}}
\def\superCommutator #1#2{\brackets{{#1},{#2}}_\involution}

\def\adjoint #1{{{{#1}^{\dag}}}}

\def\defas {:=}
\def\shallbe {\stackrel{!}{=}}
\def\equalby #1{\stackrel{\refer{#1}}{=}}

\def\isomorphic {\simeq}

\def\coordAnnihilator {\mbox{$\hat \mathrm{c}$}}
\def\coordCreator {\adjoint{\coordAnnihilator}}

\def\edgesymbol {{\mbox{\tt e}}}
\def\edge {{\vec \edgesymbol}}
\def\edgesAnnihilator {\edgesymbol}
\def\edgesCreator {{\edgesAnnihilator^\dag}}

\def\pseudoCliff {{q}}

\def\numberOperator {{\hat N}}

\def\theIntegers {{\mathbf Z\!\!\!\mathbf Z}}
\def\extd {\mathbf{{d}}}
\def\edag {{\dag_{\!\gop}}}
\def\coextd {\adjoint{\extd}}
\def\ecoextd {\extd^\edag}

\def\hextd {{\extd_H}}
\def\hcoextd {{\extd^\edag_H}}

\def\boundary {{\partial}}

\def\vol {{\bf vol}}
\def\evol {{\bf vol}_\gop}

\def\ehodge {\,\star_\gop\,}
\def\invehodge {\,\star_\gop^{-1}\,}
\def\invhodge {\,\star^{-1}\,}

\def\involution {\mathbf \iota}

\def\clifford {{\mbox{$\hat \gamma$}}}

\def\hodge {\,\star\,}

\def\Dirac {\mathbf{D}}

\def\endofproof {$\Box$\newline}

\def\eigenspace #1#2 {\mathrm{eig}\of{#1,#2}}

\def\fatomega {\mbox{\boldmath$\omega$}}

\def\fatalpha {\mbox{\boldmath$\alpha$}}
\def\fatbeta {\mbox{\boldmath$\beta$}}
\def\fatDelta {\mbox{\boldmath$\Delta$}}

\newlength{\skiplength}

\def\delgen #1{\delta_{\left\lbrace{#1}\right\rbrace}}

\def\lpartial{\raisebox{7pt}{\tiny $\leftarrow$}\hspace{-7pt}\partial}
\def\rpartial{\raisebox{7pt}{\tiny $\rightarrow$}\hspace{-7pt}\partial}
\def\lrpartial{\raisebox{7pt}{\tiny $\leftrightarrow$}\hspace{-7pt}\partial}
\def\osmotial{\Delta}

\def\formtovec {{\ast_{\!\gop}}}
\def\vectoform {{\formtovec^{-1}}}
\def\fatpartial {{\partial\hspace{-5.6pt}\partial\hspace{-5.6pt}\partial\hspace{-5.6pt}\partial}}

\def\cupprod {\smile}
\def\path #1{\ket{#1}}

\def\lgtrace {\mbox{tr}}

\title{Discrete Differential Geometry on causal graphs}
\author{Eric Forgy\thanks{The work by EF was not supported by MIT/LL.} \\ Lincoln Laboratory \\Massachusetts Institute of Technology\\
        Lexington, MA 02420, U.S.A.\\
        E-mail: \email{Forgy@ll.mit.edu}}
\author{Urs Schreiber \\ Universit{\"a}t Duisburg-Essen \\ Essen, 45117, Germany\\
   E-mail: \email{Urs.Schreiber@uni-essen.de}}
\maketitle

\abstract{
  Differential calculus on discrete spaces is studied in the manner of non-commutative
geometry by representing the differential calculus by an operator algebra 
on a suitable Krein space. The discrete analogue of a (pseudo-)Riemannian metric is 
encoded in a deformation of the inner product on that space, which is the
crucial technique of this paper. We study the general case but
find that drastic and possibly vital simplifications occur when the underlying
lattice is topologically hypercubic, in which case we explicitly construct mimetic analogues of the
volume form, the Hodge star operator, and the Hodge inner product for arbitrary discrete 
geometries.

It turns out that the formalism singles out a 
pseudo-Riemannian metric on topologically hypercubic graphs with respect to which all edges are lightlike.
We study such causal graph complexes in detail and consider some of their
possible physical applications, such as lattice Yang-Mills theory and lattice fermions.
}

\begin{document}

\newpage

\section{Introduction}

Discrete differential geometry is a subject that has been and still is attracting
attention in mathematics 
\cite{DimakisMueller-Hoissen:2002,DimakisMueller-Hoissen:1998,Sullivan:1998}, 
mathematical finance
\cite{Forgy:2002} and most notably in theoretical physics 
\cite{BecherJoos:1982,MuellerHoissen:1996,KanamoriKawamoto:2003}, 
where lattice gauge theories \cite{Langfeld:2002}
and
discretized models of gravity,
motivate the search for a discrete analogue
of the differential geometric nature of Yang-Mills theory and General Relativity.
While the adaption of
differential forms and exterior derivation to discrete spaces is well studied, 
a major stumbling block to a broad \emph{mimetic} translation
of continuum differential geometry to discrete differential geometry has always 
been the problem of how to
carry over the metric aspects of differential geometry, encoded in
notions such as the Hodge inner product of forms
$\bracket{\alpha}{\beta} = \int \alpha \wedge \star \beta$,
the Hodge star operator $\star$
and the exterior coderivative.

The existence of the Hodge inner product is essential
for all physical theories that are not purely topological, and hence
formulations of discrete differential geometry without a 
suitable notion of Hodge inner products remain unsatisfactory.

There is, however, a certain ambiguity of what operation exactly one is
willing to address as Hodge duality in discrete geometry. As with
most other objects known from continuum exterior geometry, the transit
to the discrete setting breaks some of their well known properties
while preserving others. The perhaps most instructive example of this
is the phenomenon that the existence of a nilpotent and graded-Leibnitz
exterior derivative implies that differential forms on discrete spaces
do \emph{not} in general \emph{commute} with 0-forms \cite{DimakisMueller-Hoissen:1994}. 
Therefore, in this case,
one has to choose between having a nilpotent and graded-Leinbnitz exterior
derivative and non-commuing differential forms
or having differential forms which commute with functions but no
nilpotent and graded-Leibnitz exterior derivative.
Of course these choices will reflect
the circumstances encountered in the projected application of the 
discrete formalism and no choice may be generally ``better'' than the
other.

On the other hand, the developments in \emph{Noncommutative Geometry}
\cite{Connes:1994,Varilli:1997} have shown that it proofs to be extremely fruitful to regard
geometry in general from the point of view of quantum physics \cite{FroehlichGrandjeanRecknagel:1996}, 
in the sense that all geometric objects are encoded in terms of 
operators acting on certain graded inner product spaces. In many cases
this perspective fixes the above mentioned ambiguities to some
extent. For instance, from the point of view of Noncommutative
Geometry the exterior derivative (being related to two Dirac operators
\cite{FroehlichGrandjeanRecknagel:1996}) is supposed to be odd-graded and hence in particular must
enjoy the graded Leibniz property -  which implies non-commuting
differential forms. (This is reviewed in detail
in \S\fullref{discrete differential calculus}.)

In the same sense, choices will have to be made about which properties
of the Hodge star are to be preserved in discrete differential geometry
and which are allowed to break. In particular, in this paper it is 
demonstrated that the familiar property
\begin{eqnarray}
  \label{crucial property of Hodge star operator}
  \coextd &=& \pm \star \extd \star^{-1}
\end{eqnarray}
(by which $\star$ relates the exterior derivative $\extd$ to the exterior coderivative
$\coextd$)
is (for general discrete metric ``tensors'', to be defined below)
\emph{incompatible} for instance with the relation
$\star^{-1} = \pm \star$ known from the continuum.

In this paper, discrete differential geometry is investigated from
the point of view of Noncommutative Geometry by equipping the vector space
of discrete differential forms with an inner product and representing
$\extd$ as an operator on the resulting inner product space. This
way a \emph{spectral triple} for the discrete space with $N=2$ Dirac operators
is obtained.

One of the main results of this paper is the general definition
and explicit construction of an operator $\star$ 
which does satisfy \refer{crucial property of Hodge star operator}.
(The abstract construction is in fact valid for all types of generalized
geometries seizable by the means of Noncommutative Geometry, but only
for discrete spaces is the explicit form of this $\star$ derived and
spelled out here.) This goes hand in hand with the construction of
metric geometry on discrete spaces (\cf \cite{DimakisMueller-Hoissen:1998}) 
in terms of deformed inner products
on the vector spaces which the geometric objects are represented on as
operators.

The main idea of the approach to discrete metric geometry and
Hodge duality taken here is briefly the following:

\begin{quote}
Let $\cal A$ be an associative algebra of functions. 
In the present case this will be the (complex or group-valued)
algebra of functions on a denumerable set of points of
a discrete space.

Over $\cal A$ one can consider \emph{differential calculi}
$\Omega\of{\mathcal{A},\extd}$ whose elements are generated by
elements $a \in \mathcal{A}$ (taken to be of grade 0) 
and elements $\extd a$ (of grade 1), 
where $\extd$ is a formal exterior derivative which satisfies $\extd^2 = 0$ 
as well as the graded Leibnitz rule. Elements of grade $p$ in this algebra
are the analogues of differential $p$-forms.

As a substitute for the Hodge inner product
$\bracket{\alpha}{\beta} = \int \alpha\wedge \star \beta$ known from
the continuum, define on $\Omega\of{\mathcal{A},\extd}$(regarded as
a vector space) a non-degenerate (sesquilinear) inner product
$\bracket{\cdot}{\cdot} : \Omega \times \Omega \to \C$. 

The elements
$a\in \mathcal{A}$ of the original algebra as well as $\extd$ are represented on
the resulting inner product space 
$\mathcal{H}\of{\Omega,\bracket{\cdot}{\cdot}}$
as (multiplication and differentiation) operators in the obvious way. Taking
adjoints $(\cdot)^\dagger$ of these operators with respect to $\bracket{\cdot}{\cdot}$
in particular yields $\coextd$, the analogue of the exterior 
\emph{co}derivative.

The information about the metric geometry of the discrete space is
encoded in $\bracket{\cdot}{\cdot}$ and hence inherited by $\coextd$. 
The connection to the familiar
formulation of Noncommutative Geometry is obtained by noting that
$\extd \pm \coextd$ are two odd-graded Dirac operators on $\mathcal{H}$, so
that $(\mathcal{A},\mathcal{H},\extd\pm \coextd)$ is a spectral triple.

Assume for the moment that $\bracket{\cdot}{\cdot}$ is chosen in such a way that
it describes a \emph{flat} discrete space in some sense 
(see \S\fullref{inner product for differential calculi} for details). 
Then this space
may be equipped with discrete analogues of curved metrics by \emph{deforming}
the inner product by means of any invertible, self-adjoint operator $\gop$ as
\begin{eqnarray}
  \bracket{\cdot}{\cdot}_\gop
  &\defas&
  \bra{\cdot}\gop^{-1}\ket{\cdot}
  \,.
\end{eqnarray}
In the special case of topologically hypercubic graphs we show in detail
how the operator $\hat g$ encodes the metric tensor that would be obtained 
in the continuum limit. Not all possible deformation operators $\hat g$
encode purely metric information, though. Some choices correspond
to switching on not gravitational but dilaton and Kalb-Ramond fields
(see below).

The approach outlined here a priori knows only about discrete analogues of 
differential forms, not about vector fields. With a notion of metric in hand,
however, differential forms can be mapped to analoga of vector fields, even in the
discrete setup (see \S\fullref{p-vector fields}).

The formalism then allows to straightforwardly adapt many continuum notions
to the discrete setting, as for instance Lie derivatives: Let 
$v^\sharp \in \Omega^1$ be a discrete 1-form that is dual to a discrete
vector $v$ in the sense indicated above. The usual formula 
(e.g. A.4 of \cite{Schreiber:2003a})
\begin{eqnarray}
  \mathcal{L}_v = \antiCommutator{\extd}{(v^\sharp)^\dagger}
\end{eqnarray}
for the
Lie derivative $\mathcal{L}_v$ along $v$ is directly
applicable to discrete spaces in the framework presented here. 

Now denote by $\star$ an operator on $\mathcal{H}$ 
satisfying \refer{crucial property of Hodge star operator}
and
define a \emph{volume-like form} to be an element $\ket{\rm vol}$ of 
$\Omega\of{\mathcal{A},\extd}$ that is of maximal grade and annihilated by
$\coextd$:
$
  \coextd \ket{\rm vol} = 0
$.

With these definitions the following \emph{fact} holds: 
Given the inner product space 
$\mathcal{H}\of{\Omega,\bracket{\cdot}{\cdot}}$ we have
\begin{eqnarray}
  \star\;\,\mbox{exists}
  \;\;\Leftrightarrow\;\;
  \ket{\rm vol}\;\mbox{exists}
\end{eqnarray}
and they are related by
\begin{eqnarray}
  \label{vol in terms of star}
  \ket{\rm vol} &=& \star \ket{1}
\end{eqnarray}
and
\begin{eqnarray}
  \label{star in terms of vol}
  \star \ket{a_0 \, \extd a_1\cdots \extd a_p}
  &=&
  \left(a_0 \,\extd a_1\cdots \extd a_p\right)^\dagger \ket{\rm vol}
  \,.
\end{eqnarray}

It follows in particular that $\star$ exists iff there is a solution 
$\ket{\rm vol}$ to the equation
$\bracket{\extd \alpha}{\rm vol} = 0, \;\forall\, \alpha \in \Omega\of{\mathcal{A},\extd}$.
In the case of discrete geometry on topologically hypercubic graphs 
this condition can be solved explicitly
for $\ket{\rm vol}$. 
Equation \refer{star in terms of vol} then defines
the Hodge star operator on these spaces.
\end{quote}

This way a discrete Hodge star operator in the sense of
\refer{crucial property of Hodge star operator} can be constructed
for general $\mathcal{A}$, if it exists at all. In the special
case where $\mathcal{A}$ is the algebra of functions on 
topologically hypercubic graphs it is easy to show that
it has the expected continuum limit. The main virtue of formula
\refer{star in terms of vol} is that it tells us precisely how lattice shifts
have to be included in the naive discretization of the Hodge
star in order that \refer{crucial property of Hodge star operator}
remains true. 
The details of this construction are given in 
\S\fullref{inner product for differential calculi}.

It is easy to treat group-valued functions on the discrete space in the same
spirit, which allows to formulate lattice Yang-Mills field theory on
arbitrary background geometries. This
is examined in \S\fullref{noncom function algebras}.\\

Besides this construction of the Hodge star operator, the approach
to discrete differential geometry presented here has several interesting
implications:

For instance one finds that a natural requirement on the anticommutator
of elementary edges with their adjoints that
elements of $\mathcal{A}$ cannot be represented as 
self-adjoint operators on $\mathcal{H}\of{\mathcal{A},\extd}$ 
unless the metric is flat and of Riemannian
signature.
Remarkably, the self-adjoint part of the original function
algebra with respect to a non-flat, non-Riemannian inner product 
$\bracket{\cdot}{\cdot}$ turns out to be a noncommutative
algebra even if $\mathcal{A}$ itself is commutative.

With respect to this fact it is interesting that 
on topologically hypercubic graphs there is a preferred inner product
$\bracket{\cdot}{\cdot}$
which renders all edges lightlike and induces a \emph{pseudo-Riemannian} metric
on the discrete space. A notion of time hence appears naturally in this
framework. 

This and other facts
show that topologically hypercubic spaces with pseuo-Riemannian signature 
(induced by $\bracket{\cdot}{\cdot}$) with time
flowing along the main diagonal of the hypercubes are special. These
spaces are called $n$-\emph{diamonds} here (where $n$ is the dimension)
and receive special attention. In particular using diamond complexes the
preference of our formalism for topologically hypercubic graphs is
turned into a virtue, since these appear as the time evolution
of the simplicial complexes obtained by taking spatial sections of an
$n$-diamond.

There are also close relations of the approach described here to 
the way of obtaining target space background fields by means of
deformations of superconformal field theories, as described in
\cite{Schreiber:2004,Schreiber:2004f}. There the inner product on \emph{loop space}
is deformed in precisely the same general manner as in the present paper
and background fields such as gravity, dilaton fields, Kalb-Ramond and gauge fields
are encoded in the deformation operator.

It follows in particular that, on the discrete space, some
backgrounds encoded by $\bracket{\cdot}{\cdot}$ have no
interpretation in terms of metrics (gravitation) on the discrete space but
have to be interpreted as representing the effect of dilaton and
Kalb-Ramond fields. \\

The structure of this paper is as follows:

In \S\fullref{discrete differential calculus} 
we mainly review some elements of discrete 
differential geometry follwing the work by Dimakis and M{\"u}ller-Hoissen.
\S\fullref{inner product for differential calculi} 
adds the notion of inner product to this
formalism and discusses how deformations of a default inner product 
encodes the geometry of the discrete space. Discrete flat Lorentzian geometries
are found to play a special role and are analyzed in detail in 
\S\fullref{Flat topologically hypercubic complexes}.
Finally \S\fullref{noncom function algebras} discusses aspects of the application of this
formalism to lattice Yang-Mills theory and lattice fermions.

\section{Discrete differential calculus}
\label{discrete differential calculus}

This section reviews mostly well known notions of discrete differential geometry.

\subsection{Differential calculus}
\label{section: differential calculus}

If  ${\cal A}$ is any associative 
(but not necessarily commutative)
algebra (over $\C$) with unit $1$, then
a \emph{differential calculus} over ${\cal A}$ is a
$\Z$-graded associative algebra $\Omega\of{\cal A}$ (over $\C$)
\begin{eqnarray}
  \label{graded form of a differential calculus}
  \Omega\of{\cal A}
  &=&
  \bigoplus\limits_{r \geq 0}
  \Omega^r\of{\cal A}
  \,,
\end{eqnarray}
where
$\Omega^r\of{{\cal A}}$ are $\cal A$-bimodules
(can be multiplied from the left and the right by elements of $\cal A$)
and 
$\Omega^0\of{\cal A} = {\cal A}$, together with a 
$\C$-linear map
\begin{eqnarray}
  \extd : 
  \Omega^r\of{\cal A}
  \longrightarrow
  \Omega^{r+1}\of{\cal A}
  \,,
\end{eqnarray}
which is nilpotent and obeys the graded Leibniz rule:
\begin{eqnarray}
  \label{nilpotentcy and Leibniz of general extd}
  &&\extd^2 \;=\; 0
  \nonumber\\
  &&
  \extd\left(\alpha \beta\right)
  \;=\;
  \left(\extd \alpha\right)\beta + (-1)^r\left(\extd \beta\right)
  \,,\hspace{1cm}\alpha \in \Omega^r\of{\cal A}, \beta\in \Omega\of{\cal A}
  \,.
\end{eqnarray}
(We will furthermore exclusively consider 
differential algebras $\Omega\of{\cal A}$ that are generated by the elements of 
$\Omega^0\of{\cal A}$ and $\extd \Omega^0\of{\cal A}$.)

From these definitions it follows (using $1 \cdot 1 = 1$ and the Leibiz rule) that
\begin{eqnarray}
  \label{general identity is constant}
  \extd 1 &=& 0
  \,,
\end{eqnarray}
and that every element of $\Omega^r\of{\cal A}$ can be written as a
linear combination of monomials $a_0 \extd a_1\;\extd a_2 \cdots \extd a_r$,
on which $\extd$ acts as
\begin{eqnarray}
  \label{action of extd on arbitrary form in diff calc}
  \extd \left(
    a_0 \extd a_1\;\extd a_2 \cdots \extd a_r
  \right)
  &=&
  \extd a_0\; \extd a_1\;\extd a_2 \cdots \extd a_r
  \,.
\end{eqnarray}
In general many of these terms will be equal to zero.

Elements of $\Omega^r\of{\cal A}$ are called \emph{$r$-forms} and 
we will refer to $\extd$ as the
\emph{exterior derivative}\footnote{
  Since for arbitrary $\cal A$ the
  elements of $\Omega^1\of{\cal A}$ do not generate an exterior algebra,
  as will be discussed below, this is slight abuse of language, which 
  is however justified by the fact that we will mostly be interested in
  cases where the 1-forms do anticommute.  
}.
A differential calculus for which there is a natural number $D$ such
that $\Omega^r\of{\cal A} = \left\lbrace
\begin{array}{cc} 0 & r > D\\ \neq 0 & r \leq D\end{array}\right.$ will
be said to have \emph{dimension} $D$.

We will be thinking of $\Omega\of{\cal A}$ as being represented on some vector space
and write 
\begin{eqnarray}
  \label{definition of diff algebra as operator algebra}
 {\hat \Omega}\of{\hat {\cal A}}
  &=&
  \sum\limits_{r \geq 0}
  {\hat \Omega}^r\of{\hat {\cal A}}
\end{eqnarray} 
for the respective graded operator algebra whose elements are of the form
\begin{eqnarray}
  \sum\limits_{r}
  a_{0,r} \commutator{\extd}{a_{1,r}}\commutator{\extd}{a_{2,r}} \cdots \commutator{\extd}{a_{r,r}}
  \hspace{1cm}
  \mbox{for $a_{i,j}\in \hat {\cal A}$}
  \,.
\end{eqnarray}

In particular the algebra can be represented on itself, i.e. on the vector space
\begin{eqnarray}
  {\rm V}\of{\Omega\of{{\cal A}, \extd}}
  &=&
  {\rm span}
  \of{
    \set{
      \ket{\omega} | \omega \in \Omega\of{{\cal A}, \extd} 
    }
  }
\end{eqnarray}
obtained by regarding $\Omega\of{{\cal A},\extd}$ as a vector space, 
by
\begin{eqnarray}
  \hat \extd \ket{\omega} &=& \ket{\extd \omega}
  \nonumber\\
  \hat a \ket{\omega} &=& \ket{a\omega} \hspace{1cm} a\in {\cal A}
\end{eqnarray}
(this is discussed in more detail in \S\fullref{inner product for differential calculi}). 
We will usually omit the ``hats'' $\hat {\cdot}$ on the left hand side.

The choice of any inner product  $\bracket{\cdot}{\cdot}$ on this vector space
defines an adjoint operation $(\cdot)^\dag : \hat\Omega \to \hat \Omega$.
In particular the adjoint $\coextd$ of the original differential $\extd$
is itself a nilpotent differential operator.
The total algebra of operators defined this way will be denoted as
$\hat\Omega_{(2)}\of{{\cal A}, \extd, \coextd}$. It is $\Z$-graded
with
\begin{eqnarray}
  {\rm grad}\of{\hat {\cal A}} &=& 0
  \nonumber\\
  {\rm grad}\of{\extd} &=& +1
  \nonumber\\
  {\rm grad}\of{A^\dag} &=& - {\rm grad}\of{A}
  \,.
\end{eqnarray}

The two operators 
\begin{eqnarray}
  \Dirac_\pm &\defas& \extd \pm \coextd
\end{eqnarray}
are technically \emph{Dirac operators} on 
${\rm V}\of{\Omega\of{{\cal A},\extd}}$. Since there are two of them
and because they square to the same \emph{Laplace operator}
\begin{eqnarray}
  \fatDelta &\defas& \pm \Dirac_\pm^2
\end{eqnarray}
this defines an $N=2$ differential calculus in the sense of
\cite{FroehlichGrandjeanRecknagel:1996,FroehlichGrandjeanRecknagel:1997}.

\subsection{Discrete differential calculi}
\label{discrete differential calculi}

So consider a denumerable set 
\begin{eqnarray}
  \label{the discrete set}
  G &=& \set{g_i | i\in I \subset \N}
\end{eqnarray}
 and
let $\cal A$ be the (commutative) algebra of $\C$-valued functions
over $G$:
\begin{eqnarray}
  {\cal A} \ni f : G &\to& \C
  \nonumber\\
  g_i &\mapsto& f\of{g_i}
  \,.
\end{eqnarray}
This algebra is generated by the set of discrete delta-functions
\begin{eqnarray}
  \delgen{i} : \delgen{i}\of{g_j} &=& \delta_{ij}
  \,,
  \hspace{0.8cm} i,j \in \N
  \,,
\end{eqnarray}
where $\delta_{ij}$ is the usual Kronecker symbol.
From the properties $\delgen{i}\delgen{j} = \delta_{ij}\,\delgen{i}$ 
and $\sum\limits_{i\in I} \delgen{i} = 1$
the trivial but important identities
\begin{eqnarray}
  \label{product rule for discrete deltas}
  &&\delgen{i} \,\extd \delgen{j} \;=\; -(\extd \delgen{i})\, \delgen{j} 
  + 
  \delta_{ij} \,\extd \delgen{i}
  \\
  \label{linear dependence of discrete 1-forms}
  &&
  \sum\limits_{i\in I} \extd \delgen{i} \;=\; 0
\end{eqnarray}
follow by using \refer{nilpotentcy and Leibniz of general extd} 
and \refer{general identity is constant}.\footnote{
Note that the order of the
factors is important, since $\Omega\of{\cal A}$ need not be commutative,
even if $\cal A$ is.
}
In terms of these discrete delta-functions
$\Omega^r\of{\cal A}$ may be spanned by the elements\footnote{
These do not in general constitute a basis.
That they really span the entire algebra can be seen by noting that
$\delgen{i}\extd \delgen{i}$ is linearly dependent,
  \begin{eqnarray}
    \sum\limits_{j\in I} 
    \delgen{ij}
    &=&
    \delgen{i}
    \sum\limits_{j \in I\,,j \neq i}
    \extd \delgen{j}
    \;\equalby{linear dependence of discrete 1-forms}\;
    -\delgen{i}\extd \delgen{i}
    \,,
  \end{eqnarray}
and by taking into account \refer{discrete diff calc basis by concatenation}
and \refer{concatenation of discrete simplices} below.
}
\begin{eqnarray}
  \label{definition of simplex forms}
  \delgen{i_1 i_2 \cdots i_r}
  &\defas&
  \left\lbrace
    \begin{array}{ll}
      \delgen{i_1} \extd \delgen{i_2} \extd \delgen{i_3}\cdots \extd \delgen{i_r}&
      \mbox{for $i_1 \neq i_2\,,i_2 \neq i_3 \cdots i_{r-1}\neq i_r$}\\
      0 &\mbox{otherwise}
    \end{array}
  \right\rbrace
  \,.
\end{eqnarray}
Note that by \refer{product rule for discrete deltas}
\begin{eqnarray}
  \delgen{i_1i_2 \cdots i_r}
  &=&
  \delgen{i1}
  \delgen{i_1i_2 \cdots i_r}
  \delgen{i2}
  \nonumber\\
  &=&
  \delgen{i_1} \extd \delgen{i_2} \extd \delgen{i_3}\cdots \extd \delgen{i_r} \delgen{i_r}
  \,.
\end{eqnarray}
Therefore all these objects can be obtained by concatenation of $\delgen{ij}$:
\begin{eqnarray}
  \label{discrete diff calc basis by concatenation}
  \delgen{i_1 i_2 \cdots i_r}
  &=&
  \delgen{i_1 i_2}
  \,
  \delgen{i_2 i_3}
  \cdots
  \delgen{i_{r-1} i_r}
  \,.
\end{eqnarray}
This makes it easy to check that\footnote{
One checks for $r = 1$ that
\begin{eqnarray}
  \sum\limits_{j\neq i} \left(\delgen{j} \extd \delgen{i} 
    - \delgen{i}\extd \delgen{j}\right)  
  &\equalby{linear dependence of discrete 1-forms}&
  (1-\delgen{i})\extd \delgen{i} - \delgen{i} (-\extd \delgen{i})
  \;=\;
  \extd \delgen{i}
  \,.
\end{eqnarray}
The result for all $r$ follows by induction:
\begin{eqnarray}
  \label{extd action on discrete vertices}
  \extd \delgen{i_1\cdots i_r}
  &=&
  \extd \delgen{i_1}\cdots \extd \delgen{i_{r-1}} \extd \delgen{i_r}
  \nonumber\\
  &\stackrel{\mbox{\tiny hypothesis}}{=}&
  \sum\limits_{j\in I}
  \left(
    \delgen{j \, i_1 i_2 \cdots i_{r-1}}
    -
    \delgen{i_1\,j \,  i_2 \cdots i_{r-1}}
    +
    \cdots
    +
    (-1)^{r-1}
    \delgen{i_1 i_2 \cdots i_{r-n}\, j}
  \right)
  \underbrace{
    \sum\limits_{k\neq i_r} \left(\delgen{k i_r}  - \delgen{i_r k}\right)  
  }_{\equalby{extd action on discrete vertices} \extd \delgen{i_r}}
  \nonumber\\
  &\equalby{concatenation of discrete simplices}&
  \sum\limits_{j\in I}
  \left(
    \delgen{j \, i_1 i_2 \cdots i_{r}}
    -
    \delgen{i_1\,j \,  i_2 \cdots i_{r}}
    +
    \cdots
    +
    (-1)^{r-1}
    \delgen{i_1 i_2 \cdots i_{r-n}\, j,i_r}
    +
    (-1)^r
    \delgen{i_1 i_2 \cdots i_{r-n} i_r\,j}
  \right)
  \,.
\end{eqnarray}
}
\begin{eqnarray}
  \label{action of extd on discrete microscopic forms}
  \extd \delgen{i_1 i_2 \cdots i_r}
  &=&
  \sum\limits_{j\in I}
  \left(
    \delgen{j \, i_1 i_2 \cdots i_r}
    -
    \delgen{i_1\,j \,  i_2 \cdots i_r}
    +
    \cdots
    +
    (-1)^r
    \delgen{i_1 i_2 \cdots i_r\, j}
  \right)
  \,.
\end{eqnarray}
All this shows that the $\delgen{i_1\cdots i_r}$ are related to
the connectivity of the discrete space. In particular
the non-vanishing $\delgen{i j}$
correspond to edges connection vertices $g_i$ and $g_2$ and therefore
induce the structure of an oriented graph on the set $G$. This is
further elaborated on below. 

Different differential calculi on the set $G$ are obtained from the
``universal'' one, for which all the $\delgen{ij}\,, i\neq j$ are 
non-vanishing, by eliminiting some of the corresponding edges.

\subsection{The set of edges and the graph operator.}
\label{set of edges and graph operator}
The interconnection structure of the graph is completely encoded in the
operator $\extd$. This can be made manifest by noting that the sum
of all edges of the graph
\begin{eqnarray}
  \label{definition graph operator}
  {\bf G} &\defas&
  \sum_{i,j\in G}
  \delta_{\set{i,j}}
\end{eqnarray}
acts like $\extd$ when commuted with 0-forms:
\begin{eqnarray}
  \commutator{\bf G}{\delta_{\set i}}
  &=&
  \sum_{j\in G}
  \left(
    \delta_{\set{j,i}}
    -
    \delta_{\set{i,j}}
  \right)
  \nonumber\\
  &\equalby{action of extd on discrete microscopic forms}&
  \extd \delta_{\set{i}}
  \,.
\end{eqnarray}
It follows that (\cf \refer{action of extd on arbitrary form in diff calc})
\begin{eqnarray}
  \label{general extd action in terms of G}
  \extd \left(
    a_0 \extd a_1\;\extd a_2 \cdots \extd a_r
  \right)
  &=&
  \commutator{\bf G}{a_0}\; \extd a_1\;\extd a_2 \cdots \extd a_r
  \,,
\end{eqnarray}
which shows that the set of edges of the graph completely determines the set
of $p$-forms with $p > 1$.

It should be noted, however, that ${\bf G}^p$ (which is
the sum of all ``microscopic'' $p$-forms of the differential
calculus:
$
  {\bf G}^p
  =
    \sum\limits_{i_0,i_1,\cdots,i_p \in I}
    \delta_{\set{i_0,i_1,\cdots,i_p}}
$)
does not in general vanish. But for a large class of ``nice'' graphs it does:

\subparagraph{Definition.}
We say that a graph has \emph{no intermediate edges} iff 
\begin{eqnarray}
  \label{condition on no intermediate edges}
  \delta_{\set{i,j}} \neq 0
  \;\;&\Rightarrow&\;\;
  \delta_{\set{i,k,j}} = 0
  \hspace{1cm}
  \forall\; i,j,k \in I
\end{eqnarray}
and that it has \emph{no opposite edges} iff
\begin{eqnarray}
  \label{condition on no opposite edges}
  \delta_{\set{i,j,i}} = 0
  \hspace{1cm}
  \forall\; i,j \in I
  \,,
\end{eqnarray}
or equivalently iff
\begin{eqnarray}
  \delta_{\set{i,j}} \neq 0
  \;\;&\Rightarrow&\;\;
  \delta_{\set{j,i}} = 0
  \hspace{1cm}
  \forall\; i,j \in I
  \,.
\end{eqnarray}

\subparagraph{Theorem:}
\emph{On graphs without intermediate edges}\footnote{
The bracket $\superCommutator{\cdot}{\cdot}$ is the supercommutator with respect to the
grading $\involution$.}
\begin{eqnarray}
  \label{extd action in terms of G with no intermediate edges}
  \extd \omega
  &=& 
  \superCommutator{{\bf G}}{\omega}
  \hspace{1cm}
  \forall\; \omega\in \Omega\of{\cal A}
  \,.
\end{eqnarray}
{\it Proof:}
This immediately follows from the general formula 
\refer{action of extd on discrete microscopic forms}.\footnote{
Alternatively we can argue as follows:

First note that in the absence of intermediate edges we have
\begin{eqnarray}
  \label{lemma for extd action in terms of G with no intermediate edges}
  \delta_{\set{i,j}}\neq 0
  \;\;&\Rightarrow&\;\;
  \delta_{\set{i}}
  \;{\bf G}\;
  \extd \delta_{\set{\vec j}}
  \;=\;
  -
  \delta_{\set{i}}
  \extd \delta_{\set{\vec j}}
  \;{\bf G}
  \;=\;
  -\sum\limits_{k\in I}\delta_{\set{ijk}}
  \,,
\end{eqnarray}
because the left hand side can be rewritten as
\begin{eqnarray}
  \delta_{\set{i}}
  \;{\bf G}\;
  \extd \delta_{\set{\vec j}}
  &\stackrel{\refer{definition graph operator}\refer{action of extd on discrete microscopic forms}}{=}&
  \sum\limits_{l \neq i \atop k \neq j}
  \delta_{\set{i,l}}
  \left(
    \delta_{\set{k,j}}
    -
    \delta_{\set{j,k}}
  \right)
  \nonumber\\
  &=&
  -
  \sum\limits_{k\in I}
  \delta_{\set{ijk}}
  \,,
\end{eqnarray}
where in the last step it has been used that the absence of 
intermediate edges implies $\delta_{\set{i,k,j}} = 0$.

Now we use induction on the grade ${\rm grad}\of{\omega}$ of $\omega$. For
${\rm grad}\of{\omega} = 0$ \refer{extd action in terms of G with no intermediate edges} 
is the same as \refer{general extd action in terms of G}.
For ${\rm grad}\of{\omega} = 1$ we have
\begin{eqnarray}
  \extd \left(\delta_{\set{i}} \extd \delta_{\set{j}}\right)
  &\equalby{general extd action in terms of G}&
  \commutator{\bf G}{\delta_{\set{i}}}
  \,
  \extd \delta_{\set{j}}
  \nonumber\\
  &=&
  {\bf G}\; \delta_{\set{i}}
  \,
  \extd \delta_{\set{j}}
  -
  \delta_{\set{i}}\; {\bf G} 
  \;
  \extd \delta_{\set{j}}
  \nonumber\\
  &\equalby{lemma for extd action in terms of G with no intermediate edges}&
  {\bf G}\; \delta_{\set{i}}
  \,
  \extd \delta_{\set{j}}
  +
  \delta_{\set{i}}  
  \,
  \extd \delta_{\set{j}}
  \;  
  {\bf G}
  \nonumber\\
  &=&
  \superCommutator
  {\bf G}
  {\delta_{\set{i,j}}}
  \,.
\end{eqnarray}
The graded Leibniz rule now allows to extend this to higher grades:
Let $\lambda$ be any 1-form, then
\begin{eqnarray}
  \extd\left(\lambda \omega\right)
  &=&
  (\extd \lambda)\omega - \lambda (\extd \omega)
  \nonumber\\
  &=&
  \superCommutator{\bf G}{\lambda}\omega - \lambda \superCommutator{\bf G}{\omega}
  \nonumber\\
  &=&
  \superCommutator{\bf G}{\lambda \omega}
  \,. 
\end{eqnarray}
}
\endofproof

This gives rise to some useful corollaries:

\subparagraph{Corollary I:}
{\it On graphs without intermediate edges}
\begin{eqnarray}
  \label{G^2 supercommutes with everything}
  \superCommutator{{\bf G}^2}{\omega}
  &=& 0
  \hspace{1cm}
  \forall\; \omega \in \Omega\of{\cal A}
  \,.
\end{eqnarray}
{\it Proof:}
This follows from
\begin{eqnarray}
  0 \;=\; \extd \extd \omega
  &\equalby{extd action in terms of G with no intermediate edges}&
  \superCommutator{\bf G}{\superCommutator{\bf G}{\omega}}
  \nonumber\\
  &=&
  \superCommutator{\superCommutator{\bf G}{\bf G}}{\omega}
  -
  \superCommutator{\bf G}{\superCommutator{\bf G}{\omega}}
  \,.
\end{eqnarray}
\endofproof

\subparagraph{Corollary II:}
{\it On graphs without intermediate edges}
\begin{eqnarray}
  \label{sums of paths with the same endpoints vanish}
  \sum\limits_{k\in I}
  \delta_{\set{i,k,j}}
  &=& 0
  \hspace{1cm}
  \forall\; i\neq j\in I
\end{eqnarray}
and
\begin{eqnarray}
  {\bf G}^2
  &=&
  \sum\limits_{i,j\in I}
  \delta_{\set{i,j,i}}
  \,.
\end{eqnarray}
{\it Proof}:
Let $i\neq j \in I$. Then
\begin{eqnarray}
  0 &\equalby{G^2 supercommutes with everything}& 
   \delta_{\set{i}}
  \commutator{{\bf G}^2}{\delta_{\set{j}}}
  \nonumber\\
  &=&
  \sum\limits_{k\in I}
  \delta_{\set{i,k,j}}
  \,.
\end{eqnarray}
\endofproof

In words this means that 
on graphs without intermediate edges
the sum of all ``paths of edges'' $\delta_{\set{i,k,j}}$ that connect the same
two points has to vanish. So in particular if there is only one such path from $i$
to $j$ then the corresponding 2-form $\delta_{\set{i,k,j}}$ is not an element of the calculus.
If there are precisely two paths of two edges connecting two nodes $i$ and $j$ then one
is $-1$ times the other.

\begin{figure}
\begin{center}
\begin{picture}(200,200)
\includegraphics{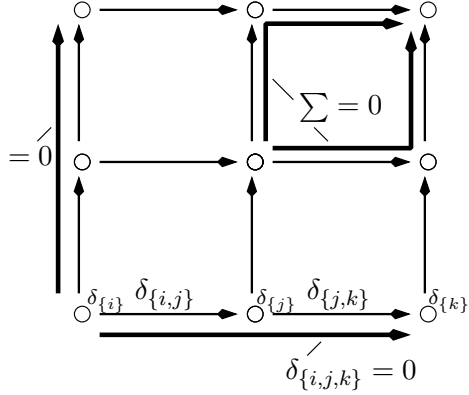}
\put(-53,92){$\sum = 0$}
\put(-58,-8){$\delta_{\set{i,j,k}}=0$}
\put(-163,73){$=0$}
\put(-133,22){$_{\delta_{\set{i}}}$}
\put(-69,22){${}_{\delta_{\set{j}}}$}
\put(-4,22){${}_{\delta_{\set{k}}}$}
\put(-115,22){$\delta_{\set{i,j}}$}
\put(-50,22){$\delta_{\set{j,k}}$}
\end{picture}
\end{center}
\caption{{\it Graph and complex.}}
\label{cubicgraph}
\end{figure}

$\,$\\

Figure \ref{cubicgraph} shows a discrete space consisting of nine nodes. Its edges are indicated
by thin arrows. In this particular example there are no
intermediate edges and no opposite edges in the sense of 
\refer{condition on no intermediate edges} and 
\refer{condition on no opposite edges}.
Four selected 2-forms are indicated by bold arrows. Two of them vanish, because they are the
unique connection between their endpoints. The two 2-forms in the upper right
quadrant are each non-vanishing and sum to zero. 

\begin{figure}[h]
\hspace{3.5cm}
\label{discdimension}
\begin{picture}(0,100)
\includegraphics{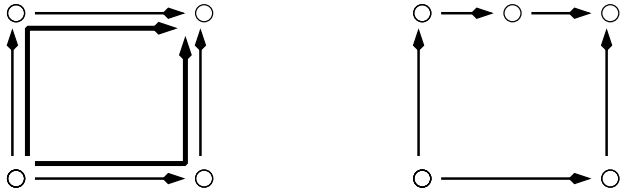}
\end{picture}
\caption{{\it Similar graphs with different dimension}}
\label{discdimension}
\end{figure}

Figure \ref{discdimension} demonstrates the effect of $\bf G$ on the \emph{dimension} of the
discrete space:
Again according to \refer{sums of paths with the same endpoints vanish}
the complex on the left contains two non-vanishing 2-forms,
but no 3-form, and is therefore 2-dimensional, while the complex on the
right contains 1-forms, but not a single 2-form. It is therefore 1-dimensional. 

\subparagraph{Corollary III:}
{\it On graphs without intermediate edges for which}
\begin{eqnarray}
  \sum\limits_{j\in I}
  \delta_{\set{i,j,i}}
  &=&
  0
  \hspace{1cm}
  \forall\; i\in I
\end{eqnarray}
{\it we have}
\begin{eqnarray}
  {\bf G}^2 &=& 0
  \,.
\end{eqnarray}
This implies in particular that
{\it on graphs without intermediate edges \emph{and} without opposite edges}
\begin{eqnarray}
  {\bf G}^2 &=& 0
  \,.
\end{eqnarray}

This corollary requires a remark: When 
${\bf G}^2 = 0$
and
$\Omega\of{{\cal A}}$
is regarded as an operator algebra $\hat\Omega\of{\hat {\cal A}}$,
as discussed in \S\fullref{section: differential calculus},
then
\begin{eqnarray}
  &&\superCommutator{\extd}{\omega}
  \;=\;
  \superCommutator{{\bf G}}{\omega}
  \nonumber\\
  &&
  \extd^2 = 0 = {\bf G}^2
\end{eqnarray}
might seem to indicate that $\extd$ and ${\bf G}$ are in fact the same object. 
This however, is not true, as becomes apparent when the vector space on which
these operators act is explicitly constructed in 
\S\fullref{vector space VOmegaA and inner products}: $\extd$ annihilates
the ``vacuum'' $\ket{1}$ while ${\bf G}$ creates a 1-form.
Also $\extd$ has a continuum limit, while ${\bf G}$ does not
(\cf the discussion concerning \refer{graph operator in terms of onb creators}).

\subparagraph{Corollary IV:}
{\it On graphs without intermediate edges}
\begin{eqnarray}
  \sum\limits_{k\in I}
  \delta_{\set{i,k,i}}
  \delta_{\set{i,j}}
	  &=& 0
  \nonumber\\
  \delta_{\set{i,j}}
  \sum\limits_{k\in I}
  \delta_{\set{j,k,j}}
  &=& 0
  \hspace{1cm}
  \forall\;
  i,j\in I
  \,.
\end{eqnarray}

\subsection{Group structures}
	
In order to get a handle on the interconnectedness of the points
$g_i \in G$ of the discrete space it is helpful to consider some
additional structure: In the standard application one is studying
fields on a lattice structure, thus having a notion of translation
along the lattice vectors. Since this amounts to making use of
the additive group structure of the lattice it is sufficient to
assume more generally that the set $G$ is equipped with a group
multiplication $g_i g_j = g_k\,, i,j,k \in I$, which
allows to ``go from $g_i$ in the direction $g_j$ and arrive at $g_k$''
on the discrete space.

A brief introduction to 
differential calculus on discrete groups can be found
in \S4.1 of \cite{DimakisMueller-Hoissen:1994}. A detailed discussion
is given in the series of papers
\cite{DimakisMueller-Hoissen:2002a,DimakisMueller-Hoissen:2002}.

The left and the right action of the group on functions of its
elements will be denoted by
\begin{eqnarray}
  (R_g f)\of{g^\prime}
  &\defas&
  f\of{g^\prime g}
  \nonumber\\
  (L_g f)\of{g^\prime}
  &\defas&
  f\of{g g^\prime}
  \,.
\end{eqnarray}
Every group element gives rise to the two 1-forms
\begin{eqnarray}
  \label{coordinate 1 forms on group lattice}
  \theta^g_{\rm L}
  &\defas&
  \sum\limits_{g^\prime \in G}
  \delgen{g^\prime} \extd \delgen{g^\prime g}
  \nonumber\\
  \theta^g_{\rm R}
  &\defas&
  \sum\limits_{g^\prime \in G}
  \delgen{g^\prime} \extd \delgen{g g^\prime}
  \,,
\end{eqnarray}
which are obviously left (right) invariant:
\begin{eqnarray}
  R_{g^\prime} \theta^g_{\rm R} &=& 0
  \nonumber\\
  L_{g^\prime} \theta^g_{\rm L} &=& 0
  \,.
\end{eqnarray}
They are related to the ``simplex-forms'' \refer{definition of simplex forms} 
by\footnote{
This is readily seen to be true for $r = 2$. For general $r$ it follows 
using \refer{discrete diff calc basis by concatenation}.
}
\begin{eqnarray}
  \delgen{g_1,g_2,\cdots g_r}
  &=&
  \delgen{g_1}
  \theta_{\rm L}^{g_1^{-1}g_2}\theta_{\rm L}^{g_2^{-1}g_3}
  \cdots
  \theta_{\rm L}^{g_{r-1}^{-1} g_r}
  \nonumber\\
  &=&
  \delgen{g_1}
  \theta_{\rm R}^{g_2 g_1^{-1}}\theta_{\rm R}^{g_3 g_2^{-1}}
  \cdots
  \theta_{\rm R}^{ g_r g_{r-1}^{-1}}
  \,.
\end{eqnarray}
The $\theta^g$ are useful, because in terms of them the
exterior derivative has the suggestive action\footnote{
We have
\begin{eqnarray}
  \label{derivation of dicsrete df}
  \extd f
  &=&
  \sum\limits_{h} f\of{h}\extd \delta_{\set{h}}
  \nonumber\\
  &\equalby{linear dependence of discrete 1-forms}&
  \sum\limits_{h\in G, g\neq e} 
  f\of{h}
  \left(
   \delta_{\set{h g^{-1}}}
  \extd \delta_{\set{h}}
  - 
   \delta_{\set{h}}
  \extd \delta_{\set{h g}}
  \right)
  \;=\;
  \sum\limits_{h\in G, g\neq e} 
  f\of{h}
  \left(
   \delta_{\set{g^{-1} h }}
  \extd \delta_{\set{h}}
  - 
   \delta_{\set{h}}
  \extd \delta_{\set{gh }}
  \right)
  \nonumber\\
  &=&
  \sum\limits_{g\neq e}
  \left(
  \left(
    R_g -1
  \right)
  f
  \right)
  \theta_L^g
  \;=\;
  \sum\limits_{g\neq e}
  \left(
  \left(
    L_g -1
  \right)
  f
  \right)
  \theta_R^g
  \,.
\end{eqnarray}
Alternatively we can use the identities
\begin{eqnarray}
  \label{extd of discrete delta in terms of discrete MC forms}
  \extd \delgen{g}
  &=&
  \sum\limits_{g^\prime \in G}
  \delgen{g g^{\prime -1}}
  \theta_{\rm L}^{g^\prime}
  \nonumber\\
  &=&
  \sum\limits_{g^\prime \in G}
  \delgen{g^{\prime -1}g}
  \theta_{\rm R}^{g^\prime}
\end{eqnarray}
and
\begin{eqnarray}
  \sum\limits_{g\neq e} \theta_{\rm L}^g
  &=&
  \sum\limits_{g^\prime \in G} \delta_{\set{g^\prime}}
  \sum\limits_{g\neq e} \extd \delta_{\set{g^\prime g}}
  \nonumber\\
  &\equalby{linear dependence of discrete 1-forms}&
  -\sum\limits_{g^\prime \in G} \delta_{\set{g^\prime}}
  \extd \delta_{\set{g^\prime}}
  \nonumber\\
  &=&
  -\theta_{\rm L}^e    
\end{eqnarray}
and write
\begin{eqnarray}
  \extd f
  &=&
  \sum\limits_{g}
  f\of{g}
  \extd \delgen{g}
  \nonumber\\
  &\equalby{extd of discrete delta in terms of discrete MC forms}&
  \sum\limits_{g,g^\prime}
  f\of{g}
  \delgen{g g^{\prime -1}}
  \theta_{\rm L}^{g^\prime}
  \;=\;
  \sum\limits_{g,g^\prime}
  f\of{g}
  \delgen{g^{\prime -1} g }
  \theta_{\rm R}^{g^\prime}    
  \nonumber\\
  &=&
  \sum\limits_{g,g^\prime}
  f\of{g g^\prime}
  \delgen{g }
  \theta_{\rm L}^{g^\prime}
  \;=\;
  \sum\limits_{g,g^\prime}
  f\of{g^\prime g}
  \delgen{g }
  \theta_{\rm R}^{g^\prime}    
  \nonumber\\
  &=&
  \sum\limits_{g^\prime}
  (R_{g^\prime} f)
  \theta_{\rm L}^{g^\prime}
  \;=\;
  \sum\limits_{g}
  (L_{g^\prime} f)
  \theta_{\rm R}^{g^\prime}        
  \nonumber\\
  &=&
  \sum\limits_{g \neq e}
  \left((R_g  - 1)f\right)
  \theta_{\rm L}^{g}
  \;=\;
  \sum\limits_{g \neq e}
  \left((L_g  - 1)f\right)
  \theta_{\rm R}^{g^\prime}          
  \,.
\end{eqnarray}
}
\begin{eqnarray}
  \extd f
  &=&
  \sum\limits_{g \neq e}
  \left((R_g  - 1)f\right)
  \theta_{\rm L}^{g}
  \;=\;
  \sum\limits_{g \neq e}
  \left((L_g  - 1)f\right)
  \theta_{\rm R}^{g^\prime}          
\end{eqnarray}
on functions $f\in \Omega^0\of{\cal A}$. 
\begin{eqnarray}
  \extd \theta_{\rm L}^g
  &=&
  \sum\limits_{g^\prime}
  \theta_{\rm L}^{g^\prime}\theta_{\rm L}^{g g^{\prime -1}}
\end{eqnarray}
A further interesting property of the forms $\theta^g$ 
(and indeed of the discrete calculus as a whole)
is that they do not
commute with functions, but instead induce translations on them:
\begin{eqnarray}
  f \theta_{\rm L}^g &=& \theta_{\rm L}^g (R_g f)
  \nonumber\\
  f \theta_{\rm R}^g &=& \theta_{\rm R}^g (L_g f)
  \,,\hspace{0.8cm}g \neq e
  \,.  
\end{eqnarray}

Different differential calculi on the set $G$ are obtained from the
``universal'' one, for which all the $\delgen{ij}\,, i\neq j$ are 
non-vanishing, by eliminiting some of the corresponding edges.

In the case where $G$ has a group structure this leads to
the concept of a
\emph{Cayley graph} or \emph{group lattice} $(G,T)$,
which is defined by the 
condition that between any two elements $g$, $g^\prime$ of $G$
there is an oriented edge $g\to g^\prime$ (equivalently: a non-vanishing
1-form $\delgen{g, g^\prime}$)
precisely if $g^{-1}g^\prime \in T$, where 
$T \subset G$ is a subset not containing the identity, $e \not\in T$.
Thus $T$ is a discrete analogue of the tangent space at a point of
the discrete space. 

\subsection{The dual approach: Paths, chains, and boundaries}

The previous sections concerned the discretization of the calculus of differential forms.
Integrals of differential forms are \emph{cochains}, the dual elements to
\emph{chains}. In \S\fullref{inner product for differential calculi} 
it is shown how cochains, chains, and their pairing
is recovered from within the the differential operator calculus. Alternatively chains
may be introduces axiomatically and the differential calculus on their dual space
be obatined as a derived entity. This is sketched in the following:

The starting point is of course again the denumerable set 
\refer{the discrete set}
$G = \set{g_i| i \in I \subset \N}$
of discrete \emph{nodes} (points) of the discrete space. 

This is given the structure of a graph by defining the vector space $P_1$ 
over the set of edges, or \emph{1-paths}, which contains an 
element $\path{i,j}$ if there is supposed to be a directed edge
from node $g_i$ to node $g_j$. 

Now let a \emph{$p$-path} be the concatenation of $p$ edges,
\begin{eqnarray}
  P_p &\defas& {\rm span}\of{\set{ \path{i_1, i_2, \cdots, i_p} | \path{i_n,i_{n+1}} \in P_1 }}
  \hspace{1cm}
  \mbox{for $p> 1$}
  \,,
\end{eqnarray}
and let $P_0$ be the vector space over the set of $0$-paths:
\begin{eqnarray}
  P_0 &\defas& {\rm span}\of{\set{ \path{i} | g_i \in G }}
  \,.
\end{eqnarray}
For convenience we set
\begin{eqnarray}
   \label{some paths may vanish}
  \path{\cdots, i_n,i_{n+1},\cdots} = 0
  &\Leftrightarrow&
  \path{i_n,i_{n+1}} \not\in P_1
  \,.
\end{eqnarray}
On the space
\begin{eqnarray}
  P &\defas&
  \bigoplus\limits_{p=0}^\infty
  P_p
\end{eqnarray}
of all paths there is an associative, non-commutative and $\R$-linear product operation 
\begin{eqnarray}
  \cupprod : P\times P  &\to& P
\end{eqnarray}
which concatenates paths that fit together:
\begin{eqnarray}
  \label{definition cup product on paths by concatenation}
  \path{i_1,\cdots,i_p} \cupprod \path{j_1, \cdots, j_q} &\defas&
  \left\lbrace
    \begin{array}{ll}
      \path{i_1,\cdots,i_p, j_2 \cdots, j_q} & \mbox{if $i_p = j_1$}\\
      0 & \mbox{otherwise}
    \end{array}
  \right.
\end{eqnarray}
(and linearly extended to all of $P$), called the \emph{cup product}.

We want to identify such paths that can be associated with $p$-dimensional cells of
the graph, thus turning the graph into a cell complex. A chain (of cells) shall be something which
has a \emph{boundary} (being another chain), such that a boundary does not have any
non-vanishing boundary itself. 

In order to model something of this sort on the space of paths first consider the
\emph{universal} case where all possible 1-paths $\path{i,j}\,,i,j\in I$ are non-
vanishing. In this case the \emph{boundary operator}
\begin{eqnarray}
  \partial : P &\to& P
\end{eqnarray}
defined by extending the relation
\begin{eqnarray}
  \label{universal action of partial}
  \partial \path{i_1,i_2,\cdots,i_p}
  &\defas&
  \path{i_2,i_3,\cdots, i_p}
  -
  \path{i_1,i_3,\cdots, i_p}
  \cdots -
  (-1)^p
  \path{i_1,i_3,\cdots, i_{p-1}}
\end{eqnarray}
linearly to all of $P$ is obviously nilpotent:
\begin{eqnarray}
  \partial^2 &=& 0
  \,.
\end{eqnarray}
The tuple
\begin{eqnarray}
  \Omega^\ast\of{P_0, \partial}
\end{eqnarray}
could be called a ``dual differential calculus'' over $P_0$ and the case where all
possible paths are non-vanishing would be the \emph{universal dual differential calculus}.

But recall from 
\refer{some paths may vanish} that not all elements in the sum \refer{universal action of partial}
need to be non-vanishing. (The only ones that are guaranteed to be non-vanishing
are the first and the last term in the sum.) 
In such cases $\partial$ won't be a nilpotent operator on $P$.
In order for $\partial$ still to act as 
a boundary operator one therefore needs to restrict it to the subspace $C$ of $P$
on which it happens to be nilpotent: This $C$ is the space of \emph{chains}:
\begin{eqnarray}
  C &\defas& \bigoplus\limits_p C_p
  \nonumber\\
  C_p &\defas& {\rm ker}\of{(\partial^2)_{|P_p}}
  \,.
\end{eqnarray}
The cup product \refer{definition cup product on paths by concatenation} obviously
does not respect $C$. It can be turned into a proper product on $C$ simply by 
projection onto C:
\begin{eqnarray}
  \cdot \;:\; C &\to& C
  \nonumber\\
  S_p \cdot T_q &\mapsto& \frac{p!q!}{pq}{\cal P}_C\of{S \cupprod T}
  \,.
\end{eqnarray}
(Here ${\cal P}_C : P \to P$ is the projector onto $C$.) 

The space $C^\ast$ dual to the space of ($p$-)chains is that of ($p$-)cochains:
\begin{eqnarray}
  C^\ast = \bigoplus\limits_p C^{\ast p}
  &\defas&
  \set{
    \bra{\fatomega}: C \to \R\,, |\,\bra{\fatomega}\of{\ket{S}} = \bracket{\fatomega}{S}\,,\; S\in C
  }
  \,.
\end{eqnarray}
Any operator $A : C \to C$ defines an operator $A^{\rm t} : C^\ast \to C^\ast$, called the
\emph{transpose} of $A$ by
\begin{eqnarray}
  \bracket{A^{\rm t}\fatomega}{S}
  &\defas&
  \bracket{\fatomega}{A S}
  \,.
\end{eqnarray}
The transpose of the boundary operator
\begin{eqnarray}
  \extd 
  &\defas&
  \partial^{\rm t}
\end{eqnarray}
is called the \emph{cobaundary operator}. It is obviously nilpotent
\begin{eqnarray}
  \extd^2 &=& 0
  \,.
\end{eqnarray}
We will now show that it indeed induces a differential calculus on $C^\ast$:

For the elements dual to $0$-paths write
\begin{eqnarray}
  \label{dual 0-paths}
  \bracket{\delgen{i}}{j}
  &\defas&
  \bracket{i}{j}
  \nonumber\\
  &\defas&
  \delta_{ij}
\end{eqnarray}
and for those dual to 1-paths similarly
\begin{eqnarray}
  \bracket{\delgen{ii'}}{j'j}
  &\defas&
  \bracket{ii'}{j'j}
  \nonumber\\
  &\defas&
  \delta_{i,j}\delta_{i',j'}
  \,.
\end{eqnarray}
From
\begin{eqnarray}
  \sum\limits_{i\in I} \bracket{\extd\delgen{i}}{ij}
  &=& 
  \sum\limits_{i\in I}
  \langle{\delgen{i}}|\partial \ket{jk}
  \nonumber\\
  &=&
  0  
\end{eqnarray}
it follows that
\begin{eqnarray}
  \sum\limits_{i\in I}
  \bra{\extd \delgen{i}}
  &=& 0
  \,.
\end{eqnarray}
Furthermore
\begin{eqnarray}
  \bracket{\extd \delgen{k}}{ij}
  &=&
  \langle{\delgen{k}}|\partial \ket{ij}
  \nonumber\\
  &=&
  \bracket{\delgen{k}}{i} - \bracket{\delgen{k}}{j}
  \nonumber\\
  &=&
  \delta_{ki} - \delta_{kj}
\end{eqnarray}
implies that
\begin{eqnarray}
  \label{boundary approach extd of a delta-function}
  \bra{\extd \delgen{k}}
  &=&
  \sum\limits_{j\in I \atop j \neq k}
  \bra{\delgen{jk}
  -
  \delgen{jk}}
  \,.
\end{eqnarray}
Now for $\bra{\fatalpha}$ any element of $C^\ast$ we \emph{define} the products
$\fatalpha \cdot \delgen{i}$ and $\delgen{i}\cdot \fatalpha$ by
\begin{eqnarray}
  \label{product on C* from product on C}
  \bracket{
    \fatalpha \cdot \delgen{i}
  }
  {
    S
  }  
  &\defas&
  \bracket{
    \fatalpha
  }
  {
    i \cdot S
  }
  \nonumber\\
  \bracket{
    \delgen{i}\cdot
    \fatalpha 
  }
  {
    S
  }  
  &\defas&
  \bracket{
    \fatalpha
  }
  {
    S \cdot i
  }
  \,.
\end{eqnarray}
In particular this implies
\begin{eqnarray}
  \delgen{i}\cdot \delgen{j} &=& \delta_{i,j}\delgen{i}
  \,.
\end{eqnarray}
Furthermore it follows from \refer{boundary approach extd of a delta-function} that
\begin{eqnarray}
    \delgen{ij}  
  &=&
  -
    (\extd \delgen{i})\cdot \delgen{j}
  \nonumber\\
  &=&
  \delgen{i}\cdot (\extd \delgen{j})
\end{eqnarray}
and that
\begin{eqnarray}
  (\extd \delgen{i})\cdot \delgen{i} &=&
  \sum\limits_{j\neq i}
  \delgen{ji}
  \nonumber\\
   \delgen{i}\cdot
  (\extd \delgen{i})  &=&
  -\sum\limits_{j\neq i}
  \delgen{ij}  
\end{eqnarray}
which together gives the Leibniz rule on $0$-cochains:
\begin{eqnarray}
  \extd(\delgen{i}\cdot\delgen{j})
  &=&
  (\extd \delgen{i})\cdot\delgen{j}
  +
  \delgen{i}\cdot
  (\extd \delgen{j})
  \,.
\end{eqnarray}

\subparagraph{Dual paths.}
The objects
\begin{eqnarray}
  \delgen{i_1,i_2,\cdots i_p}
  &\defas&
  \delgen{i_1}\cdot
  \left(
  \extd\left(\delgen{i_2}\cdot \left(\extd\left(\delgen{i_2}\cdot \left(\extd\left(\cdots \delgen{i_p}\right)\right)\right)
  \right) \right)\right)
  \,,
  \hspace{1cm}
  i_1\neq i_2\,, i_2 \neq i_3, \cdots
\end{eqnarray}
constitute, up to a sign, the dual basis to the path basis, because
\begin{eqnarray}
  \bracket{\delgen{i_1,\cdots i_p}}{i_p^\prime \cdots i_1^\prime}
  &=&
  -(-1)^{p(p+1)/2}\;
  \delta_{i_1,i_1^\prime}\cdots \delta_{i_p,i_p^\prime}
  \,.
\end{eqnarray}

{\it Proof:}
We use induction over $p$. For $p=1$
\begin{eqnarray}
  \bracket{\delgen{i}}{i^\prime} &=&  \delta_{i,i^\prime}
\end{eqnarray}
by definition \refer{dual 0-paths}.
Then
\begin{eqnarray}
  \bracket{
  \delgen{i_1}\cdot \left(\extd \delgen{i_2,\cdots,i_p} \right)
  }{
    (i_p^\prime\cdots i_2^\prime i_1^\prime)
  }
  &\equalby{product on C* from product on C}&
  \delta_{i_1, i_1^\prime}
  \bracket{
  \extd \delgen{i_2,\cdots,i_p}
  }{
    (i_p^\prime\cdots i_2^\prime i_1^\prime)
  }  
  \nonumber\\
  &\equalby{universal action of partial}&
  \delta_{i_1,i_1^\prime}
  \bracket{
  \delgen{i_2,\cdots,i_p}
  }{
    (i_{p-1}^\prime\cdots i_2^\prime i_1)
    -
    (i_p^\prime,i_{p-2}^\prime\cdots i_2^\prime i_1)
    -
    (-1)^p(i_p^\prime\cdots i_2^\prime)
  }    
  \nonumber\\
  &\stackrel{i_1\neq i_2}{=}&
  -(-1)^{p}
  \delta_{i_1,i_1^\prime}
  \bracket{\delgen{i_2,\cdots,i_p}}
  {i_p,\cdots, i_2}
  \nonumber\\
  &\stackrel{\mbox{\tiny hypothesis}}{=}&
  -(-1)^{p}
  \delta_{i_1,i_1^\prime}
  (-1)^{p-1}
  \delta_{i_2,i_2^\prime}
  \bracket{\delgen{i_3,\cdots,i_p}}
  {i_p,\cdots, i_3}
  \nonumber\\
  &=&
  \cdots
  \nonumber\\
  &=&
  -
  (-1)^{p(p+1)/2}\delta_{i_1,i_1^\prime}\cdots\delta_{i_p,i_p^\prime}  
  \,.
\end{eqnarray}
\endofproof

\subparagraph{Coboundary of dual paths.}
It follows that
\begin{eqnarray}
  \label{extd of dual paths}
  \extd \delgen{i_1,i_2,\cdots,i_p}
  &=&
  \sum\limits_{j\in I}
  \left(
    \delgen{j,i_1,\cdots,i_p}
    -
    \delgen{i_1,j,i_2,\cdots,i_p}
    +
    \cdots
    +
    (-1)^p
    \delgen{i_1,i_2,\cdots,i_p,j}
  \right)
  \,.
\end{eqnarray}

{\it Proof:}
By shifting $\extd$ over to the other side
\begin{eqnarray}
  \bracket{\extd \delgen{i_1,\cdots i_p}}{j_{p+1},\cdots , j_1}
  &=&
  \bra{\delgen{i_1,\cdots i_p}}\partial\ket{j_{p+1},\cdots , j_1}
  \nonumber\\
  &=&
  \bracket{\delgen{i_1,\cdots i_p}}
  {j_{p},\cdots , j_1}
  -
  \bracket{\delgen{i_1,\cdots i_p}}
  {j_{p+1},j_{p-1}\cdots , j_1}
  \cdots
  -
  (-1)^p
  \bracket{\delgen{i_1,\cdots i_p}}
  {j_{p+1},\cdots , j_2}
  \nonumber\\
  &=&
  \cdots
  -
  (-1)^{(p+1+1)(p+1)/2}
  \delta_{i_p,j_{p+1}}
  \cdots
  \delta_{i_1,j_2}
\end{eqnarray}
one obtains the same result as from \refer{extd of dual paths}.
\endofproof

Next we again \emph{define} the product of two dual paths by
\begin{eqnarray}
  \delgen{i_1,\cdots, i_p}\cdot \delgen{j_1,\cdots,j_q}
  &\defas&
  \delta_{i_p,j_2}
  \delgen{i_1, \cdots,i_p,j_2,\cdots,j_q}
  \,.
\end{eqnarray}
It follows from \refer{extd of dual paths}
that $\extd$ is graded-Leibniz on the resulting algebra:
\begin{eqnarray}
  \extd (\left(\delgen{i_1,\cdots,i_p}\delgen{j_1,\cdots,j_q}\right))
  &=&
  \left(\extd \delgen{i_1,\cdots,i_p}\right)\delgen{j_1,\cdots,j_q}
  +
  (-1)^p
  \delgen{i_1,\cdots,i_p}
  (\extd
  \delgen{j_1,\cdots,j_q}
  )
  \,.
\end{eqnarray}

Identifying $C^{\ast 0}$ with $\cal A$ and the ``$\cdot$''-product with the implicit algebra product in
\S\fullref{discrete differential calculi}
one thus finds that indeed 
\begin{eqnarray}
  C^\ast &\isomorphic& \Omega\of{{\cal A},\extd}
\end{eqnarray}
is a differential calculus over $\cal A$.

\paragraph{Simplices.}

Ordinarily the study of discrete geometry is concerned with simplices. The space $P$ of
paths that has been introduced above, together with the boundary map $\partial$,
actually provides a generalization which contains simplicial geometry as a subcase.

\subparagraph{Definition:} 
Any $0$-path $\ket{i_1}$ is a $0$-simplex.
A $p$-\emph{simplex} is a $(p+1)$-path 
$\ket{i_1,\cdots ,i_{p+1}}$ such that every sub-path
$\ket{i_1,\cdots,i_{n-1},i_{n+1},\cdots, i_{p+1}}$ is non-vanishing and a simplex itself.

In other words a simplex is a path all whose nodes are pairwise connected by $1$-paths.

Let
\begin{eqnarray}
  S &=& \bigoplus\limits_{p=0,1,2,\cdots} S_p
\end{eqnarray}
be the vector space of simplices, then obviously the boundary operator $\partial$
is nilpotent on $S$
\begin{eqnarray}
  \partial^2 S &=& 0
  \,.
\end{eqnarray}

While simplices feature prominently in the literature on discrete spaces, it is somewhat
remarkable that the discrete differential calculus over them lacks many of the nice
features that we will find for discrete differential calculus on \emph{cubic complexes}.

Most notably, at any given node of a simplicial complex there will be more than
$D$ independent 1-forms. As discussed in \S\fullref{the continuum limit} this brings with
it certain problems with the continuum limit of this differential calculus.\footnote{
  See \cite{KanamoriKawamoto:2003} for an example of how the presence of more than $D$
  independent 1-forms leads to problems.
}

A way out of this problem was suggested in \cite{ForgyChew:2000}. In the approach
suggested there one essentially defines discrete differential calculi on
$D$-dimensional spaces consisting of single $D$-simplices. Since a
$p$-simplex has $p$ edges originating at every node this removes the problem
of the surplus edges, but of course at the cost that now the different
simplices have to be ``glued together'' again by methods that essentially
lie outside those of differential calculus, e.g. by averaging functions over
those nodes which should be idenified.

\begin{figure}[h]
\begin{center}
\begin{picture}(200,50)
\includegraphics{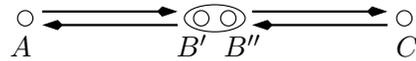}
\put(-153,-9){$A$}
\put(-90,-9){$B^\prime$}
\put(-72,-9){$B^{\prime\prime}$}
\put(-7,-9){$C$}
\end{picture}
\end{center}
\caption{{\it A ``glued'' 1-complex}}
\label{glued1simplex.eps}
\end{figure}

As an example consider a 1-dimensional simplicial complex consisting of two 
1-simplices as in figure \ref{glued1simplex.eps}.
Let
\begin{eqnarray}
  f 
  &\defas&
  f\of{A}\delgen{A} + f\of{B}\left(\delgen{B^\prime} + \delgen{B^{\prime\prime}}\right)
  + f\of{C}\delgen{C} 
\end{eqnarray}
be a function on this space, regarded as the disjoint union of two 1-simplices.
The ``gluing'' of these two 1-simplices 
is achieved by restricting attention to such functions which take the same value
at the two \emph{different} nodes $B^\prime$ and $B^{\prime\prime}$.

Now compute the action of the lattice Laplace operator on this function
\begin{eqnarray}
  \coextd \extd f
  &=&
  \coextd
  \left(
    \left(
    f\of{A}
    - f\of{B}
    \right)
    \left(
      \delgen{B^\prime A} - \delgen{AB^\prime}
    \right)
    +
    \left(
    f\of{B}
    - f\of{C}
    \right)
    \left(
      \delgen{C B^{\prime\prime}} - \delgen{B^{\prime\prime}C}
    \right)
  \right)
\end{eqnarray}

But even differential calculus on a single $D$-simplex has undesireable features.
For instance volume forms 
in the sense of
\S\fullref{volume form and hodge duality}
don't exist (except for $D=1$) .

\subsection{Edges and adjoint edges.}
\label{edges and adjoint edges}

Ordinary continuous manifolds are handled by mapping them locally,
by introduction of coordinate functions, to $\R^D$ and 
then working on this convenient space. Similarly a discrete space can be mapped locally
to a rectangular lattice $\Z^D$. Here the abelian group structure can be used to
induce a notion of translation along coordinate lines on the former discrete space.

On $\Z^D$ we naturally label the points of the discrete space by vectors 
$\vec x = [x^0,x^1,\cdots,x^{D-1}]^{\rm T}\in \Z^D$.  
The coordinate functions then read
\begin{eqnarray}
  \label{discrete coordinates}
  X^\mu &=& \epsilon\sum_{\vec x} \, x^\mu \delta_{\set{\vec x}}
  \,.
\end{eqnarray}
The real number $\epsilon$ is something like the lattice spacing as measured in these coordinates.
Even though it is more or less arbitrary and could in principle be set to unity,
we'll keep it around as a convenient label for those terms that will vanish in the
continuum limit, as for instance in formula 
\refer{general commutator of discrete difgferential with coordinates} below.
In the expression \refer{discrete coordinates} it serves the purpose of
accounting for the fact that $\Delta x^\mu \geq 1$, so that without 
letting $\epsilon\to 0$ as the number of nodes is increased in the continuum limit
the coordinate difference $X^\mu\of{i}-X^\mu\of{j}$ between any two nodes would grow
without bounds.

We consider the special case where from each point $\vec x$ of the discrete space there originate
precisely $D$ edges $e_a$ which as vectors in $\Z^D$ have components
\begin{eqnarray}
  \vec e_a\of{\vec x} &\defas& [e_a{}^0\of{\vec x}, e_a{}^1\of{\vec x},
    \cdots, e_a{}^{D-1}\of{\vec x}]^{\rm T}
  \hspace{1cm}
  a\in\set{0,\cdots,D-1}
  \,.
\end{eqnarray}

In other words we set
\begin{eqnarray}
  \label{graph operator in terms of vielbein edges}
  {\bf G}
  &=&
  \sum\limits_{\vec x,a}
  \delta_{\set{\vec x, \vec x + \vec e_a\of{\vec x}}}
\end{eqnarray}
and equivalently
\begin{eqnarray}
  {\bf G}^\dag
  &=&
  \sum\limits_{\vec x,a}
  \delta^\dag_{\set{\vec x + \vec e_a\of{\vec x}, \vec x}}
  \,,
\end{eqnarray}
where
\begin{eqnarray}
  \label{definition of adjoint edges}
\delta^\dag_{\set{\vec x + \vec e_a\of{\vec x},\vec x}}
&\defas&
\left(\delta_{\set{\vec x, \vec x + \vec e_a\of{\vec x}}}\right)^\dag
\end{eqnarray}
are the \emph{adjoint edges}.
Because of
\begin{eqnarray}
  \delta_{\set{\vec x,\vec x + \vec e_a\of{\vec x}}}
  &=&
  \delta_{\set{\vec x}}
  \commutator{\extd}{\delta_{\set{\vec x + \vec e_a\of{\vec x}}}}
  \nonumber\\
  &=&
  \delta_{\set{\vec x}}
  \circ
  \extd
  \circ
  \delta_{\set{\vec x + \vec e_a\of{\vec x}}}
  \,,
\end{eqnarray}
where ``$\circ$'' is the implicit operator product which we write here for clarity, we have\footnote{
Note that 
$\delta_{\set{\vec x}}^\dag = \delta_{\set{\vec x}}$.
Later we will consider modified adjoint relations whch do not enjoy this property.
}
\begin{eqnarray}
  \delta^\dag_{\set{\vec x + \vec e_a\of{\vec x}}}
  &=&
  \delta_{\set{\vec x + \vec e_a\of{\vec x}}}
  \circ
  \coextd
  \circ
  \delta_{\set{\vec x}}  
  \nonumber\\
  &=&
  \delta_{\set{\vec x + \vec e_a\of{\vec x}}}
  \commutator{
  \coextd}
  {
  \delta_{\set{\vec x}}
  }    
  \,.
\end{eqnarray}
The resulting graph typically looks as shown in figure \ref{disccalc.eps}:
\begin{figure}[h]
\begin{picture}(330,200)
\includegraphics{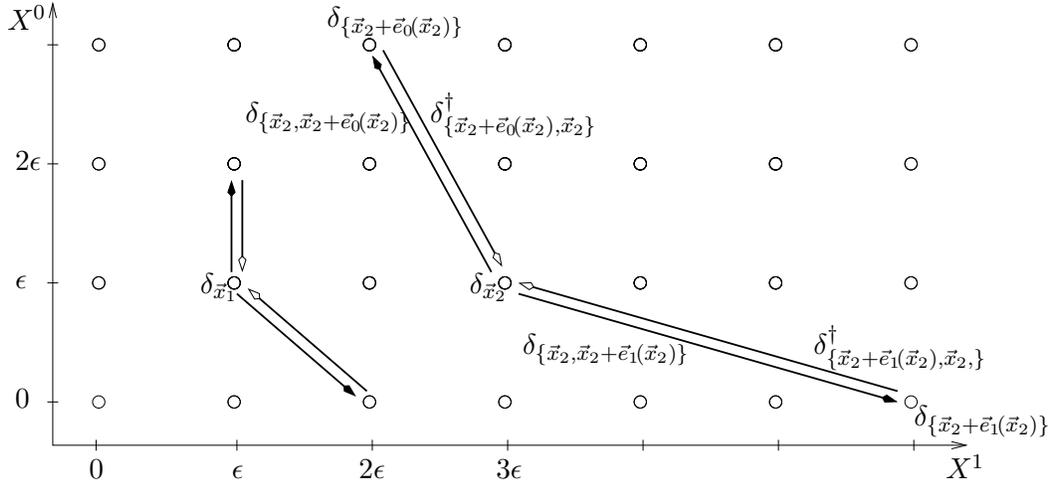}
\put(-10,-10){$X^1$}
\put(-365,160){$X^0$}
\put(-362,17){$0$}
\put(-362,62){$\epsilon$}
\put(-362,107){$2\epsilon$}
\put(-334,-10){$0$}
\put(-280,-10){$\epsilon$}
\put(-232,-10){$2\epsilon$}
\put(-180,-10){$3\epsilon$}
\put(-292,60){$\delta_{\vec x_1}$}
\put(-190,60){$\delta_{\vec x_2}$}
\put(-170,37){$\delta_{\set{\vec x_2,\vec x_2 + \vec e_1\of{\vec x_2}}}$}
\put(-60,37){$\delta^\dag_{\set{\vec x_2 + \vec e_1\of{\vec x_2},\vec x_2,}}$}
\put(-275,125){$\delta_{\set{\vec x_2,\vec x_2 + \vec e_0\of{\vec x_2}}}$}
\put(-205,125){$\delta^\dag_{\set{\vec x_2 + \vec e_0\of{\vec x_2},\vec x_2}}$}
\put(-22,10){$\delta_{\set{\vec x_2 + \vec e_1\of{\vec x_2}}}$}
\put(-244,161){$\delta_{\set{\vec x_2 + \vec e_0\of{\vec x_2}}}$}
\end{picture}
$\,$\\
\caption{{\it Edges and adjoint edges on a coordinate patch}}
\label{disccalc.eps}
\end{figure}
The little circles represent the nodes of the graph and are labelled by the 
discrete $\delta$-functions which are supported at a given node. 
For clarity only the edges originating at node 
$\vec x_1 = 
\epsilon
\left[
  1 \atop 1
\right]
$ and those originating at 
$
\vec x_2
=
\epsilon
\left[
  1 \atop 3
\right]
$ 
have been drawn, together with the corresponding adjoint edges, which are characterized by
their open arrow heads. The $\vec e_a$ at these two points have been chosen arbitrarily.
$\,$\\

We will frequently need the ``macroscopic'' 1-forms 
\begin{eqnarray}
  \label{discrete ONB creators}
  \edgesCreator^a
  &\defas&
  \epsilon
  \sum_{\vec x}
  \delta_{\set{\vec x, \vec x + \vec e_a\of{\vec x}}}
  \nonumber\\
  \edgesAnnihilator^a
  &\defas&
  (\edgesCreator^a)^\dag
  \nonumber\\
  &=&
  \epsilon
  \sum_{\vec x}  
  \delta^\dag_{\set{\vec x + \vec e_a\of{\vec x},\vec x}}
  \,.
\end{eqnarray}
(Note that these differ from the objects 
\refer{coordinate 1 forms on group lattice} in that here the $\vec e_a$ are
functions of $\vec x$.)

\subsection{Coordinate differentials}
In terms of the ``microscopic'' edges the coordinate differentials can be written as
\begin{eqnarray}
  \extd X^\mu
  &=&
  \sum_{\vec x} \epsilon x^\mu \extd \delta_{\set{\vec x}}
  \nonumber\\
  &\equalby{derivation of dicsrete df}&
  \sum_{\vec x,a} \epsilon x^\mu 
  \left(
    \delta_{\set{\vec x - \vec e_a\of{\vec x}}}
    \extd \delta_{\set{\vec x}}  
    -
    \delta_{\set{\vec x}}
    \extd
    \delta_{\set{\vec x + \vec e_a\of{\vec x}}}     
  \right)
  \nonumber\\
  &=&
  \sum_{\vec x,a}
  \epsilon  
  e_a{}^\mu\of{\vec x}
  \delta_{\set{\vec x}}
  \extd  
  \delta_{\set{\vec x + \vec e_a}}
  \nonumber\\
  &=&
  e_a{}^\mu \edgesCreator^a
  \,.
\end{eqnarray}
(Unless indicated otherwise we are using implicit summation on repeated indices.)

(This formula is the reason why a factor $\epsilon$ has been included in the
definition \refer{discrete ONB creators}.
Since $\extd X^\mu$ is an object with a well behave continuum limit
the objects $\edgesCreator^a$ must have a continuum limit, too. Incidentally
we have
\begin{eqnarray}
  \label{graph operator in terms of onb creators}
  {\bf  G}
  &=&
  \frac{1}{\epsilon}
  \sum\limits_a \edgesCreator^a
  \,.
\end{eqnarray}
This implies that ${\bf G}$ itself is not well defined for $\epsilon \to 0$, which makes
sense, since of course in the continuum theory there is no 1-form with the same anticommutators
as $\extd$.)

By taking adjoints we similarly obtain
\begin{eqnarray}
  \coextd X^\mu
  &=&
  -
  \sum_{\vec x,a}
  \epsilon  
  e_a{}^\mu\of{\vec x}
  \delta_{\set{\vec x + \vec e_a\of{\vec x}}}
  \coextd  
  \delta_{\set{\vec x}}
  \nonumber\\
  &=&
  \edgesAnnihilator^a e_a{}^\mu
\end{eqnarray}
with
\begin{eqnarray}
  \edgesAnnihilator^a 
  &\defas&
  \sum_{\vec x}
  \epsilon  
  \delta_{\set{\vec x + \vec e_a}}
  \coextd  
  \delta_{\set{\vec x}}  
  \,.
\end{eqnarray}
It follows that at a given point the coordinate forms are related to the ``microscopic forms'' by
\begin{eqnarray}
  \delta_{\set{\vec x}}\extd X^\mu 
  &=&
  \sum_{a}
  \epsilon  
  e_a{}^\mu\of{\vec x}
  \delta_{\set{\vec x}}
  \extd  
  \delta_{\set{\vec x + \vec e_a}}
  \,.
\end{eqnarray}
Assuming that the matrix $e_a{}^\mu\of{\vec x}$ is invertible with inverse $e_\mu{}^a\of{\vec x}$, 
i.e. $e_a{}^\mu\of{\vec x} e_\mu{}^b\of{\vec x} = \delta_a^b$, this relation may be inverted:
\begin{eqnarray}
  \label{microscopic forms in terms of local coordinate forms}
  \delta_{\set{\vec x}}
  \extd  
  \delta_{\set{\vec x + \vec e_a}}
  &=&
  \frac{1}{\epsilon}
  e_\mu{}^a\of{\vec x} \delta_{\set{\vec x}}\extd X^\mu  
  \,.
\end{eqnarray}
Using this and the local commutation relation
\begin{eqnarray}
  \commutator
  {X^\mu}
  {\delta_{\set{\vec x, \vec x + \vec e_a}}}
  &=&
  -
  \epsilon e_a{}^\mu\of{\vec x} \delta_{\set{\vec x, \vec x + \vec e_a}}
  \nonumber\\
  \label{cubic lattice local forms functions commutator}
  \commutator
  {X^\mu}
  {\delta^\dag_{\set{\vec x + \vec e_a, \vec x }}}
  &=&
  +
  \epsilon e_a{}^\mu\of{\vec x} \delta^\dag_{\set{\vec x + \vec e_a, \vec x }}
\end{eqnarray}
we find for the commutator of a coordinate form with a coordinate function the 
expression
\begin{eqnarray}
  \label{general commutator of discrete difgferential with coordinates}
  \commutator{X^\mu}{\extd X^\nu}
  &=&
  \commutator
  {X^\mu}
  {
  \sum_{\vec x,a}
  \epsilon  
  e_a{}^\nu\of{\vec x}
  \delta_{\set{\vec x}}
  \extd  
  \delta_{\set{\vec x + \vec e_a}}
  }
  \nonumber\\
  &=&
  -\sum_{\vec x,a}
  \epsilon^2  
  e_a{}^\mu\of{\vec x}
  e_a{}^\nu\of{\vec x}
  \delta_{\set{\vec x}}
  \extd  
  \delta_{\set{\vec x + \vec e_a}}
  \nonumber\\
  &\equalby{microscopic forms in terms of local coordinate forms}&
  -\sum_{\vec x,a}
  \epsilon\,
  e_a{}^\mu\of{\vec x}
  e_a{}^\nu\of{\vec x}
  e_\lambda{}^a\of{\vec x}
  \delta_{\set{\vec x}}
  \extd  
  X^\lambda  
  \nonumber\\
  &\defas&
  \epsilon\,
  C^{\mu\nu}{}_\lambda
  \extd X^\lambda
  \,.
\end{eqnarray}
In the last line we have identified the ``structure functions'' as
\begin{eqnarray}
  \label{strucure functions of discrete lattice manifold}
  C^{\mu\nu}{}_\lambda
  &\defas&
  -
  \sum_{a}
  e_a{}^\mu
  e_a{}^\nu
  e_\lambda{}^a
  \,.
\end{eqnarray}
By acting with $\extd$ on both sides of the above equation one obtains
\begin{eqnarray}
  \antiCommutator{\extd X^\mu}{\extd X^\nu}
  &=&
  \epsilon
  \extd C^{\mu\nu}{}_\lambda
  \, \extd X^\lambda
  \,.
\end{eqnarray}
Taking the adjoint of everything yields analogously
\begin{eqnarray}
  \commutator{X^\mu}{\coextd X^\nu}
  &=&
  -\epsilon\, \left(\coextd X^\lambda\right) C^{\mu\nu}{}_\lambda
\end{eqnarray}
and
\begin{eqnarray}
  \antiCommutator{\extd X^\mu}{\extd X^\nu}
  &=&
  \epsilon
  \coextd X^\lambda
  \,
  \coextd C^{\mu\nu}{}_\lambda 
  \,.  
\end{eqnarray}

To conform with the conventions in the continuum case we introduce for the coordinate differentials
the notation
\begin{eqnarray}
  \label{coordinate differentials in the discrete case}
  \coordCreator^\mu &=& \commutator{\extd}{ X^\mu}
  \nonumber\\
  \coordAnnihilator^\mu &=& -\commutator{\coextd}{ X^\mu}
  \,.  
\end{eqnarray}

\subsection{Component derivation}

The 1-form $\extd f$ (for $f$ any element of $\Omega^0$) can be written as
\begin{eqnarray}
  \label{definition lpartial and rpartial}
  \extd f 
  &\defas&
  (\lpartial_\mu f)\extd X^\mu
  \nonumber\\
  &=&
  (\extd X^\mu)(\rpartial_\mu f)
  \,.
\end{eqnarray}

One finds (see \cite{Forgy:2002}) that, while the objects $\lpartial$ and $\rpartial$
satisfy
\begin{eqnarray}
  \lpartial_\mu X^\nu &=& \delta_\mu^\nu
  \nonumber\\
  \rpartial_\mu X^\nu &=& \delta_\mu^\nu
\end{eqnarray}
they don't quite act like true partial
derivatives because their product rule receives a lattice correction:\footnote{
  As shown in \cite{Forgy:2002} one has
\begin{eqnarray}
  \extd (fg)
  &\equalby{nilpotentcy and Leibniz of general extd}&
  (\extd f)g + f (\extd g)
  \nonumber\\
  &\equalby{definition lpartial and rpartial}&
  (\lpartial_\lambda f)\extd X^\lambda \, g
  +
  f (\lpartial_\lambda g)\extd X^\lambda
  \nonumber\\
  &=&
  (\lpartial_\lambda f)\left(
    g\, \extd X^\lambda + 
   \underbrace{\commutator{\extd X^\lambda}{g}}_{= \commutator{\extd g}{X^\lambda}
   = (\lpartial_\nu g)\commutator{\extd X^\nu}{X^\lambda} }
  \right)
  +
  f(\lpartial_\lambda g)\extd X^\lambda
  \nonumber\\
  &=&
  \left(
    (\lpartial_\lambda f)g + f(\lpartial_\lambda g)
  \right)
  \extd X^\lambda
  -
  \epsilon
  (\lpartial_\lambda f)(\lpartial_\nu g)
  C^{\nu\lambda}{}_\mu \extd X^\mu
  \nonumber\\
  &=&
  \left(
    (\lpartial_\lambda f)g + f(\lpartial_\lambda g) 
    - 
    \epsilon(\lpartial_\mu f)(\lpartial_\nu g)C^{\mu\nu}{}_\lambda
  \right)
  \extd X^\lambda
  \nonumber\\
  &=&
  \lpartial_\lambda (fg)\extd X^\lambda
  \,.
\end{eqnarray}
}
\begin{eqnarray}
  \label{product rule for discrete almost partial derivatives}
  \lpartial_\lambda(fg)
  &=&
  (\lpartial_\lambda f)g
  +
  f(\lpartial g)
  -
  \epsilon (\lpartial f)(\lpartial g)C^{\mu\nu}{}_\lambda
  \,.
\end{eqnarray}

\paragraph{Component derivative in preferred coordinates.}

With respect to the \emph{preferred coordinates}  
that satisfy $\extd X^\mu = \edgesCreator^{(a=\mu)}$
(these exist globally on the \emph{cubic graphs} discussed in \S\fullref{subsection: cubic graphs})
the $\lpartial_\mu$ and $\rpartial_\mu$
objects have a simple expression in terms of finite differences
(assume $\vec e_a = {\rm const}$ for notational convenience):
\begin{eqnarray}
  \extd f
  &=&
  \extd
  \sum\limits_{\vec x}
  f\of{\vec x} \delta_{\set{\vec x}}
  \nonumber\\
  &=&
  \sum\limits_{\vec x}
  f\of{\vec x} \extd \delta_{\set{\vec x}}
  \nonumber\\
  &=&
  \sum\limits_a
  \sum\limits_{\vec x}
  f\of{\vec x} 
  \left(
    \delta_{\set{\vec x-\vec e_a,\vec x}}  
    -
    \delta_{\set{\vec x, \vec x + \vec e_a}}
  \right)
  \nonumber\\
  &=&
  \sum\limits_a
  \frac{
    T_{-\vec e_a}[f] -f
  }{\epsilon}
  \extd X^{a}
  \nonumber\\
  &=&
  \sum\limits_a
  \extd X^{a}
  \frac{
    f - T_{\vec e_a}[f] 
  }{\epsilon}
  \,,
\end{eqnarray}
where $T$ is the translation operator acting as
\begin{eqnarray}
  T_{\vec y}[f]\of{\vec x}
  &\defas&
  f\of{\vec x - \vec y}
  \,.
\end{eqnarray}
Therefore
\begin{eqnarray}
  \label{forward discrete derivative}
  (\lpartial_a f)\of{\vec x}
  &=&
  \frac{f\of{\vec x + \vec e_a}-f\of{\vec x}}{\epsilon}
\end{eqnarray}
is the \emph{forward finite difference quotient}
and
\begin{eqnarray}
  \label{backward discrete derivative}
  (\rpartial_a f)\of{\vec x}
  &=&
  \frac{f\of{\vec x} - f\of{\vec x - \vec e_a}}{\epsilon}
\end{eqnarray}
the \emph{backward finite difference quotient} 
of the function $f$ with respect to $\vec e_a$.

Note that all forward and backward differences mutually commute:\footnote{
  \begin{eqnarray}
    \lpartial_a\lpartial_b f\of{\vec x}
    &=&
    \lpartial_a
    \left(
      f\of{\vec x + \vec e_b} - f\of{\vec x}
    \right)
    \nonumber\\
    &=&
    f\of{\vec x +\vec e_a +\vec e_b} - f\of{\vec x + \vec e_a}
    -
    f\of{\vec x + \vec e_b} + f\of{\vec x}
  \end{eqnarray}

  \begin{eqnarray}
    \rpartial_a\rpartial_b f\of{\vec x}
    &=&
    \lpartial_a
    \left(
      f\of{\vec x } - f\of{\vec x - \vec e_b}
    \right)
    \nonumber\\
    &=&
    f\of{\vec x } - f\of{\vec x - \vec e_b}
    -
   f\of{\vec x -\vec e_a} + f\of{\vec x - \vec e_a - \vec e_b}  
\end{eqnarray}

\begin{eqnarray}
  \lpartial_a \rpartial_b f\of{\vec x}
  &=& 
  \lpartial_a 
   \left(
      f\of{\vec x} - f\of{\vec x - \vec e_b}   
   \right)
  \nonumber\\
  &=&
  f\of{\vec x + \vec e_a} - f\of{\vec x + \vec e_a - \vec e_b}
  - f\of{\vec x} + f\of{\vec x - \vec e_b}
\end{eqnarray}

\begin{eqnarray}
  \rpartial_b \lpartial_a f\of{\vec x}
  &=& 
  \rpartial_b 
   \left(
      f\of{\vec x + \vec e_a} - f\of{\vec x}   
   \right)
  \nonumber\\
  &=&
  f\of{\vec x + \vec e_a} - f\of{\vec x}
  -
  f\of{\vec x + \vec e_a - \vec e_b} + f\of{\vec x - \vec e_b}
\end{eqnarray}
}
\begin{eqnarray}
  \label{discrete cubic forward and backward differences commute}
  \lpartial_a \lpartial_b  f &=& \lpartial_b \lpartial_a f \nonumber\\
  \rpartial_a \rpartial_b  f &=& \rpartial_b \rpartial_a f \nonumber\\
  \lpartial_a \rpartial_b  f &=& \rpartial_b \lpartial_a f
  \,.
\end{eqnarray}

Furthermore they are interrelated by translations
\begin{eqnarray}
  \label{interchanging discrete derivatives by translation}
  T_{-\edge_a}[\rpartial_a f] &=& \lpartial_a f
  \nonumber\\
  T_{+\edge e_a}[\lpartial_a f] &=& \rpartial_a f
  \,,  
\end{eqnarray}
with which they commute:
\begin{eqnarray}
  T_{\vec y} \circ \rpartial_a &=& \rpartial_a \circ T_{\vec y}
  \nonumber\\
  T_{\vec y} \circ \lpartial_a &=& \lpartial_a \circ T_{\vec y}
  \,.
\end{eqnarray}

The arithmetic mean of left and right discrete derivative gives the
symmetric difference quotient
\begin{eqnarray}
  \lrpartial_a f\of{\vec x}
  &\defas&
  \frac{1}{2}(\lpartial_a + \rpartial_a)f\of{\vec x}
  \nonumber\\
  &=&
  \frac{1}{2\epsilon}
  \left(
    f\of{\vec x + \vec e_a}
    -
    f\of{\vec x - \vec e_a}
  \right)
\end{eqnarray}
while their difference (which is sometimes called the \emph{osmotic} derivative in
the theory of stochastic processes) 
is $\epsilon$ times a second order discrete
derivative:
\begin{eqnarray}
  \osmotial_a
  &\defas& 
  (\lpartial_a - \rpartial_a)f\of{\vec x}
  \nonumber\\
  &=&
  \frac{1}{\epsilon}
  \left(
    f\of{\vec x + \vec e_a}
    -
    2f\of{\vec x}
    +
    f\of{\vec x - \vec e_a}
  \right)
  \nonumber\\
  &=&
  \epsilon\,
  \lpartial_a \rpartial_a f\of{\vec x}
  \,.
\end{eqnarray}
The resulting formulas
\begin{eqnarray}
  \lpartial_a &=& \lrpartial_a + \frac{1}{2}\osmotial_a
  \nonumber\\
  \rpartial_a &=& \lrpartial_a - \frac{1}{2}\osmotial_a
\end{eqnarray}
for the forward and backward derivative
are reminiscent of those used in stochastic calculus,  \cf \S\fullref{Example: Hyper diamond model and stochastic calculus.}.

The symmetric difference quotient is obviously sensitive only to ``wavelengths''
$\lambda \geq 2\epsilon$. 

\paragraph{Anticommutator of 1-forms.}

In the case of trivial structure functions \refer{trivial structure functions of discrete calculus} 
the anticommutator
of any two 1-forms is
\begin{eqnarray}
  \label{anticommutator of 1-form}
  \antiCommutator{\alpha_\mu \extd X^\mu}{\beta_\nu \extd X^\nu}
  &=&
  \alpha_\mu \commutator{\extd X^\mu}{\beta_\nu}\extd X^\nu
  -
  \beta_\nu \commutator{\extd X^\nu}{\alpha_\mu}\extd X^\mu
  \nonumber\\
  &=&
  \alpha_\mu \commutator{\extd \beta_\nu}{X^\mu}\extd X^\nu
  -
  \beta_\nu \commutator{\extd \alpha_\mu}{X^\nu}\extd X^\mu
  \nonumber\\
  &=&
  \alpha_\mu (\lpartial_\lambda \beta_\nu)\commutator{\extd X^\lambda}{X^\mu}\extd X^\nu
  -
  \beta_\nu (\lpartial_\lambda \alpha_\mu)\commutator{\extd X^\lambda}{X^\nu}\extd X^\mu  
  \nonumber\\
  &=&
  \epsilon
  \sum_\lambda
  \left(
  \alpha^\lambda (\lpartial_\lambda \beta_\mu) 
  -
  \beta^\lambda (\lpartial_\lambda \alpha_\mu) 
  \right)
  \extd X^\lambda\extd X^\mu
  \,.
\end{eqnarray}
Hence the anticommutator vanishes precisely if
\begin{eqnarray}
  \label{condition for anticommutation of 1-forms in flat discrete space}
  &&\antiCommutator{\alpha_\mu \extd X^\mu}{\beta_\nu \extd X^\nu} \;=\;0
  \nonumber\\
  &\Leftrightarrow&
  \alpha^\lambda (\lpartial_\lambda \beta_\mu) 
  -
  \beta^\lambda (\lpartial_\lambda \alpha_\mu) 
  \;=\; 0
  \hspace{1cm}
  \mbox{(no sum over $\lambda$ and for all $\lambda\neq \mu$)}
  \,.
\end{eqnarray}

\subsection{Example: Stochastic calculus.}
\label{Example: Hyper diamond model and stochastic calculus.}

Consider the ``2-dimensional'' graph with edges
\begin{eqnarray}
  e_\pm &\defas& 
  \left[
    \begin{array}{c}
      1 \\
      \pm \sigma/\sqrt{\epsilon}
    \end{array}
  \right]
  \,,
\end{eqnarray}
i.e. the graph given by the vielbein
\begin{eqnarray}
  \label{2d hyperdiamond vielbein}
  e_a{}^\mu &=&
  \left[
    \begin{array}{cc}
      1 & \sigma/\sqrt{\epsilon} \\
      1 & -\sigma/\sqrt{\epsilon}
    \end{array}
  \right]
  \hspace{1cm}
  a = +,-\,;\; \mu = 0,1 
\end{eqnarray}
with inverse
\begin{eqnarray}
  \label{2d inverse hyper diamond vielbein}
  e_\mu{}^a
  &=&
  \frac{1}{2}
  \left[
    \begin{array}{cc}
      1 & 1 \\
      \sqrt{\epsilon}/\sigma & -\sqrt{\epsilon}/\sigma
    \end{array}
  \right]
  \hspace{1cm}
  a = +,-\,;\; \mu = 0,1
  \,. 
\end{eqnarray}
Here $\sigma$ is a real constant and obviously we need to assume that $\sigma/\sqrt{\epsilon}$
is an integer. 

We want to model a simple physical system on this graph and for this purpose we think
of $X^0$ as a time-coordinate and $X^1$ as a space coordinate. 

For $\sigma = 2\sqrt{\epsilon}$ the resulting graph is indicated in figure \ref{discstoch}:

\begin{figure}[h]
\begin{center}
\begin{picture}(230,130)
\includegraphics{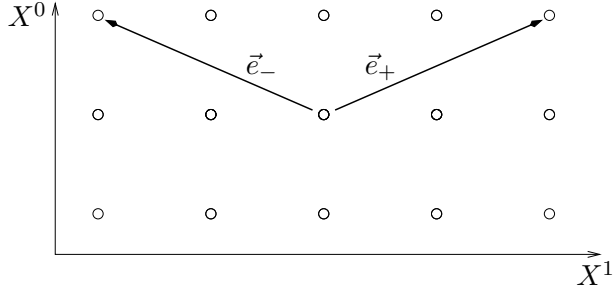}
\put(-10,-10){$X^1$}
\put(-225,88){$X^0$}
\put(-90,70){$\vec e_+$}
\put(-135,70){$\vec e_-$}
\end{picture}

$\,$\\
\caption{{\it A graph that induces stochastic calculus.}}
\label{discstoch}
\end{center}
\end{figure}

$\,$\\

For $\sigma = \sqrt{\epsilon}$ we get (2 copies of) the 2-dimensional ``hyper diamond'',
which (as soon as the structure is equipped with some notion of metric, as will be done below)
can be interpeted as having two future directed lightlike edges at every node. A ``particle''
constrained to move along edges of this graph will perform a locally lightlike zig-zag walk
as known from Feynman's checkerboard model.  

The non-relativistic limit of this motion is 
obviously modeled by taking $\sigma/\sqrt{\epsilon} \gg 1$: There is no longer a limiting
velocity and the ``particle'' performs numerous ``jumps'' of spatial distance 
$\Delta X^1 = \sigma \sqrt{\epsilon} = \sigma \sqrt{\Delta X^0}$. Intuitively, this
should be related to a random walk. Interestingly, this stochastic
process is automatically reproduced by the discrete differential calculus related
to the above graph, to which we now turn:

Plugging \refer{2d hyperdiamond vielbein} and 
\refer{2d inverse hyper diamond vielbein} 
into the formula \refer{strucure functions of discrete lattice manifold} 
for the ``structure functions'' yields
\begin{eqnarray}
  C^{\mu\nu}{}_\lambda
  &=&
  -\frac{1}{2}
  \left(
  \left[
   \begin{array}{c}
     1 \\ \sigma/\sqrt{\epsilon}
   \end{array}
  \right]^\mu
  \left[
   \begin{array}{c}
     1 \\ \sigma/\sqrt{\epsilon}
   \end{array}
  \right]^\nu
  \left[
   \begin{array}{c}
     1 \\ \sqrt{\epsilon}/\sigma
   \end{array}
  \right]_\lambda
  +
  \left[
   \begin{array}{c}
     1 \\ -\sigma/\sqrt{\epsilon}
   \end{array}
  \right]^\mu
  \left[
   \begin{array}{c}
     1 \\ -\sigma/\sqrt{\epsilon}
   \end{array}
  \right]^\nu
  \left[
   \begin{array}{c}
     1 \\ -\sqrt{\epsilon}/\sigma
   \end{array}
  \right]_\lambda
  \right)
  \nonumber\\
\end{eqnarray}
or equivalently
\begin{eqnarray}
  -\epsilon C^{(\mu=1)(\nu=1)}{}_\lambda
  &=&
  \left[
    \begin{array}{c}
      \sigma^2 \\
      0
    \end{array}
  \right]_\lambda
  \nonumber\\
  -\epsilon C^{(\mu=1)(\nu=0)}{}_\lambda
  &=&
  \epsilon
  \left[
    \begin{array}{c}
      0 \\
      1
    \end{array}
  \right]_\lambda  
  \nonumber\\
  -\epsilon C^{(\mu=0)(\nu=0)}{}_\lambda
  &=&
  \epsilon
  \left[
    \begin{array}{c}
      1 \\
      0
    \end{array}
  \right]_\lambda
  \,.  
\end{eqnarray}
According to \refer{general commutator of discrete difgferential with coordinates} 
this translates into the following commutation relations between coordinate differentials
and coordinates:
\begin{eqnarray}
  \label{2d hyper diamond cartesian differential/coordinates commutation relations}
  \commutator{\extd X^0}{X^0} &=& \epsilon\, \extd X^0
  \nonumber\\
  \commutator{\extd X^0}{X^1} &=& \epsilon\, \extd X^1
  \nonumber\\
  \commutator{\extd X^1}{X^1} &=&  \sigma^2\, \extd X^0
  \,.
\end{eqnarray}

\paragraph{Non-relativistic case.}
In the continuum limit $\epsilon \to 0$ the structure functions become
\begin{eqnarray}
  \lim\limits_{\epsilon\to 0}
  \;
  -\epsilon C^{\mu\nu}{}_\lambda
  &=&
  \left\lbrace
    \begin{array}{cl}
      \sigma^2 & \mbox{for $\mu=\nu=1$ and $\lambda = 0$}\\ 
       0 & \mbox{otherwise}
    \end{array}
  \right.
  \,.
\end{eqnarray}
Precisely these values of the structure functions had been noticed in 
\cite{Forgy:2002}
to induce a differential calculus that mimics properties of Ito's stochastic
calculus:

Using \refer{product rule for discrete almost partial derivatives} 
we obtain in the limit $\epsilon \to 0$ the left and right
derivatives
\begin{eqnarray}
  \lpartial_1 (fg)
  &=&
  (\lpartial_1 f)g
  + f(\lpartial_1 g)
  \nonumber\\
  \lpartial_0(fg)
  &=&
  (\lpartial_0 f)g + f(\lpartial g)
  +
  \sigma^2 (\lpartial_1 f) (\lpartial_1 g)
  \,.
\end{eqnarray}
It is readily checked that the proper partial derivatives $\partial$ are related
to $\lpartial$ by
\begin{eqnarray}
  \partial_1 &=& \lpartial_1
  \nonumber\\
  \partial_0 &=& \lpartial_0 - \frac{\sigma^2}{2}\partial_1 \partial_1
  \,.
\end{eqnarray}
In other words we have
\begin{eqnarray}
  \extd f
  &=&
  \left(
    \partial_0 f + \frac{\sigma^2}{2}\partial_1 \partial_1 f
  \right)
  \extd X^0
  +
  (\partial_1 f)\extd X^1
  \nonumber\\
  &=&
  \extd X^0
  \left(
    \partial_0 f - \frac{\sigma^2}{2}\partial_1 \partial_1 f
  \right)
  +
  \extd X^1 (\partial_1 f)
  \,.
\end{eqnarray}
(In the last line the analogous derivation for $\rpartial$ instead of $\lpartial$ has been used,
see \cite{Forgy:2002}). This shows that $\lpartial_0$ is the \emph{forward} and
$\rpartial_0$ the \emph{backward} temporal derivative of drift-free a Wiener process.

\subsection{The continuum limit}
\label{the continuum limit}

\subparagraph{Opposite edges and continuum limit.}
On a $D$-dimensional manifold there are preciely $D$ linearly independent 1-forms at every point. 
Since this is not true for discrete complexes with opposite edges (\cf \refer{condition on no opposite edges}),
these cannot approximate ordinary differential geometry on manifolds in the continuum limit.

As an example consider the ``1-dimensional'' complex with a unit edge and opposite edge at every node:
\begin{eqnarray}
  {\bf G}
  &=&
  \sum\limits_{x\in \Z} \left(\delta_{\set{x,x+1}} + \delta_{\set{x,x-1}}\right)
  \,.
\end{eqnarray} 
The coordinate differential, which is the only independent differential in the continuum theory, is
\begin{eqnarray}
  \extd X
  &=&
  \epsilon
  \sum\limits_x \delta_{\set{x,x+1}}
  -
  \epsilon
  \sum\limits_x \delta_{\set{x,x-1}}
  \,.
\end{eqnarray}
But 
\begin{eqnarray}
  \commutator{\extd X}{X}
  &=&
  \epsilon \, S
  \,,
\end{eqnarray}
where
\begin{eqnarray}
  S 
  &\defas&
  \epsilon
  \sum\limits_x \delta_{\set{x,x+1}}
  +
  \epsilon
  \sum\limits_x \delta_{\set{x,x-1}}  
\end{eqnarray}
is a 1-form that has no counterpart in the differential geometry of the real line
and does not approximate any continuum object. 

In this sense the continuum limit of complexes with opposite edges does not exist. 
This is the motivation for concentrating in \S\fullref{edges and adjoint edges} 
on $D$-dimensional graphs that have precisely
$D$ outgoing edges at every node. Another indication that complexes without
opposite edges (and without intermediate edges) are ``nice'', is that for
them the graph operator is nilpotent (\cf \S\fullref{set of edges and graph operator}).

\subparagraph{Dimensionality.}

One would expect from a well-behaved discrete space that its dimension does not vary from node
to node. In particular, a 
well behaved $D$-dimensional differential complex (i.e. $D$-forms are top forms),
should have a non-vanishing $D$ form supported at each seperate node.

\begin{figure}[h]
\begin{center}
\begin{picture}(200,200)
\includegraphics{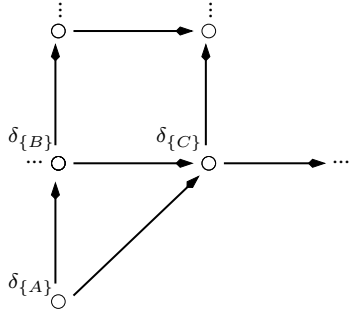}
\put(-130,10){${}_{\delta_{\set{A}}}$}
\put(-130,65){${}_{\delta_{\set{B}}}$}
\put(-74,65){${}_{\delta_{\set{C}}}$}
\end{picture}
\end{center}
\caption{{\it A graph with intermediate edges.}}
\label{intermediateedges.eps} 
\end{figure}

But this condition is rather stringent: Consider for example a 2-dimensional complex
with intermediate edges as in figure \ref{intermediateedges.eps}. At node $B$ it 
supports a 2-form (\cf figure \ref{cubicgraph}), but at node $A$, where the 
``intermediate'' edges $\delta_{\set{A,C}}$ originates, it does not, because
\begin{eqnarray}
  0 &=&
  \left(\extd \extd \delta_{\set{A}}\right) \delta_{\set{C}}  
  \nonumber\\
  &=&
  \left(
    \extd (-\delta_{\set{A,B}} - \delta_{\set{A,C}})
  \right)\delta_{\set{C}}
  \nonumber\\
  &=&
  \delta_{\set{A,B,C}}
\end{eqnarray}
and similarly for all other 2-forms originating at $A$.

\begin{figure}[h]
\hspace{2cm}
\begin{picture}(200,200)
\includegraphics{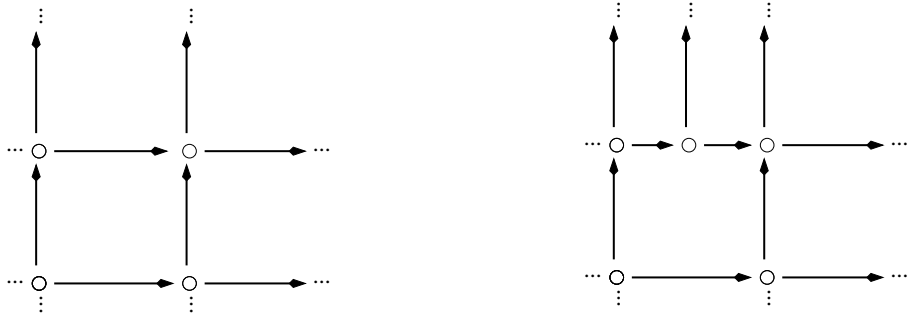}
\end{picture}
\caption{Two typical graphs with neither opposite nor intermediate edges and
with $D$ edges originating at every node}
\label{nooppositeedges.eps} 
\end{figure}

If one therefore restricts oneself to graphs with neither opposite nor intermediate edges and
with precisely $D$-edges originating at every node, then one is left with examples such
as depicted in figure \ref{nooppositeedges.eps}. But the graph on the right 
of this figure does not have 
$D=2$-forms at each node, as was discussed in the text to figure \ref{discdimension}.

Therefore only cubic graphs as on the left of figure \ref{nooppositeedges.eps}
seem to have a well-behaved dimensionality as well as a manifold continuum limit.

\newpage
\section{Inner product, integration, and metric}
\label{inner product for differential calculi}

So far we have an abstract adjoint operation $(\cdot)^\dag$. 
We want to construct a vector space with inner product on which
the elements in $\hat\Omega_{(2)}\of{\hat A}$ act as operators and with respect to which 
the formal $(\cdot)^\dag$-operation is the true adjoint operation:

\subsection{The vector space ${\rm V}\of{\Omega\of{\cal A}}$ and inner products on it.}
\label{vector space VOmegaA and inner products}

{\bf Definition: Statelike and null operators.} An operator $A$ is called \emph{statelike} if it lies in 
$\hat \Omega\of{\hat {\cal A}}$ (\cf \refer{definition of diff algebra as operator algebra})
It is called \emph{null} if it ``annihilates the constant 0-form'', i.e.
\begin{eqnarray}
  \mbox{$A$ is null}
  &\Leftrightarrow&
  A \;=\; \sum_i A_i Q_i
  \hspace{1cm}
  Q_i \in \set{\extd,\coextd,\commutator{\coextd}{X^\mu}}
  \,.
\end{eqnarray}

Obviously every operator $A$ can uniquely be written as the sum of a statelike operator 
$A_{\cal S}$
with a null operator $A_{\cal N}$
by appropriately commuting $\extd$ and $\coextd$ to the right:
\begin{eqnarray}
  A &=& A_{\cal S} + A_{\cal N}
  \,.
\end{eqnarray}

\subparagraph{Definition:}
On the space $\hat {\Omega}_{(2)}\of{\hat {\cal A}}$ define the symmetric bilinear map
\begin{eqnarray}
  \begin{array}{cccl}
    (\cdot,\cdot) \;: & \hat {\Omega}_{(2)} \times \hat {\Omega}_{(2)} &  \rightarrow & \hat {\cal A} \\
                      & (A,B)                                          &  \mapsto     & 
   \left\lbrace
  \begin{array}{ll}
   \left(
    A^\dag B
    +
    B^\dag A
  \right)_{\cal S}
  & 
  \mbox{if ${\rm grad}\of{A} = {\rm grad}\of{B}$}
  \\
  0 & \mbox{otherwise}
  \end{array}
  \right.
  \end{array}
\end{eqnarray}
Note that the term in parentheses on the right hand side is indeed in 
${\hat \Omega}^0\of{\hat{\cal A}} = \hat {\cal A}$.

By construction $A^\dag$ is the adjoint of $A$ with respect to $(\cdot,\cdot)$:
\begin{eqnarray}
  (A\cdot,\cdot) &=& (\cdot, A^\dag \cdot)
  \,.
\end{eqnarray}

But we want a global inner product on the space $\cal H$ of \emph{statelike} operators.
With the equivalence relation
\begin{eqnarray}
  A \sim B &\Leftrightarrow& (A-B)_{\cal S} = 0
\end{eqnarray}
the space ${\cal H}$ can be characterized as
\begin{eqnarray}
  {\cal H}
  &\defas&
  \hat {\Omega}_{(2)}\of{\hat {\cal A}}/\sim
  \,.
\end{eqnarray}
As an element of $\cal H$ any operator $A$ 
(or rather the equivalence class it represents)
will be alternatively written as
\begin{eqnarray}
  [A]_\sim
  &\defas&
  \ket{(A)_{\cal S}}
  \,.
\end{eqnarray}
For instance
\begin{eqnarray}
  \left[\;\commutator{\extd}{X^\mu} + A \extd + B \coextd + C \commutator{\coextd}{X^\nu}\;\right]_\sim
  &=&
  \ket{\commutator{\extd}{X^\mu}}
\end{eqnarray}
corresponds to the 1-form $\extd X^\mu$. For convenience we will often use the $\Omega\of{\cal A}$-notation
instead of the $\hat \Omega\of{\hat{\cal A}}$-notation inside kets:
\begin{eqnarray}
  \ket{a_0 \commutator{\extd}{a_1} \commutator{\extd}{a_2}\cdots \commutator{\extd}{a_p}}
  &\defas&
  \ket{a_0 \extd a_1 \extd a_2 \cdots \extd a_p}
  \,.  
\end{eqnarray}

A special state is the \emph{constant 0-form} or
\emph{vacuum state}, which is associated with the identity operator,
\begin{eqnarray}
  [1]_\sim &=& \ket{1}
  \,,
\end{eqnarray}
and which can be used to characterize null operators:
\begin{eqnarray}
  A\ket{1} \;=\; 0 \;&\Leftrightarrow&\; A = (A)_{\cal N}
  \,.
\end{eqnarray}

\subparagraph{Definition: Integration.}
For some yet to be specified $dV\in {\cal A}$ let 
\begin{eqnarray}
  \label{definition discrete integral}
  \int A \; dV&\defas&
  \sum\limits_{\vec x}
  A\of{\vec x} dV\of{\vec x}
  \hspace{1cm}
  \mbox{for $A \in {\cal A}$}
  \,.
\end{eqnarray}

\subparagraph{Definition: Pre-Inner product.}
A bilinear form on the space of all operators is given by
\begin{eqnarray}
  \label{discrete pre-inner product}
  \bracket{A}{B}
  &\defas&
  \int 
  (A,B)\; dV
  \,.
\end{eqnarray}

This gives an inner product on ${\cal H}$ only if it vanishes on null operators. It is
readily checked that
\begin{eqnarray}
  \bracket{A}{B \coextd} &=& 0
  \nonumber\\
  \bracket{A}{B \commutator{\coextd}{a}} &=& 0
  \hspace{1cm}
  \forall\; a \in \hat{\cal A}
  \,.  
\end{eqnarray}
The only possibly problematic case is therefore 	
\begin{eqnarray}
  \bracket{A}{B \extd}
  &=&
  \left\lbrace
  \begin{array}{ll}
  \frac{1}{2}
  \int
  \left(
  \coextd
  B^\dag A
  \right)_{\cal S}
  \; dV
  & \mbox{for ${\rm grad}\of{A} = {\rm grad}\of{B} + 1$ }  \\
  0 & \mbox{otherwise}
  \end{array}
  \right.
  \,.
\end{eqnarray}
 Because $A$ and $B$ are
arbitrary the bilinear form $\bracket{\cdot}{\cdot}$ gives an inner product on
${\cal H}$ iff
\begin{eqnarray}
  \label{discrete divergence condition}
  \int (\coextd A)_{\cal S}\; dV &=& 0
  \hspace{1cm}
  \forall\; A\in {\hat \Omega}^1\of{\hat {\cal A}}
  \,,
\end{eqnarray} 
which, given the $N=2$ differential calculus with all its supercommutation relations,
is a condition on $dV\of{\vec x}$, i.e. on the value of
discrete volume associated with the node $\vec x$.

Note that in the continuum theory this is the well known relation
\begin{eqnarray}
  0 &=& 
  \int \extd \ast A_\mu dx^\mu
  \nonumber\\
  &=&
  \int \sqrt{|{\rm det}g|} \nabla_\mu A^\mu\; d^Dx
  \,.
\end{eqnarray}

If this condition is satisfied, then the $(\cdot)^\dag$ defined in the
$\hat \Omega_{(2)}$ differential calculus is truly the adjoint operation with
respect to \refer{discrete pre-inner product}.

\paragraph{Generality of the inner product.}

Consider any graded inner product on $\Omega\of{\cal A}$ and assume the adjoints
of all operators with respect to this inner product are known.
Then obviously the inner product is completely specified by its action on
$\Omega^0$: Namely let $A, B$ be any two operators of the same grade on $\Omega$
and recall $\ket{A} \defas A\ket{1}$ and $\ket{B} \defas B \ket{1}$.
It follows that
\begin{eqnarray}
  \label{axc}
  \bracket{A}{B}
  &=&
  \bra{1}A^\dag B \ket{1}
  \nonumber\\
  &=&
  \bracket{1}{f}
  \,,
\end{eqnarray}
where
$f = A^\dag B \ket{1} \in \cal A$.
Therefore the inner product is known completely when the linear form
\begin{eqnarray}
  \bracket{1}{\cdot} : \cal A \to \C
\end{eqnarray}
is known.
Any such map must necessarily be of the form
\begin{eqnarray}
  \bracket{1}{f} &=& \sum_{\vec x} f\of{\vec x}\, dV\of{\vec x}
\end{eqnarray}
for some $dV\in \cal A$. 
Plugging this into the symmetrized \refer{axc} yields \refer{discrete pre-inner product}:
\begin{eqnarray}
  \bracket{A}{B}
  &=&
  \int
  \frac{1}{2}
  \left(
    A^\dag B + B^\dag A
  \right)_{\cal S}\of{\vec x}
  \;
  dV
  \,.
\end{eqnarray}
This shows that the class of inner products
\refer{discrete pre-inner product} is in fact the most general class of
inner products on $\Omega\of{\cal A}$ with respect to which elements in $\Omega^0$
are self-adjoint.
The consistency condition
\begin{eqnarray}
  0 &=& \bra{1}\coextd A\ket{1}
\end{eqnarray}
gives a constraint on the otherwise arbitrary function $dV$ in terms of the
supercommutator structure of the $N=2$ calculus of operators acting on
$\Omega\of{\cal A}$.

Because this condition 
can be expressed in terms of elementary edges as
\begin{eqnarray}
  \label{condition that integral over divergence of discrete edge vanishes}
  \int
  (\coextd \delta_{\set{\vec x,\vec x + \vec e_a\of{\vec x}}})_{\cal S}
  \;
  dV
  &=&
  0
  \hspace{1cm}
  \forall\; 
  \vec x,a 
\end{eqnarray}
we turn to an analysis of the diveregence of an elementary edge
in \S\fullref{length, volume and divergence}.

\subsection{Length, volume, and divergence}
\label{length, volume and divergence}

\subparagraph{Theorem:}
{\it
When the result of acting with an elementary adjoint edge on a given elementary edge is taken to be 
in $\hat \Omega^0\of{\hat A} = \hat A$ then the algebra demands that}
\begin{eqnarray}
  \label{statelike part of adjoint edge/edge product}
  \left(
    \delta^\dag_{\set{\vec y + \vec e_b\of{\vec y}, \vec y}}
    \delta_{\set{\vec x, \vec x + \vec e_a\of{\vec x}}}
  \right)_{\cal S}
  &=&
  \delta_{\set{\vec x}}\of{\vec y}
  \delta_{a,b}
  \frac{\ell^2\of{a,\vec x+ \vec e_a\of{\vec x}}}{\epsilon^2}
  \delta_{\set{\vec x + \vec e_a\of{\vec x}}}
\end{eqnarray}
{\it 
for some function $\ell^2\of{a,\vec x,\vec e_a\of{\vec x}}$ that vanishes when 
$\delta_{\set{\vec x,\vec x + \vec e_a\of{\vec x}}} = 0$.}

{\it Proof:}
First of all,
the left hand side obviously vanishes when $\vec x \neq \vec y$, therefore
the factor $\delta_{\set{\vec x}}\of{\vec y}$ on the right. Next, the
left hand side must be an element of $\hat \Omega^0\of{\hat A} = \hat A$.
This implies that it must commute with all other elements in 
$\hat A$, which again implies that $\vec y + \vec e_b\of{\vec y} = \vec x + \vec e_a\of{\vec x}$,
which, since $\vec x = \vec y$ and $e_a{}^\mu$ is invertible, can only be true for
$a = b$. This is enforced by the second factor on the right. So the left hand side
is $\delta_{\set{\vec x + \vec e_a\of{\vec x}}}$ times a real number, which we
encode in the function $\ell^2$. \endofproof.

\subparagraph{Theorem:} 
{\it The divergence of an elementary edge is}
\begin{eqnarray}
  \label{divergence of elementary edge}
  \left(
  \coextd \delta_{\set{\vec x,\vec x + \vec e_a\of{\vec x}}} 
  \right)_{\cal S}
  &=&
  -
  \left(
    \frac{
    {\rm div}\of{e^a}\of{\vec x}
    }{\epsilon} + \frac{\ell^2\of{a,\vec x}}{\epsilon^2}
  \right)
  \delta_{\set{\vec x}}
  +
  \frac{\ell^2\of{a,\vec x + \vec e_a\of{\vec x}}}{\epsilon^2}
  \delta_{\set{\vec x + \vec e_a\of{\vec x}}}
  \,,
\end{eqnarray}
where
\begin{eqnarray}
  {\rm div}\of{e^a}
  &\defas&
  -\left(\coextd \edgesCreator^a\right)_{\cal S}
\end{eqnarray}
is the divergence of the ``macroscopic'' 1-form \refer{discrete ONB creators}.

{\it Proof}:
First we will show that
\begin{eqnarray}
  \label{some lemma in some proof abc}
  \coextd \delta_{\set{\vec x,\vec x + \vec e_a\of{\vec x}}} \ket{1}
  &=&
  f\of{a,\vec x}
  \delta_{\set{\vec x}}
  \ket{1}
  +
  g\of{a,\vec x + \vec e_a\of{\vec x}}
  \delta_{\set{\vec x + \vec e_a\of{\vec x}}}
  \ket{1}
\end{eqnarray}
for some $f$ and $g$. It then follows that 
\begin{eqnarray}
  f &=& 
  -
  \left(
    \frac{{\rm div}\of{e^a}}{\epsilon} + \frac{\ell^2\of{a}}{\epsilon^2}
  \right)
  \nonumber\\
  g &=&
  \frac{\ell^2\of{a}}{\epsilon^2}
  \,.
\end{eqnarray}
The first assertion is proven by multiplying with $\delta_{\set{\vec y}}$:
\begin{eqnarray}
  \delta_{\set{\vec y}}
  \coextd \delta_{\set{\vec x,\vec x + \vec e_a\of{\vec x}}} \ket{1}
  &=&
  \commutator{\delta_{\set{\vec y}}}{\coextd} \delta_{\set{\vec x,\vec x + \vec e_a\of{\vec x}}} \ket{1}
  +
  \coextd \delta_{\set{\vec y}}\delta_{\set{\vec x,\vec x + \vec e_a\of{\vec x}}} \ket{1}
  \,.  
\end{eqnarray}
The first term on the right can only give a contribution if one of the adjoint edges that is produced by
$\commutator{\delta_{\set{\vec y}}}{\coextd}$ has a non-vanishing statelike product with the
original edge. According to \refer{statelike part of adjoint edge/edge product}
this implies that the first term on the right gives a contribution only for 
$\vec y = \vec x$ or
$\vec y = \vec x + \vec e_a\of{\vec x}$. Furthermore the second term on the right is
obviously non-vanishing only for $\vec y = \vec x$. This proofs \refer{some lemma in some proof abc}.

Next we write
\begin{eqnarray}
  \label{some formula in some proof}
  \coextd \delta_{\set{\vec x,\vec x + \vec e_a\of{\vec x}}} \ket{1}
  &=&
  \coextd \left( \delta_{\set{\vec x}} 
    \commutator{\extd}
   { 
   \delta_{\set{\vec x + \vec e_a\of{x}}}
   }
  \right) 
   \ket{1}
  \nonumber\\
  &=&
  \commutator{\coextd}{\delta_{\set{\vec x}}}
  \commutator{\extd}{\delta_{\set{\vec x + \vec e_a\of{x}}}}
  \ket{1}
  +
  \delta_{\set{\vec x}}
  \coextd \commutator{\extd} {\delta_{\set{\vec x + \vec e_a\of{\vec x}}}}
  \ket{1}
  \,.
\end{eqnarray}
The first term on the right gives
\begin{eqnarray}
  \commutator{\coextd}{\delta_{\set{\vec x}}}
  \commutator{\extd}{\delta_{\set{\vec x + \vec e_a\of{x}}}}
  \ket{1}
  &=&
  \delta^\dag_{\set{\vec x + \vec e_a\of{\vec x},\vec x}}
  \delta_{\set{\vec x, \vec x + \vec e_a\of{\vec x}}}
  \ket{1}    
  \nonumber\\
  &\equalby{statelike part of adjoint edge/edge product}&
  \frac{\ell^2\of{a,\vec x + \vec e_a\of{\vec x}}}{\epsilon^2}
  \delta_{\set{\vec x + \vec e_a\of{\vec x}}}
  \ket{1}
  \,,
\end{eqnarray}
because according to 
\refer{statelike part of adjoint edge/edge product}
all other products of adjoint edges with edges that might appear
annihilate the vacuum state and hence give no contribution. 

Due to \refer{some lemma in some proof abc}
the second term on the right of \refer{some formula in some proof} is  
equal to $\delta_{\set{\vec x}}$ times
the left hand side of \refer{some formula in some proof}:
\begin{eqnarray}
  \delta_{\set{\vec x}}
  \coextd \commutator{\extd} {\delta_{\set{\vec x + \vec e_a\of{\vec x}}}}
  \ket{1}
  &=&
  \delta_{\set{\vec x}}
  \coextd 
  \delta_{\set{\vec x}}
  \commutator{\extd} {\delta_{\set{\vec x + \vec e_a\of{\vec x}}}}
  \ket{1}
\end{eqnarray}
so that \refer{some formula in some proof} becomes
\begin{eqnarray}
  \coextd \delta_{\set{\vec x,\vec x + \vec e_a\of{\vec x}}} \ket{1}
  &=&
  \frac{\ell^2\of{a,\vec x + \vec e_a\of{\vec x}}}{\epsilon^2}
  \delta_{\set{\vec x + \vec e_a\of{\vec x}}}
  \ket{1}
  +
  \delta_{\set{\vec x}}
  \coextd \delta_{\set{\vec x,\vec x + \vec e_a\of{\vec x}}} \ket{1}
  \,. 
\end{eqnarray}
In other words there is a function $f\of{a,\vec x}$ such that
\begin{eqnarray}
  \label{step in proof of discrete divergence relation}
  \coextd \delta_{\set{\vec x,\vec x + \vec e_a\of{\vec x}}} \ket{1}
  &=&
  \frac{\ell^2\of{a,\vec x + \vec e_a\of{\vec x}}}{\epsilon^2}
  \delta_{\set{\vec x + \vec e_a\of{\vec x}}}
  \ket{1}
  +
  f\of{a,\vec x}
  \delta_{\set{\vec x}}
  \ket{1}
  \,. 
\end{eqnarray}
It follows that 
\begin{eqnarray}
  &&
  (\coextd \edgesCreator^a)_{\cal S}
  \;=\;
  \frac{\ell^2\of{a}}{\epsilon}
  + 
  \epsilon \,f
  \nonumber\\
  &\Rightarrow&
  f \;=\; 
  -
  \left(
  \frac{
  {\rm div}\of{e^a}
  }{\epsilon	}
  +
  \frac{\ell^2\of{a}}{\epsilon^2}
  \right)
  \,.
\end{eqnarray}
Plugging this into \refer{step in proof of discrete divergence relation} completes the proof.
\endofproof

\subparagraph{Integral over total divergence.}
Equation \refer{condition that integral over divergence of discrete edge vanishes}
used the integral over the divergence of a fundamental edge. This can now
be evaluated more explicitly:
\begin{eqnarray}
  \int (\coextd \delta_{\set{\vec x,\vec x + \vec e_a\of{\vec x}}})_{\cal S}
  \; dV
  &\equalby{divergence of elementary edge}&
    -\left(    
    \frac{{\rm div}\of{e^a}\of{\vec x}}{\epsilon}
    +
    \frac{\ell^2\of{a,\vec x}}{\epsilon^2}
    \right)
    dV\of{\vec x}
    + \frac{\ell^2\of{a,\vec x + \vec e_a\of{x}}}{\epsilon^2}
    dV\of{\vec x + \vec e_a\of{\vec x}}
  \,.
  \nonumber\\
\end{eqnarray}
The vanishing of this expression is equivalent to
\begin{eqnarray}
  \label{relation of discrete divergence to length and volume}
  {\rm div}\of{e_a}\of{\vec x}\; dV\of{\vec x}
  &=&
  \frac{1}{\epsilon}
  \left(
  \ell^2\of{a,\vec x + \vec e_a\of{x}}
    dV\of{\vec x + \vec e_a\of{\vec x}}
  -
  \ell^2\of{a,\vec x}dV\of{\vec x}
  \right)
   \,.
\end{eqnarray}

Given the functions $dV$ and $\ell^2$ this can be used to solve for 
${\rm div}\of{e^a}$ such that \refer{discrete pre-inner product} does indeed define an
inner product on the space $\cal H$ of states with respect to which
$(\cdot)^\dag$ is the adjoint operation.

Note that in the continuum theory the equation\footnote{
This follows from
\begin{eqnarray}
  {\rm div}\of{e^a}\; {\bf vol}
  &=&
  \ast \ast \extd \ast e^a{}_\mu \extd x^\mu
  \nonumber\\
  &=&
  \extd \left({\bf e}^a\rightharpoonup {\bf vol}\right)
  \nonumber\\
  &=&
  \antiCommutator{\extd}{{\bf e}^a	\rightharpoonup} {\bf vol}
  \nonumber\\
  &=&
  {\cal L}_{e^a} {\bf vol}
  \,.
\end{eqnarray}
}
\begin{eqnarray}
  {\rm div}\of{e_a} {\bf vol}
  &=&
  {\cal L}_{e_a} {\bf vol}
\end{eqnarray}
holds, where ${\bf vol}$ is the volume form and ${\cal L}_{e_a}$ the Lie derivative
on forms along $e_a$.

Inserting the consistency condition \refer{relation of discrete divergence to length and volume}
into \refer{divergence of elementary edge} 
fixes the divergence of a fundamental edge in terms of $\ell^2$ and $dV$:
\begin{eqnarray}
  \label{fixed divergence of elementary edge}
  \left(
  \coextd \delta_{\set{\vec x,\vec x + \vec e_a\of{\vec x}}} 
  \right)_{\cal S}
  &=&
  \frac{\ell^2\of{a,\vec x + \vec e_a\of{x}}}{\epsilon^2}
  \left(
   \delta_{\set{\vec x + \vec e_a\of{\vec x}}}
   -
    \frac{dV\of{\vec x + \vec e_a\of{\vec x}}}{dV\of{\vec x}}
    \delta_{\set{\vec x}}
  \right)
  \,.
\end{eqnarray}

\paragraph{Non-orthogonal and lightlike edges.}

The restriction, found above, for the edges $e_a$ at a given node to 
be mutually orthogonal with respect to a metric encoded in the adjoint
operator $(\cdot)^\dag$ has its merits and its disadvantages: On the one hand
side every restriction of possible arbitrariness by formal restrictions is
a relief, on the other hand objects like 
for instance lightlike edges,
which are desireable from the physical point of view,
are also also ruled out by this requirement.  

The obvious way out is to use adjointness relations which violate the
assumption that 0-forms are strictly self-adjoint with respect to them.
It is perfectly sensible that 0-forms should be self-adjoint only up to
a lattice correction proportional to $\epsilon$. 

A large class of
useful relations of this form are found by introducing any
$(\cdot)^\dag$-self-adjoint invertible operator $\gop$ 
and the modified inner product
\begin{eqnarray}
  \bracket{\cdot}{\cdot}_\gop
  &\defas&
  \bra{\cdot}\gop^{-1}\ket{\cdot}
\end{eqnarray}
induced by it.
This gives us a new adjointness relation
\begin{eqnarray}
  A^{\dag_\gop}
  &=&
  \gop
  A^\dag
  \gop^{-1}
  \,.
\end{eqnarray}
The 0-forms will in general not be self-adjoint with respect to this new 
inner product.

Probably the most important of such operators $\gop$ are those
that are \emph{Krein space operators} which turn the discrete Riemannian metric
into a semi-Riemannian one (see \cite{Strohmaier:2001} for a discussion of
semi-Riemannian non-commutative geometry with Krein space operators). 
An example for such a case is discussed in
\S\fullref{Example: Hyper diamond model and stochastic calculus.}.

\subsection{Volume form and Hodge duality}
\label{volume form and hodge duality}

\subparagraph{Volume-like forms.}

On a $D$ dimensional $N=2$ complex 
with inner product $\bra{\cdot}\gop^{-1}\ket{\cdot}$ and
adjoint $(\cdot)^{\dag_\gop}$
a \emph{volume-like form}
is a $D$-form $\ket{\evol} = {\evol}\ket{1}$ that is annihilated by $\ecoextd$:
\begin{eqnarray} 
  \label{definition volume-like form}
  \ecoextd  \ket{\evol} &=& 0
  \,.
\end{eqnarray}

\subparagraph{Volume form.}
A volume form is a volume-like form that furthermore satisfies
\begin{eqnarray}
  \label{definition of proper volume form}
  {\evol}^\edag {\evol}\ket{1} &=& \ket{1}
  \,.
\end{eqnarray}
(This condition, familiar from the continuum, will fix the as yet undetermined
``volume element'' $dV$ that enters the definition of any inner product, \cf
\S\fullref{vector space VOmegaA and inner products}.

At least for sufficiently non-pathologic
$D$-dimensional complexes the volume form should exist and be unique.
This is certainly the case for the concrete examples that we will study
(e.g. cubic graphs), as well as for the continuum theory.)

Because of $\ecoextd = \gop\coextd \gop^{-1}$ 
a volume-like form $\ket{\evol}$ is related to the respective
volume-like form $\ket{\vol}$ by
\begin{eqnarray}
  \label{relation evol to vol}
  \ket{\evol}
  &=&
  \gop\ket{\vol}
  \,.
\end{eqnarray}
Because $\gop$ is grade preserving it must simply multiply $\ket{\vol}$ by
a function. 
Motivated by the construction in \S\fullref{metrics on cubic graphs}
this function (multiplied from the right) will formally be called ${\rm det}\of{g}$, i.e.
\begin{eqnarray}
  \label{definition det eta}
  \ket{\evol}
  &\defas&
  \vol \,{\rm det}\of{g}\,\ket{1}
  \,.
\end{eqnarray}
We note again, that this equation \emph{defines} ${\rm det}\of{g} \in {\cal A}$.

Next, using the volume form the familiar notion of ``integrating a $D$-form over
a $D$-dimensional space'' can be formalized by means of the following definition:

\subparagraph{Integrating a $D$ form over a $D$-dimensional complex.}
\begin{eqnarray}
  \label{integral over discrete D form}
  \int f \,{\evol}
  &\defas&
  \int f\; dV
  \hspace{1cm}
  f \in {\cal A}
  \,.
\end{eqnarray}

\paragraph{Hodge duality.}

The volume form serves as a ``vacuum'' with respect to $\ecoextd$, just
as $\ket{1}$ is the vacuum with respect to $\extd$.

The operator 
$\ehodge : \Omega \to \Omega$ with $\ehodge \Omega^p = \Omega^{D-p}$
which intertwines the sequences
\begin{eqnarray}
  &&
  0 \stackrel{\extd}{\longrightarrow}
  \Omega^0 \stackrel{\extd}{\longrightarrow} \Omega^1 
  \stackrel{\extd}{\longrightarrow}
  \cdots \stackrel{\extd}{\longrightarrow} \Omega^D
  \stackrel{\extd}{\longrightarrow} 0
  \nonumber\\
  &&0 \stackrel{\ecoextd}{\longleftarrow}
  \Omega^0 \stackrel{\ecoextd}{\longleftarrow} \Omega^1 
  \stackrel{\ecoextd}{\longleftarrow}
  \cdots \stackrel{\ecoextd}{\longleftarrow} \Omega^D
  \stackrel{\ecoextd}{\longleftarrow} 0
\end{eqnarray}
so that 
\begin{eqnarray}
  \label{the definition of Hodge duality a}
  \ehodge \extd &=& \;\ecoextd \ehodge\, (-1)^{\numberOperator+1}
\end{eqnarray}
is called the \emph{Hodge star operator}. (Here $\numberOperator$ is the 
\emph{number operator} defined by
$\numberOperator \ket{A} = p\ket{a} \;\Leftrightarrow\; A\in \Omega^p$.)

When $\ehodge$ is invertible this may be rewritten as
\begin{eqnarray}
  \label{ecoextd in terms of ehodge and extd}
  \ecoextd
  &=&
  \ehodge \extd \invehodge (-1)^{D - \numberOperator+1}
  \,.
\end{eqnarray} 
The action of the Hodge star operator can be expressed explicitly
in terms of the volume form and the adjoint operation:

$\,$\\

\subparagraph{Theorem:} The action of the Hodge star operator is given by
\begin{eqnarray}
  \label{explicit Hodge action}
  \ehodge \ket{A} &=& (A_{\cal S})^\edag \ket{\evol}
  \,.
\end{eqnarray}
{\it Proof:}
Let $A \defas a_0 \commutator{\extd}{a_1} \commutator{\extd}{a_2} 
    \cdots 
  \commutator{\extd}{a_p}$ be an element in $\hat\Omega\of{\hat {\cal A}}$, then
\begin{eqnarray}
  \label{proof of Hodge star construction}
  \ecoextd  \ehodge  \ket{A}
  &\equalby{explicit Hodge action}&
  \ecoextd  
  \commutator{a_p^\edag}{\ecoextd}\commutator{a_{p-1}^\edag}{\ecoextd} 
   \cdots \commutator{a_1^\edag}{\ecoextd} a_0^\edag 
  \ket{\evol}
  \nonumber\\
  &=&
  (-1)^{p+1}
  \commutator{a_p^\edag}{\ecoextd}\commutator{a_{p-1}^\edag}{\ecoextd} \cdots 
  \commutator{a_1^\edag}{\ecoextd} 
  \commutator{a_0^\edag}{\ecoextd} 
  \ket{\evol}
  \nonumber\\
  &\equalby{explicit Hodge action}&
  (-1)^{p+1}\ehodge \commutator{\extd}{a_0} \commutator{\extd}{a_1}\cdots \commutator{\extd}{a_p} \ket{1}
  \nonumber\\
  &=&
  (-1)^{p+1}\ehodge \extd \; a_0 \commutator{\extd}{a_1}\cdots \commutator{\extd}{a_p} \ket{1}
  \nonumber\\
  &=&
  (-1)^{p+1}\ehodge\extd \ket{A}
  \,.
\end{eqnarray}

\subsection{Cubic graphs}
\label{subsection: cubic graphs}

So far from \refer{statelike part of adjoint edge/edge product} 
we only know the statelike part of the product of
an adjoint edge with an edge. Further information of the form of this 
operator is obtained by using the structure of the $\Omega\of{{\cal A},\extd}$
calculus. For instance if $\delta_{\set{\vec x, \vec x + \vec e_a\of{\vec x}}} A = 0$
then
\begin{eqnarray}
  0 &=& 
  \delta^\dag_{\vec x+\vec e_b\of{\vec x},\vec x}
  \delta_{\set{\vec x, \vec x + \vec e_a\of{\vec x}}}
  A
  \nonumber\\
  &=&  
  \left(
  \left(
  \delta^\dag_{\vec x+\vec e_b\of{\vec x},\vec x}
  \delta_{\set{\vec x, \vec x + \vec e_a\of{\vec x}}}
  \right)_{\cal S}
  +  
  \left(
  \delta^\dag_{\vec x+\vec e_b\of{\vec x},\vec x}
  \delta_{\set{\vec x, \vec x + \vec e_a\of{\vec x}}}
  \right)_{\cal N}  
  \right)
  A
\end{eqnarray}
gives a condition on the possible values of 
$
  \left(
  \delta^\dag_{\vec x+\vec e_b\of{\vec x},\vec x}
  \delta_{\set{\vec x, \vec x + \vec e_a\of{\vec x}}}
  \right)_{\cal N}  
$.
These conditions must be solved for each possible graph, i.e. for each
possbible ${\bf G}$ in \refer{graph operator in terms of vielbein edges},  seperately.

\subparagraph{Definition:} A \emph{cubic graph} is a graph of the form
\refer{graph operator in terms of vielbein edges} for which
\begin{eqnarray}
  \label{definition of cubic graph}
  \delta_{\set{\vec x,\vec x + \vec e_a}}
  \delta_{\set{\vec x + \vec e_a,\vec x + \vec e_a + \vec e_a}}
  &=& 0
  \nonumber\\
  \delta_{\set{\vec x,\vec x + \vec e_a}}
  \delta_{\set{\vec x + \vec e_a,\vec x + \vec e_a + \vec e_b}}
   &=&
  -
  \delta_{\set{\vec x,\vec x + \vec e_b}}
  \delta_{\set{\vec x + \vec e_b,\vec x + \vec e_b + \vec e_a}}
  \,.  
\end{eqnarray}

On cubic graphs the objects \refer{discrete ONB creators} obviously satisfy
\begin{eqnarray}
  \label{anticommutation relations creators/creators on cubic graphs}
  \edgesCreator^a\edgesCreator^b + \edgesCreator^b \edgesCreator^a &=& 0
  \nonumber\\
  \edgesAnnihilator^a\edgesAnnihilator^b + \edgesAnnihilator^b \edgesAnnihilator^a &=& 0
  \hspace{1cm}
  \forall\; a,b \in \set{1,\cdots, D}
  \,.
\end{eqnarray}
But mixed anticommutators are generally more complex:

\subparagraph{Theorem:} On a cubic graph it is possible to have the following
anticommutation relations
\begin{eqnarray}
  \label{discrete edge anticommutators for cubic graphs}
  \delta^\dag_{\set{\vec x + \vec e_a ,\vec x}}
  \delta_{\set{\vec x,\vec x + \vec e_a}}
  &=&
  \frac{\ell^2\of{a,\vec x + \vec e_a}}{\epsilon^2}
    \delta_{\set{\vec x + \vec e_a}}
    -
\frac{\ell^2\of{a,\vec x + \vec e_a}}{\ell^2\of{a,\vec x + \vec e_a + \vec e_a}}
    \delta_{\set{\vec x + \vec e_a, \vec x + \vec e_a + \vec e_a}}
    \delta^\dag_{\set{\vec x + \vec e_a + \vec e_a,\vec x + \vec e_a}}
  \nonumber\\
  \delta^\dag_{\set{\vec x + \vec e_a, \vec x}}
  \delta_{\set{\vec x,\vec x + \vec e_b}}
  &=&
  -
  \frac{\ell^2\of{a,\vec x + \vec e_a}}{\ell^2\of{a,\vec x + \vec e_a + \vec e_b}}
  \delta_{\set{\vec x + \vec e_a, \vec x + \vec e_a + \vec e_b}}
  \delta^\dag_{\set{\vec x + \vec e_a + \vec e_b, \vec x + \vec e_b}}
  \hspace{1cm}
  \mbox{for $a \neq b$}
  \,.
\end{eqnarray}

(We \emph{conjecture} that thes relations for cubic graphs are in fact unique, i.e.
no consistent modification is possible.)

{\it Proof:}
As mentioned above, we essentially need to check associativity of the product
of an adjoint edge with the product of two edges. For instance the expression
\begin{eqnarray}
  \delta^\dag_{\set{\vec x + \vec e_a, \vec x}}
  \delta_{\set{\vec x, \vec x + \vec e_a}}
  \delta_{\set{\vec x + \vec e_a, \vec x + \vec e_a + \vec e_a}} 
\end{eqnarray}
needs to vanish, because the product of the last two factors does so by assumption of 
a cubic graph structure.
We now check if, when evaluating the product of the first two factors first
(using \refer{discrete edge anticommutators for cubic graphs}) and
then multiplying with the last factor, the result still vanishes. And indeed:
\begin{eqnarray}
  \delta^\dag_{\set{\vec x + \vec e_a, \vec x}}
  \delta_{\set{\vec x, \vec x + \vec e_a}}
  \delta_{\set{\vec x + \vec e_a, \vec x + \vec e_a + \vec e_a}} 
  &\equalby{discrete edge anticommutators for cubic graphs}&
    \frac{\ell^2\of{a,\vec x + \vec e_a}}{\epsilon^2}
    \delta_{\set{\vec x + \vec e_a, \vec x + \vec e_a + \vec e_a}} 
   \nonumber\\
    &&
    -
    \frac{\ell^2\of{a,\vec x + \vec e_a}}{\ell^2\of{a,\vec x + \vec e_a + \vec e_a}}
    \delta_{\set{\vec x + \vec e_a, \vec x + \vec e_a + \vec e_a}}
    \delta^\dag_{\set{\vec x + \vec e_a + \vec e_a,\vec x + \vec e_a}}
  \delta_{\set{\vec x + \vec e_a, \vec x + \vec e_a + \vec e_a}} 
  \nonumber\\
  &\equalby{discrete edge anticommutators for cubic graphs}&
  0 
  +
  \frac{\ell^2\of{a,\vec x + \vec e_a}}{\ell^2\of{a,\vec x + \vec e_a + \vec e_a}}
  \delta_{\set{\vec x + \vec e_a, \vec x + 2\vec e_a }}
  \delta_{\set{\vec x + 2\vec e_a , \vec x + 3\vec e_a }}
  \delta^\dag_{\set{\vec x + 3\vec e_a ,\vec x + 2\vec e_a }}
  \nonumber\\
  &\equalby{definition of cubic graph}& 
  0 
  \,.
\end{eqnarray}
Similarly, to check the second relation we write
\begin{eqnarray}
  \label{first part of second relation}
  \delta^\dag_{\set{\vec x + \vec e_a, \vec x}}
  \left(
    \delta_{\set{\vec x, \vec x  + \vec e_b}}
    \delta_{\set{\vec x + \vec e_b, \vec x + \vec e_b + \vec e_a}}
  \right)
  &\equalby{discrete edge anticommutators for cubic graphs}&
  -
  \frac{\ell^2\of{a,\vec x + \vec e_a}}{\ell^2\of{a,\vec x + \vec e_a + \vec e_b}}
  \delta_{\set{\vec x + \vec e_a, \vec x + \vec e_a + \vec e_b}}
  \delta^\dag_{\set{\vec x + \vec e_a + \vec e_b, \vec x + \vec e_b}}
  \delta_{\set{\vec x + \vec e_b, \vec x + \vec e_b + \vec e_a}}
  \nonumber\\
  &\equalby{discrete edge anticommutators for cubic graphs}&
  -
  \frac{\ell^2\of{\vec x + \vec e_a}}{\epsilon^2}
  \delta_{\set{\vec x + \vec e_a, \vec x + \vec e_a + \vec e_b}}
  \nonumber\\
  &&  
  +
  \frac{\ell^2\of{a,\vec x + \vec e_a}}{\ell^2\of{a,\vec x + 2\vec e_a + \vec e_b}}
  \delta_{\set{\vec x + \vec e_a, \vec x + \vec e_a + \vec e_b}}
  \delta_{\set{\vec x + \vec e_a + \vec e_b, \vec x + 2\vec e_a + \vec e_b}}
  \delta^\dag_{\set{\vec x + 2\vec e_a + \vec e_b,\vec x + \vec e_a + \vec e_b}}
  \nonumber\\
\end{eqnarray}
and
\begin{eqnarray}
  \label{second part of second relation}
  \delta^\dag_{\set{\vec x + \vec e_a, \vec x}}
  \left(
    \delta_{\set{\vec x, \vec x  + \vec e_a}}
    \delta_{\set{\vec x + \vec e_a, \vec x + \vec e_b + \vec e_a}}
  \right)
  &\equalby{discrete edge anticommutators for cubic graphs}&
  \frac{\ell^2\of{a,\vec x + \vec e_a}}{\epsilon^2}
    \delta_{\set{\vec x + \vec e_a, \vec x + \vec e_b + \vec e_a}}
  \nonumber\\
  &&
    -
  \frac{\ell^2\of{a,\vec x + \vec e_a}}{\ell^2\of{a,\vec x + \vec e_a + \vec e_a}}
    \delta_{\set{\vec x + \vec e_a, \vec x + \vec e_a + \vec e_a}}
    \delta^\dag_{\set{\vec x + \vec e_a + \vec e_a,\vec x + \vec e_a}}
   \delta_{\set{\vec x + \vec e_a, \vec x + \vec e_b + \vec e_a}}
    \nonumber\\
  &\equalby{discrete edge anticommutators for cubic graphs}&
  \frac{\ell^2\of{a,\vec x + \vec e_a}}{\epsilon^2}
    \delta_{\set{\vec x + \vec e_a, \vec x + \vec e_b + \vec e_a}}
  \nonumber\\
  &&
  +
  \frac{\ell^2\of{a,\vec x + \vec e_a}}{\ell^2\of{\vec x + 2\vec e_a + \vec e_b}}
  \delta_{\set{\vec x + \vec e_a, \vec x + \vec e_a + \vec e_a}}
  \delta_{\set{\vec x + \vec e_a + \vec e_a, \vec x + 2\vec e_a + \vec e_b}}
  \delta^\dag_{\set{\vec x + 2\vec e_a + \vec e_b, \vec x + \vec e_a + \vec e_b}}
  \,.
  \nonumber\\
\end{eqnarray}
Here \refer{first part of second relation} and \refer{second part of second relation}
are indeed equal up to a sign, the way it should be.
\endofproof

\subparagraph{Corollary:} 
For the vielbein forms  and adjoint forms \refer{discrete ONB creators}
it follows that on cubic graphs
\begin{eqnarray}
  \edgesAnnihilator^a\edgesCreator^a
  &\equalby{discrete edge anticommutators for cubic graphs}&
  \ell^2\of{a}
  -
  \frac{\ell^2\of{a,\cdot}}{\ell^2\of{a,\cdot + \vec e_a}}
  \edgesCreator^a\edgesAnnihilator^a
  \nonumber\\
  \edgesAnnihilator^a
  \edgesCreator^b
  &\equalby{discrete edge anticommutators for cubic graphs}&
  -
  \frac{\ell^2\of{a,\cdot}}{\ell^2\of{a,\cdot + \vec e_b}}
  \edgesCreator^b\edgesAnnihilator^a
  \,.
\end{eqnarray}
Therefore on cubic graphs these satisfy the canonical anticommutation relations 
without lattice corrections precisely
if the ``lengths'' $\ell$ of all elementary edges are identical:
\begin{eqnarray}
  \label{condition for discrete canonical anticommutators}
  \antiCommutator{\edgesCreator^a}{\edgesAnnihilator^b}
  =
  \delta^{a,b}
  \ell^2
  &\Leftrightarrow&
  \ell^2\of{a,\vec x} = \ell^2 = {\rm const}
  \,.
\end{eqnarray}

{\it For the rest of this section we consider cubic graphs with 
\begin{eqnarray}
  \ell^2 = 1
  \,.
\end{eqnarray}
}

\newpage
\subparagraph{Divergence on cubic graphs.} {\it
  The divergence of an arbitrary $p$-form on a cubic graph with $\ell = 1$ is
  determined by
 }
\begin{eqnarray}
  \label{discrete divergence on cubic graphs}
  &&\coextd
  \edgesCreator^{a_1}\edgesCreator^{a_2}\cdots \edgesCreator^{a_{p+1}}\delta_{\vec y}
  \ket{1}
  \nonumber\\
  &=& 
  \frac{(-1)^p}{\epsilon\,p!}
  \sum\limits_\pi
  \sigma\of{\pi}
  \;
    \edgesCreator^{a_{\pi\of{1}}}\edgesCreator^{a_{\pi\of{2}}}\cdots 
     \edgesCreator^{a_{\pi\of{p}}}    
  \;
  \left(
    \delta_{\set{\vec y}}
    -
    \delta_{\set{\vec y - \vec e_{a_{\pi\of{p+1}}}}}
    \frac{dV\of{\vec y}}{dV\of{\vec y - \vec e_{a_{\pi\of{p+1}}}}}
  \right)
  \ket{1}
  \,.
  \nonumber\\
\end{eqnarray}
(Here $\pi$ is a permutation of $\set{1,2,\cdots,p}$ and $\sigma\of{\pi}$ is its signature,
which ensures that the right hand side has the same behaviour under index permutation
as the left hand side.)

Note that for $p=1$ this does indeed reproduce \refer{statelike part of adjoint edge/edge product},
which was derived for general graphs.

{\it Proof:}
The ``matrix elements'' of $\coextd$ with respect to elementary $p$-forms  
(which constitute a basis of $\Omega\of{\cal A}$) are (note that we can
always bring the $\edgesCreator^a$ in both factors into the given order)
\begin{eqnarray}
  &&
  \bracket{
    \edgesCreator^{a_1}\edgesCreator^{a_2}\cdots \edgesCreator^{a_p}
    \delta_{\set{\vec x}}
  }{
    \coextd
    \edgesCreator^{a_1}\edgesCreator^{a_2}\cdots \edgesCreator^{a_p}\edgesCreator^{a_{p+1}}\delta_{\set{\vec y}}
  }
  \nonumber\\
  &=&
  \bracket{
    \extd
    \edgesCreator^{a_1}\edgesCreator^{a_2}\cdots \edgesCreator^{a_p}
    \delta_{\set{\vec x}}
  }{
    \edgesCreator^{a_1}\edgesCreator^{a_2}\cdots \edgesCreator^{a_p}\edgesCreator^{a_{p+1}}\delta_{\set{\vec y}}
  }  
  \nonumber\\
  &\equalby{extd action in terms of G with no intermediate edges}&
  (-1)^p
  \bracket{
    \edgesCreator^{a_1}\edgesCreator^{a_2}\cdots \edgesCreator^{a_p}
    \left(
      \delta_{\set{\vec x - \vec e_{a_{p+1}},\vec x}}
      -
      \delta_{\set{\vec x, \vec x + \vec e_{a_{p+1}}}}
    \right)
  }{
    \edgesCreator^{a_1}\edgesCreator^{a_2}\cdots \edgesCreator^{a_p}\edgesCreator^{a_{p+1}}\delta_{\set{\vec y}}
  }    
  \nonumber\\
  &=&
  (-1)^p
  \frac{1}{\epsilon}
  \bracket{
    \edgesCreator^{a_1}\edgesCreator^{a_2}\cdots \edgesCreator^{a_p}\edgesCreator^{a_{p+1}}
    \left(
      \delta_{\set{\vec x}}
      -
      \delta_{\set{\vec x + \vec e_{a_{p+1}} }}
    \right)
  }{
    \edgesCreator^{a_1}\edgesCreator^{a_2}\cdots \edgesCreator^{a_p}\edgesCreator^{a_{p+1}}\delta_{\set{\vec y}}
  }    
  \nonumber\\
  &\equalby{condition for discrete canonical anticommutators}&
  (-1)^p
  \frac{1}{\epsilon}
  \bracket{
      \delta_{\set{\vec x}}
      -
      \delta_{\set{\vec x + \vec e_{a_{p+1}} }}
  }{
    \delta_{\set{\vec y}}
  }
  \nonumber\\
  &=&
  (-1)^p
  \frac{1}{\epsilon}
  \left(
    \delta_{\set{\vec y}}\of{\vec x}\, 
    -
    \delta_{\set{\vec y}}\of{\vec x + \vec e_{a_{p+1}}}\,     
  \right)
  dV\of{\vec y}
  \,.
\end{eqnarray}
This is indeed reproduced by \refer{discrete divergence on cubic graphs}:
\begin{eqnarray}
  &&
  \bracket{
    \edgesCreator^{a_1}\edgesCreator^{a_2}\cdots \edgesCreator^{a_p}
    \delta_{\set{\vec x}}
  }{
    \coextd
    \edgesCreator^{a_1}\edgesCreator^{a_2}\cdots \edgesCreator^{a_p}\edgesCreator^{a_{p+1}}\delta_{\set{\vec y}}
  }
  \nonumber\\
  &\equalby{discrete divergence on cubic graphs}&
  \frac{(-1)^p}{\epsilon}
  \bracket{
    \edgesCreator^{a_1}\edgesCreator^{a_2}\cdots \edgesCreator^{a_p}
    \delta_{\set{\vec x}}
  }{
   \edgesCreator^{a_1}\edgesCreator^{a_2}\cdots \edgesCreator^{a_p}
   \left(
     \delta_{\set{\vec y}}
     -
     \delta_{\set{\vec y - \vec a_{a_{\pi\of{p+1}}}}}
     \frac{dV\of{\vec y}}{dV\of{\vec y - \vec e_{a_{p+1}}}}
   \right)
 }
 \nonumber\\
  &=&
  \frac{(-1)^p}{\epsilon}
  \left(
    \delta_{\set{\vec y}}\of{\vec x}\, 
    -
    \delta_{\set{\vec y}}\of{\vec x + \vec e_{a_{p+1}}}\,     
  \right)
  dV\of{\vec y}
\end{eqnarray}
\endofproof

\subparagraph{Corollary:}  For $dV = {\rm const}$
\begin{eqnarray}
  \label{relation coextd and lpartial}
  \coextd \ket{\edgesCreator^{a_1}\cdots \edgesCreator^{d_p} f}
  &=&
  -
  \sum\limits_r
  \ket{
  \edgesCreator^{a_1}\cdots\edgesCreator^{a_{r-1}}\edgesCreator^{a_{r+1}}
  \cdots \edgesCreator^{a_D} 
  (\lpartial_{a_r} f)
  }
  \,,
\end{eqnarray}
which involves the forward discrete derivative \refer{forward discrete derivative}.
Compare this with
\begin{eqnarray}
  \label{relation extd and rpartial}
  \extd \ket{\edgesCreator^{a_1}\cdots \edgesCreator^{d_p} f}
  &=&
  (-1)^p
  \ket{
  \edgesCreator^{a_1}\cdots\edgesCreator^{a_D} 
  \edgesCreator^b
  (\rpartial_{b} f)
  }
  \,,
\end{eqnarray}
which involves the backward discrete derivative \refer{backward discrete derivative}.
This phenomenon is further studied in \S\fullref{partial derivative operator}. 

\subparagraph{Example: Divergence in terms of elementary edges.}

Equation \refer{discrete divergence on cubic graphs} 
owes its relative concise form to the fact that the equivalence
\refer{sums of paths with the same endpoints vanish} 
of ``paths of two edges'' with the same source and target nodes is encoded in
the anticommutativity of the $\edgesCreator^a$. When this is reexpressed in terms
of elementary edges, using
\begin{eqnarray}
  \edgesCreator^a \delgen{\vec x+ \edge_a}
  &=&
  \delgen{\vec x, \vec x+ \edge_a}
  \nonumber\\
  \edgesCreator^a \edgesCreator^b\delgen{\vec x + \edge_a + \edge_b}
  &=&
  \delgen{\vec x,\vec x + \edge_a,\vec x + \edge_a + \edge_b}
\end{eqnarray}
 one obtains formulas such as the following (for $dV={\rm const}$):
\begin{eqnarray}
  \label{example formulas for divergence of elementary edges}
  \coextd \delgen{\vec x , \vec x + \edge_a}\ket{1}
  &=&
  \frac{1}{\epsilon}
  \left(
    \delgen{\vec x + \edge_a} - \delgen{\vec x}
  \right)\ket{1}
  \nonumber\\
  \coextd \delgen{\vec x , \vec x + \edge_a, \vec x + \edge_a + \edge_b}\ket{1}
  &=&
  \frac{1}{\epsilon}
  \left(
    \edgesCreator^b\left(
      \delgen{\vec x + \edge_a + \edge_b} - \delgen{\vec x + \edge_b}
    \right)
    -
    \edgesCreator^a\left(
      \delgen{\vec x + \edge_a + \edge_b} - \delgen{\vec x + \edge_a}
    \right)
  \right)\ket{1}  
  \nonumber\\
  &=&
  \frac{1}{\epsilon}
  \left(
  \delgen{\vec x + \edge_a,\vec x + \edge_a + \edge_b}
  + 
  \delgen{\vec x, \vec x + \edge_a}
  -
  \delgen{\vec x + \edge_b, \vec x + \edge_a + \edge_b}
  -
  \delgen{\vec x,\vec x + \edge_b}
  \right)
  \,.
  \nonumber\\
\end{eqnarray}
In words: The divergence of a ``path'' of fundamental edges is the 
(weighted) sum of paths obtained
by removing an edge from the beginning or from the end of the original path
as well as of all paths that the original one is equal to up to a sign.

In terms of chains this translates to the usual boundary map, 
\cf \refer{explicit action of discrete boundary operator} and
figure \ref{boundary.eps} on page \pageref{boundary.eps}.

\subparagraph{Volume form on cubic graphs.}
{\it
Volume-like forms (\cf \refer{definition volume-like form}) on cubic graphs 
are precisely of the form
\begin{eqnarray}
  \label{volume-like forms on cubic graphs}
  \ket{\evol}
  &=& 
  c\, \edgesCreator^{1}\edgesCreator^2 \cdots \edgesCreator^{D} \frac{{\rm det}\of{g}}{dV}\ket{1}
  \hspace{1cm}
  \mbox{ for $c \in \C$}
\end{eqnarray}
(where ${\rm det}\of{g}$ was defined in \refer{definition det eta}).
These become proper volume forms (\cf \refer{definition of proper volume form}) iff
\begin{eqnarray}
  \label{relation dV to det(eta) in discrete geometry}
  dV
  &=& 
  c \sqrt{{\rm det}\of{g}}
  \,,
\end{eqnarray}
which then read
\begin{eqnarray}
  \ket{\evol}
  &=&
  \edgesCreator^{1}\edgesCreator^2 \cdots \edgesCreator^{D} \sqrt{{\rm det}\of{g}} \ket{1}
  \,.
\end{eqnarray}
}

{\it Proof:}
First we check the defining condition \refer{definition volume-like form} for volume-like forms
for the case $\gop = {\rm id}$ (which is sufficient, due to \refer{relation evol to vol}):
\begin{eqnarray}
  \coextd
  \ket{\vol}
  &=&
  c
  \coextd
  \edgesCreator^{1}\edgesCreator^2 \cdots \edgesCreator^{D} \frac{1}{dV}\ket{1}
  \nonumber\\
  &=&
  c \sum\limits_{\vec y}
  \coextd\edgesCreator^{1}\edgesCreator^2 \cdots \edgesCreator^{D} 
  \delta_{\set{\vec y}}
  \frac{1}{dV\of{\vec y}}
  \ket{1}
  \nonumber\\
  &\equalby{discrete divergence on cubic graphs}&
  \frac{1}{\epsilon}
  \frac{1}{p!}
  \sum\limits_\pi
  \sigma\of{\pi}
  \;
    \edgesCreator^{a_{\pi\of{1}}}\edgesCreator^{a_{\pi\of{2}}}\cdots 
     \edgesCreator^{a_{\pi\of{p}}}    
  \;
  \underbrace{
  \sum\limits_{\vec y}
  \left(
    \delta_{\set{\vec y}}
    -
    \delta_{\set{\vec y - \vec a_{a_{\pi\of{p+1}}}}}
    \frac{dV\of{\vec y}}{dV\of{\vec y - \vec e_{a_{\pi\of{p+1}}}}}
  \right)
  \frac{1}{dV\of{\vec y}}
  }_{=\frac{1}{dV} - \frac{1}{dV}= 0}
  \ket{1}
  \nonumber\\
  &=& 0
  \,.
\end{eqnarray}
Uniqueness of \refer{volume-like forms on cubic graphs} is immediate: Any other
volume-like form must differ from 
\refer{volume-like forms on cubic graphs}
by a non-constant factor. Commuting $\coextd$ past
this factor gives an operator of the form $f_a \edgesAnnihilator^a$, which does
not annihilate $\ket{\vol}$.

Assertion \refer{relation dV to det(eta) in discrete geometry}
is proved by evaluating \refer{definition of proper volume form}:
\begin{eqnarray}
  \ket{1} &\shallbe&
  \evol^\edag
  \evol
  \ket{1}
  \nonumber\\
  &=&
  c^2\;
  \gop(\gop\vol)^\dag \gop^{-1}\;
  \gop
  \vol 
  \ket{1}
  \nonumber\\
  &=&
  c^2\;
  \gop\vol^\dag \gop \vol \ket{1}
  \nonumber\\
  &=&
  c^2\;
  \gop \vol^\dag \vol \;{\rm det}\of{g}\ket{1}
  \nonumber\\
  &=&
  c^2\;
  \gop
  \frac{{\rm det}\of{g}}{dV^2}\ket{1}
  \nonumber\\
  &\stackrel{\gop_{|\cal A} = {\rm id}}{=}&
  c^2\;
  \frac{{\rm det}\of{g}}{dV^2}\ket{1}
  \,.
\end{eqnarray}
\endofproof

\paragraph{Hodge star on cubic graphs with self-adjoint 0-forms} 

According to \refer{volume-like forms on cubic graphs} the volume-like form is 
\begin{eqnarray}
  \ket{\bf vol} &=&  \edgesCreator^1 \edgesCreator^2\cdots \edgesCreator^D \frac{1}{dV} \ket{1}
   \nonumber\\
  &=&
  \frac{1}{D!}\, \epsilon_{a_1 a_b \cdots a_p}\edgesCreator^{a_1} \edgesCreator^{a_2}\cdots \edgesCreator^{a_D}
  \frac{1}{dV}
  \ket{1}
\end{eqnarray}
for some $v\in {\cal A}$.

Let
\begin{eqnarray}
  A &=& 
   \alpha_{a_1 a_2 \cdots a_p}\edgesCreator^{a_1}\edgesCreator^{a_2}\cdots \edgesCreator^{a_p}\ket{1}
  \nonumber\\
  B &=& \beta_{b_1 b_2 \cdots b_p}\edgesCreator^{b_1}\edgesCreator^{b_2}\cdots \edgesCreator^{b_p}\ket{1}
\end{eqnarray}
and
\begin{eqnarray}
  A^\prime &=& 
   \edgesCreator^{a_1}\edgesCreator^{a_2}\cdots \edgesCreator^{a_p} \alpha_{a_1 a_2 \cdots a_p}\ket{1}
  \nonumber\\
  B^\prime &=& \edgesCreator^{b_1}\edgesCreator^{b_2}\cdots \edgesCreator^{b_p}\beta_{b_1 b_2 \cdots b_p}\ket{1}
  \,.
\end{eqnarray}

The action of the Hodge star (with respect to $\gop = {\rm id}$) is explicitly
\begin{eqnarray}
  \label{action of discrete hodge}
  \hodge \edgesCreator^{a_1}\edgesCreator^{a_2}\cdots\edgesCreator^{a_p}\alpha_{a_1 a_2 \cdots a_p}\ket{1}
  &=&
  \frac{1}{(D-p)!}
  \alpha_{a_1 a_2 \cdots a_p}
  \epsilon^{a_1 a_2 \cdots a_p}{}_{a_{p+1}\cdots a_D}
  \edgesCreator^{a_{p+1}}\edgesCreator^{a_{p+2}}  \cdots
  \edgesCreator^{a_D}
  \frac{1}{dV}
  \ket{1}
  \,,
  \nonumber\\
\end{eqnarray}
and that of its inverse is
\begin{eqnarray}
  \hodge^{-1} 
  \alpha_{a_1 a_2 \cdots a_p}
  \edgesCreator^{a_1}\edgesCreator^{a_2}\cdots\edgesCreator^{a_p}\ket{1}
  &=&
  \frac{(-1)^{p(D-p)}}{(D-p)!}
  \epsilon^{a_1 a_2 \cdots a_p}{}_{a_{p+1}\cdots a_D}
  \edgesCreator^{a_{p+1}}\edgesCreator^{a_{p+2}}  \cdots
  \edgesCreator^{a_D}
  \alpha_{a_1 a_2 \cdots a_p}
  T_{\vec e_{a_1}+\cdots \vec e_{a_{p}}}[dV]
  \ket{1}  
  \,.
  \nonumber\\
\end{eqnarray}
Here $\epsilon$ is the totally antisymmetric unit symbol with $\epsilon_{123\cdots D} = 1$,
indices have been shifted with $\delta_{ab}$ and summation over index pairs is implied.

This is clearly a discrete approximation to the Hodge star in flat Euclidean space.
The Hodge star for general signature and geometry is derived in \refer{Hodge for general metrics}.

\subparagraph{Expressing the inner product by means of the Hodge star.}
One finds
\begin{eqnarray}
  \label{discrete inner product in terms of Hodge, ordinary case}
  A \hodge B \ket{1} 
  &=& 
  B \hodge A \ket{1}
  \nonumber\\
  &=&
  p!\;
  \alpha_{a_1 a_2 \cdots a_p}\beta^{a_1 a_2 \cdots a_p}
  \ket{\bf vol}
\end{eqnarray}
and hence
\begin{eqnarray}
  \bracket{A}{B}
  &=&
  \int (A^\prime \hodge B^\prime)_{\cal S}
  \,.
\end{eqnarray}
(The integral over a $D$-form was defined in \refer{integral over discrete D form}.)

{\it Proof:}
The implication
\begin{eqnarray}
  \mbox{$(a_i)$ is not a permutation of $(b_i)$}
  \;&\Rightarrow&\;
  \edgesCreator^{a_1}\edgesCreator^{a_2}\cdots\edgesCreator^{a_p}
  \edgesAnnihilator^{b_1}\edgesAnnihilator^{b_2}\cdots \edgesAnnihilator^{b_p}
  f \ket{{\bf vol}}
  = 0
\end{eqnarray} 
(for $f\in \cal A$)
is crucial. It implies that
\begin{eqnarray}
  \label{some relation in some proof aslkf}
  \edgesCreator^{a_1}\edgesCreator^{a_2}\cdots\edgesCreator^{a_p}
  \edgesAnnihilator^{b_1}\edgesAnnihilator^{b_2}\cdots \edgesAnnihilator^{b_p}
  f \,v\,\edgesCreator^1 \cdots \edgesCreator^D\ket{1}
  &=&
  f \, v\,
  \edgesCreator^{a_1}\edgesCreator^{a_2}\cdots\edgesCreator^{a_p}
  \edgesAnnihilator^{b_1}\edgesAnnihilator^{b_2}\cdots \edgesAnnihilator^{b_p}
  \edgesCreator^1 \cdots \edgesCreator^D\ket{1}
  \nonumber\\
  &=&
  f \delta^{a_1 a_2 \cdots a_p}_{b_1 b_2 \cdots b_p} \ket{\bf vol}
\end{eqnarray}
because $f\in \cal A$ commutes with $\edgesCreator^a\edgesAnnihilator^a$ (no sum).
Therefore

\begin{eqnarray}
  A \hodge B \ket{1}
  &=&
  \alpha_{a_1 a_2 \cdots a_p}\edgesCreator^{a_1}\edgesCreator^{a_2}\cdots \edgesCreator^{a_p}
  \edgesAnnihilator^{b_p} \edgesAnnihilator^{b_{p-1}}\cdots \edgesAnnihilator^{b_1}
  \beta_{b_1 b_2 \cdots b_p}
  \ket{\bf vol}
  \nonumber\\
  &=&
  p! \; 
  \alpha_{a_1 a_2 \cdots a_p}\beta^{a_1 a_2 \cdots a_p}
  \ket{\bf vol}
  \,.
\end{eqnarray}
\endofproof

\subparagraph{Hodge star on cubic graphs for $\gop$-inner products.}

The Hodge star $\ehodge$ which is associated with  the $\gop$-inner product 
$\bra{\cdot}\gop^{-1}\ket{\cdot}$
has a simple relation to the Hodge star $\hodge$ associated with the unmodified
inner product $\bracket{\cdot}{\cdot}$:

Assume that $\ehodge$ and $\hodge$ are both invertible. Then, according to
\refer{ecoextd in terms of ehodge and extd},
\begin{eqnarray}
  \label{another equation for relation ehodge to hodge}
  \ecoextd
  &=&
  \ehodge
  \extd
  \invehodge
  (-1)^{D-\numberOperator + 1}	
\end{eqnarray}
and
\begin{eqnarray}
  \label{some equation for relation ehodge to hodge}
  \coextd &=& 
  \hodge \extd \invhodge (-1)^{D-\numberOperator + 1}
  \,.
\end{eqnarray}
Using the identity $A^\edag = \gop A^\dag \gop^{-1}$ the left hand side of the
former equation can be rewritten as
\begin{eqnarray}
  \ecoextd
  &=&
  \gop \coextd \gop^{-1}
  \nonumber\\
  &\equalby{some equation for relation ehodge to hodge}&
  \gop \left(\hodge \extd \invhodge \right)\gop^{-1}
  \;(-1)^{D-\numberOperator + 1}
  \nonumber\\
  &=&
  (\gop \hodge) \extd (\gop \hodge)^{-1}
  \;(-1)^{D-\numberOperator + 1}
  \,.
\end{eqnarray}
Comparison with \refer{another equation for relation ehodge to hodge} yields
the result.
\begin{eqnarray}
  \label{relation hodge to ehodeg}
  \ehodge
  &=&
  \gop \circ \hodge
  \,.
\end{eqnarray}

\subparagraph{Corollary.} For \emph{diagonal} operators $\gop$, i.e. those that satisfy
 $\gop \edgesCreator^{a_1}\cdots \edgesCreator^{a_p}\ket{1} 
\propto \edgesCreator^{a_1}\cdots \edgesCreator^{a_p}\ket{1}$,
equation \refer{discrete inner product in terms of Hodge, ordinary case} generalizes to
\begin{eqnarray}
  \bra{A}\gop^{-1}\ket{B}
  &=&
  \int (A^\prime \ehodge B^\prime)_{\cal S}
  \,.
\end{eqnarray}

\subsection{Metrics on cubic graphs}
\label{metrics on cubic graphs}

We have seen that with respect to the ordinary adjoint operation $(\cdot)^\dag$
on $\hat \Omega_{(2)}$ (which is defined as leaving elements in $\hat \Omega_{(2)}^0$ invariant)
the elementary edges have to be orthogonal in the sense of 
\refer{statelike part of adjoint edge/edge product} and that the anticommutation
relations \refer{discrete edge anticommutators for cubic graphs} 
become quite unnatural when $\ell^2$ is not a constant. Even though \emph{a priori}
one might have expected that a metric on the complex could be encoded
in relations such as \refer{statelike part of adjoint edge/edge product}, this is now
seen to be unrealistic. 

To handle non-flat spaces
we therefore need to drop the assumption that 0-forms are self-adjoint. Instead of
developing the corresponding formalism completely from scratch again, the existing
inner product $\bracket{\cdot}{\cdot}$ \refer{discrete pre-inner product} 
on a flat space can simply be modified by deforming it with a \emph{metric operator}
$\gop$ which is of 0 grade, self adjoint with respect to 
$(\cdot)^\dag$ and invertible:
\begin{eqnarray}
  \bracket{\cdot}{\cdot}_\gop
  &\defas&
  \bra{\cdot}\gop^{-1}\ket{\cdot}
  \,.
\end{eqnarray}
Note that
\begin{eqnarray}
  \label{adjoint of modified inner product}
  A^\edag &=& \gop A^\dag \gop^{-1}
  \,.
\end{eqnarray}

For the purpose of presenting the following considerations it is convenient to 
let $X^\mu$ be the \emph{preferred coordinate system} defined by\footnote{
For instance $X^3$ increases by $\epsilon$ when one moves from $\vec x$ to $\vec x + \vec e_3$.
It remains constant when moving into the direction of any other $\vec e_a,\,a \neq 3$.
}
\begin{eqnarray}
  \label{preferred coordinates on cubic spaces}
  \commutator{\extd}{X^\mu} &=& \edgesCreator^{(a=\mu)}
  \,.
\end{eqnarray}

Now let $g = (g_{\mu\nu})$ be a $D\times D$ matrix that is supposed to play the role
of the metric tensor with respect to these preferred coordinates. 

We are looking for an operator $\hat g$ such that
\begin{eqnarray}
  \commutator{\extd}{X^\nu}^\edag\commutator{\extd}{X^\mu} \ket{1}
  &=&
  (g^{-1})^{\mu\nu} \ket{1}
\end{eqnarray}
and more generally
\begin{eqnarray}
  &&\commutator{\extd}{X^\nu}^\edag
   \commutator{\extd}{X^{\mu_1}} 
   \commutator{\extd}{X^{\mu_2}}
   \cdots
   \commutator{\extd}{X^{\mu_p}}
   \, f
  \ket{1}
  \nonumber\\
  &=&
  -
  \sum\limits_{j=1}^p
   (-1)^j
   \commutator{\extd}{X^{\mu_1}} 
   \cdots
   \commutator{\extd}{X^{\mu_{j-1}}}
   \commutator{\extd}{X^{\mu_{j+1}}}
   \cdots  
   \commutator{\extd}{X^{\mu_p}}
  (g^{-1})^{\nu\mu_j}
  \, f 
\ket{1}
  \,.
  \nonumber\\
\end{eqnarray}

Note that for constant $g_{\mu\nu}$ this is equivalent to
\begin{eqnarray}
  \label{canonical anticommutation relations for constant discrete metrics}
  \antiCommutator
  {\commutator{\extd }{X^\mu}}
  {\commutator{\extd}{X^\nu}^\edag}
  &=&
  (g^{-1})^{\mu\nu}
  \hspace{1cm}
  \mbox{(for constant $g_{\mu\nu}$)}
  \,.
\end{eqnarray}

\subparagraph{Action of the metric operator.}
This operator $\hat g$ must act as
\begin{eqnarray}
  \label{action of gop}
  \gop \; \edgesCreator^{a_1}\cdots \edgesCreator^{a_p} \,f\ket{1}
  &=&
  \sum\limits_{b_1,\cdots b_p}
  \edgesCreator^{b_1}\cdots \edgesCreator^{b_p}
  g_{b_1 a_1}
  \cdots
  g_{b_p a_p}
  \, f
  \ket{1}
  \,,
\end{eqnarray}
and equivalently its inverse has the action
\begin{eqnarray}
  \label{action of inverse gop}
  \gop^{-1} \; \edgesCreator^{a_1}\cdots \edgesCreator^{a_p} \,f\ket{1}
  &=&
  \sum\limits_{b_1,\cdots b_p}
  \edgesCreator^{b_1}\cdots \edgesCreator^{b_p}
  (g^{-1})^{b_1 a_1}
  \cdots
  (g^{-1})^{b_p a_p}
  \, f
  \ket{1}
  \,.
\end{eqnarray}

{\it Proof:}
\begin{eqnarray}
  &&
  \commutator{\extd}{X^\nu}^\edag\;
   \commutator{\extd}{X^{\mu_1}} 
   \commutator{\extd}{X^{\mu_2}}
   \cdots
   \commutator{\extd}{X^{\mu_p}}
   \, f
  \ket{1}
  \nonumber\\
  &\equalby{preferred coordinates on cubic spaces}&
   (\edgesCreator^\nu)^\edag\;
   \edgesCreator^{\mu_1} 
   \cdots
   \edgesCreator^{\mu_2}
   \, f
  \ket{1}
  \nonumber\\
  &\equalby{adjoint of modified inner product}&
   \gop\edgesAnnihilator^\nu \gop^{-1}\;
   \,
   \edgesCreator^{\mu_1} 
   \cdots
   \edgesCreator^{\mu_2}
   \, f
  \ket{1}
  \nonumber\\
  &\equalby{action of gop}&
   \gop\edgesAnnihilator^\nu \;
   \,
   \sum\limits_{\lambda_1,\cdots,\lambda_p}
   \edgesCreator^{\lambda_1} 
   \cdots
   \edgesCreator^{\lambda_2}
   (g^{-1})^{\lambda_1\mu_1}
   \cdots
   (g^{-1})^{\lambda_p\mu_p}
   \, f
  \ket{1}    
  \nonumber\\
  &\equalby{condition for discrete canonical anticommutators}&
   -
   \gop^{-1} 
   \;
   \sum\limits_{\lambda_1,\cdots,\lambda_p}
   \sum\limits_{j=1}^p
   (-1)^j
   \edgesCreator^{\lambda_1} 
   \cdots
   \edgesCreator^{\lambda_{j-1}}
   \edgesCreator^{\lambda_{j+1}}
   \cdots
   \edgesCreator^{\lambda_p} 
    \delta^{\nu,\lambda_j}
   (g^{-1})^{\lambda_1\mu_1}
   \cdots
   (g^{-1})^{\lambda_p\mu_p}
   \, f
  \ket{1}      
  \nonumber\\
  &\equalby{action of inverse gop}&
   -
   \sum\limits_{j=1}^p
   (-1)^j
   \commutator{\extd}{X^{\mu_1}} 
   \cdots
   \commutator{\extd}{X^{\mu_{j-1}}}
   \commutator{\extd}{X^{\mu_{j+1}}}
   \cdots
   \commutator{\extd}{X^{\mu_p}} 
   (g^{-1})^{\nu\mu_j}
   \, f
  \ket{1}        
  ,.
\end{eqnarray}
\endofproof

\subparagraph{Explicit form of the metric operator.}

The metric operator $\gop$ may be written in terms of the $\edgesCreator$ and $\edgesAnnihilator$
as
\begin{eqnarray}
  \gop
  &=&
  \sum\limits_p
  \frac{1}{p!}
  \sum\limits_{a_1, \cdots, a_p \atop b_1, \cdots,b_p}
  \edgesCreator^{a_1}\edgesCreator^{a_2}\cdots \edgesCreator^{a_p}
  \;\;g_{a_1 b_1} g_{a_2 b_2}\cdots g_{a_p b_p}\;\;
  \edgesAnnihilator^{b_1}\edgesAnnihilator^{b_2}\cdots \edgesAnnihilator^{b_p}\;\;
  P_p
  \,,
\end{eqnarray}
where $P_p$ is the projector on $\Omega^p$.

Obviously this $\gop$ is indeed of 0 grade, is self-adjoint when
$g^{\mu\nu}$ is symmetric and is invertible if $g^{\mu\nu}$ is.

{\it Proof:} Just apply this operator on any $p$-form. \endofproof

In the continuum, or for constant metric tensors, there
is a more concise way to write the metric operator, namely
\begin{eqnarray}
   \gop
   &=&
   \exp\of{
     \edgesCreator^a \left(\ln g\right)_{ab} \edgesAnnihilator^b
   }
  \,,
\end{eqnarray}
where $\left(\ln g\right)$ is the logarithm of the matrix $g$ in the sense that
\begin{eqnarray}
  g_{ab} &=& \exp\of{\ln g}_{ab} \;=\; \delta_{ab} + (\ln g)_{ab} + 
  \frac{1}{2}\sum\limits_c (\ln g)_{ac} (\ln g)_{cb} + \cdots
  \,.
\end{eqnarray}
{\it Proof:}
Since by assumption the scalar coefficient of the metric operator 
commutes with all form creation and annihilation
operators we have
\begin{eqnarray}
   \gop
   \;\edgesCreator^c\;
   \gop^{-1}
   &=&
   \exp\of{\edgesCreator^a (\ln g)_{ab}\edgesAnnihilator^b}
   \;\edgesCreator^c\;
   \exp\of{-\edgesCreator^a (\ln g)_{ab}\edgesAnnihilator^b}
   \nonumber\\
   &=&
   \edgesCreator^c +
   \commutator{\edgesCreator^a (\ln g)_{ab}\edgesAnnihilator^b}{\edgesCreator^c}
   +
   \frac{1}{2}
  \commutator{\edgesCreator^a (\ln g)_{ab}\edgesAnnihilator^b}{
  \commutator{\edgesCreator^a (\ln g)_{ab}\edgesAnnihilator^b}{\edgesCreator^c}}
  +
  \cdots
  \nonumber\\
  &=&
  \left(
   \delta_{ca}
   +
   (\ln g)_{ca}
   +
   \frac{1}{2}(\ln g)^2_{ca}
   +
   \cdots
  \right)
  \edgesCreator^a
  \nonumber\\
  &=&
  g_{ca}\edgesCreator^a
  \,.
\end{eqnarray}
Using $\gop^{-1} f \ket{1} = f \ket{1}$ it follows that
\begin{eqnarray}
  \gop  \edgesCreator^{a_1}\edgesCreator^{a_2}\cdots \edgesCreator^{a_p} f \ket{1}
  &=&
  \gop  \edgesCreator^{a_1} \gop^{-1}
  \;
  \gop  \edgesCreator^{a_2} \gop^{-1}
  \cdots
  \gop  \edgesCreator^{a_p} \gop^{-1}
  f \ket{1}
  \nonumber\\
  &=&
  \edgesCreator^{b_1}\cdots \edgesCreator^{b_p}g_{b_1 a_1}\cdots g_{b_p a_p}  f \ket{1}
  \,.    
\end{eqnarray}
\endofproof

Using the metric operator \refer{action of inverse gop} 
we can finally write down the explicit action
of the discrete Hodge star operator on cubic graphs with arbitrary discrete
metrics:

\subparagraph{Hodge star on cubic graphs for general metrics.}

Combination of \refer{relation hodge to ehodeg} and \refer{action of discrete hodge} yields
\begin{eqnarray}
  \label{Hodge for general metrics}
  &&\ehodge \edgesCreator^{a_1}\edgesCreator^{a_2}\cdots\edgesCreator^{a_p}\alpha_{a_1 a_2 \cdots a_p}\ket{1}
  \nonumber\\
  &=&
  \frac{1}{(D-p)!}
  \epsilon^{a_1 \cdots a_D}
  \edgesCreator^{b_{p+1}}  \cdots\edgesCreator^{b_D}
  g_{a_{p+1} b_{p+1}}\cdots g_{a_D b_D}
  T_{\vec e_{p+1}+\cdots \vec e_{D}}[\alpha_{a_1 a_2 \cdots a_p}]
  \frac{1}{dV}
  \ket{1}
  \,.
  \nonumber\\
\end{eqnarray}
For $dV = \sqrt{{\rm det}\of{g}}$ (\cf \refer{relation dV to det(eta) in discrete geometry})
this is seen to be formally the same equation as in the continuum theory, the only
difference being the shift $T_{\vec e_{p+1}+\cdots \vec e_{D}}$ of the coefficient
function. This lattice effect vanishes in the continuum.

\subparagraph{Component form of the modified inner product}
Two $p$ forms
$\fatalpha = 
  \commutator{\extd}{X^{\mu_1}}
  \cdots
  \commutator{\extd}{X^{\mu_p}}
  \alpha_{\mu_1  \cdots \mu_p}
   \ket{1}
$
and
$\fatbeta = 
  \commutator{\extd}{X^{\nu_1}}
  \cdots
  \commutator{\extd}{X^{\nu_p}}
  \alpha_{\nu_1  \cdots \nu_p}
  \ket{1}
$
have ordinary inner product
\begin{eqnarray}
  \label{gop inner product in component form}
  \bracket{\fatalpha}{\fatbeta}
  &=&
  \int 
  \delta^{\mu_1 \nu_1} \cdots \delta^{\mu_p \nu_p}
  \alpha_{\mu_1 \cdots \mu_p}\beta_{\nu_1 \cdots \nu_p}
  \; dV
\end{eqnarray}
and $\gop$-modified inner product
\begin{eqnarray}
  \label{gop-modified inner product in components}
  \bra{\fatalpha}\gop^{-1}\ket{\fatbeta}
  &=&
  \int 
  (g^{-1})^{\mu_1 \nu_1} \cdots (g^{-1})^{\mu_p \nu_p}
  \alpha_{\mu_1 \cdots \mu_p}\beta_{\nu_1 \cdots \nu_p}
  \; dV
\end{eqnarray}
(summation over upper and lower indices is implicit).

When $dV$ is chosen such that a proper volume form exists, so that
$dV = \sqrt{{\rm det}\of{g}}$ \refer{relation dV to det(eta) in discrete geometry} then
equation \refer{gop-modified inner product in components} is of the same form as
the Hodge inner product of the continuum theory. Note that this
requires 
that the components of the $p$ forms entering the product are written
\emph{on the right} and \emph{with respect to the preferred coordinates}
\refer{preferred coordinates on cubic spaces}.

\subsection{$p$-vector fields}
\label{p-vector fields}

The space $\Omega\of{{\cal A},\extd}$ is \emph{dual} to the space
\begin{eqnarray}
  \Omega^\ast\of{\cal A} &=& \bigoplus\limits_p \Omega^\ast_p\of{\cal A}
  \nonumber\\
\end{eqnarray}
of \emph{$p$-vector fields}. If we 
suggestively write $\fatpartial_\mu$ for the element dual to
$\extd X^\mu$ (where $X^\mu$ are the preferred coordinates \refer{preferred coordinates on cubic spaces}) then
\begin{eqnarray}
  \Omega^\ast_p\of{\cal A}
  &\defas&
  {\rm span}\of{
  \set{
    v^{\mu_1\mu_2\cdots \mu_p}\fatpartial_{\mu_1}\fatpartial_{\mu_2}\cdots \fatpartial_{\mu_p}
   |
   v^{\mu_1\cdots \mu_p} = v^{[\mu_1\cdots \mu_p]} \in {\cal A}
  }
  }
  \,.
\end{eqnarray}
Given any $p$-form $\fatalpha$ and a $p$-vector $v$ with components
\begin{eqnarray}
  \fatalpha &\defas&
  \extd X^{\mu_1}\extd X^{\mu_2}\cdots \extd X^{\mu_p}
  \;\alpha_{\mu_1\mu_2\cdots \mu_p}
  \nonumber\\
  v &\defas&
  v^{\mu_1\mu_2\cdots \mu_p}
  \fatpartial_{\mu_1}\fatpartial_{\mu_2}\cdots\fatpartial_{\mu_p}  
\end{eqnarray}
the evaluation of $\fatalpha \in \Omega$ on $v \in \Omega^\ast$
can be defined by
\begin{eqnarray}
  \Omega^p \times \Omega^\ast_p &\longrightarrow& {\cal A}
  \nonumber\\
  \fatalpha\of{v} &\defas& \alpha_{\mu_1 \cdots \mu_p}v^{\mu_1 \cdots \mu_p}
\end{eqnarray}
and
\begin{eqnarray}
  \Omega^p \times \Omega^\ast_p &\longrightarrow& \C
  \nonumber\\
  \int \fatalpha\of{v}\; 
  &=&
  \int 
  \alpha_{\mu_1 \cdots \mu_p}v^{\mu_1 \cdots \mu_p}
  \,,
\end{eqnarray}
using the integral \refer{definition discrete integral}.

In the presence of a metric as in \S\fullref{metrics on cubic graphs} both
spaces can be identified using the invertible map
\begin{eqnarray}
  \formtovec : \Omega &\to& \Omega^\ast
\end{eqnarray}
defined by
\begin{eqnarray}
  \label{formtovec and vectoform}
  \formtovec
  \left(
  \extd X^{\mu_1}\extd X^{\mu_2}\cdots \extd X^{\mu_p}
  \;\alpha_{\mu_1\mu_2\cdots \mu_p}
  \right)	
  &\defas&
  dV\,
  \alpha_{\mu_1\mu_2\cdots \mu_p}(g^{-1})^{\mu_1\nu_1}(g^{-1})^{\mu_2\nu_2}\cdots (g^{-1})^{\mu_p\nu_p}
  \fatpartial_{\nu_1}\fatpartial_{\nu_2}\cdots\fatpartial_{\nu_p}
  \nonumber\\
  \vectoform
  \left(
  v^{\mu_1\mu_2\cdots \mu_p}
  \fatpartial_{\mu_1}\fatpartial_{\mu_2}\cdots\fatpartial_{\mu_p}
  \right)	
  &\defas&
  \extd X^{\nu_1}\extd X^{\nu_2}\cdots \extd X^{\nu_p}
  \;v^{\mu_1\mu_2\cdots \mu_p}g_{\mu_1 \nu_1}g_{\mu_2\nu_2}\cdots g_{\mu_p\nu_p}\,\frac{1}{dV}
  \,.
\end{eqnarray}

Due to \refer{gop inner product in component form}
the evaluation map of forms on vectors is then related to the inner product 
$\bracket{\cdot}{\cdot}$ \refer{discrete pre-inner product} on forms by
\begin{eqnarray}
  \int 
  \fatalpha\of{v}
  &=&
  \langle \fatalpha \;|\; \vectoform v \rangle_\gop
  \nonumber\\
  \bracket{\fatalpha}{\fatbeta}_\gop
  &=&
  \int 
  \fatalpha\of{\,\formtovec \fatbeta}
  \,.
\end{eqnarray}

Using $\formtovec$ the difference between $p$-forms and $p$-vectors becomes immaterial and
for $\fatalpha \in \Omega^p$ and $v\in \Omega_p$ we define for notational convenience:
\begin{eqnarray}
  \bracket{\fatalpha}{v} &\defas& \frac{1}{p!}\bracket{\fatalpha}{\vectoform v}_\gop
  \,.
\end{eqnarray}
(Both sides are of course independent of the metric.)

\subsection{$p$-chains and the boundary operator}

We can identify discrete $p$-vector fields with $p$-\emph{chains} of the discrete complex.
For instance $\delta_{\set{\vec x}}\fatpartial_a$ describes the 1-chain 
$[\vec x, \vec x + \vec e_a]$, and so on.

To emphasize this fact we introduce the familiar notation
\begin{eqnarray}
  \int_S \alpha &\defas& \frac{1}{p!}\int \alpha\of{S} \;=\;\bracket{\fatalpha}{S}\; 
\end{eqnarray}
for the integral of the $p$-form ($p$-cochain) $\fatalpha$ over the $p$-vector ($p$-chain) 
$S = S^{\mu_1\cdots\mu_p}\fatpartial_{\mu_1}\cdots\fatpartial_{\mu_p}$.

\subparagraph{Operators on chains and transposition.}
Every operator $A: \Omega \to \Omega$ corresponds to an operator $A^{\rm t} : \Omega^\ast \to \Omega^\ast$,
called the \emph{transpose} of $A$, which is
defined by
\begin{eqnarray}
  \label{switching from operators on Omega to those on Omega^ast}
  \bracket{A\,\fatalpha}{S}
  &\defas&
  \bracket{\fatalpha}{A^{\rm t}\, S}
  \,.
\end{eqnarray}
(Obviously the operation $(\cdot)^{\rm t}$ is independent of the metric.)
By writing
\begin{eqnarray}
  \label{switching from operators on Omega to those on Omega^ast}
  \int_S A \fatalpha
  &=&
  \frac{1}{p!}
  \int (A \fatalpha)\of{S}
  \nonumber\\
  &=&
  \frac{1}{p!}
  \bra{A \fatalpha}{\vectoform S }\rangle_\gop
  \nonumber\\
  &=&
  \frac{1}{p!}
  \bra{\fatalpha}{A^\edag \vectoform S}\rangle_\gop
  \nonumber\\
  &=&
  \frac{1}{p!}
  \int \fatalpha\of{\formtovec A^\edag \vectoform S }\; dV
\end{eqnarray}
one finds
\begin{eqnarray}
  \label{definition of transpose}
  A^\edag &=& \vectoform \circ A^{\rm t} \circ \formtovec
  \,.
\end{eqnarray}

\subparagraph{The boundary operator.}
In particular 
\begin{eqnarray}
  \label{definition of discrete boundary operator}
  \boundary &\defas& \extd^{\rm t}
\end{eqnarray}
is the \emph{boundary operator} that maps any chain to its boundary. In our formulation
its nilpotency is inherited from that of $\extd$:
\begin{eqnarray}
  \boundary^2 &=& 0
  \,.
\end{eqnarray}
As a special case of \refer{switching from operators on Omega to those on Omega^ast} 
\emph{Stoke's theorem} is obtained:
\begin{eqnarray}
  \int_S \extd \fatalpha &=& \int_{\boundary S} \fatalpha
  \,.
\end{eqnarray}

The explicit action of $\partial$ is found by applying the various definitions:
\begin{eqnarray}
  \label{explicit action of discrete boundary operator}
  &&\partial\; \left( \delgen{\vec x} \fatpartial_{a_1} \cdots \fatpartial_{a_{p+1}} \right)
  \nonumber\\
  &\equalby{definition of transpose}&
  \formtovec \ecoextd \vectoform \left(\delgen{\vec x} \fatpartial_{a_1} \cdots \fatpartial_{a_{p+1}}\right)
  \nonumber\\
  &\stackrel{\refer{formtovec and vectoform}\refer{action of inverse gop}}{=}&
  \formtovec\, \gop\, \coextd\, \edgesCreator^{a_1} \cdots \edgesCreator^{a_{p+1}}\delgen{\vec x}
  \frac{1}{dV\of{\vec x}}\ket{1}
  \nonumber\\
  &\equalby{discrete divergence on cubic graphs}&
  \formtovec\, \gop\,
  \frac{(-1)^p}{\epsilon p!}
  \sum\limits_\pi
  \sigma\of{\pi}
  \edgesCreator^{a_{\pi\of{1}}}\cdots \edgesCreator^{a_{\pi\of{p}}}
  \left(
    \delgen{\vec x} - \delgen{\vec x - \edge_{a_{\pi\of{p+1}}}}
    \frac{dV\of{\vec x}}{dV\of{\vec x - \edge_{a_{\pi\of{p+1}}}}}
  \right)
  \frac{1}{dV\of{\vec x}}
  \ket{1}
  \nonumber\\
  &\equalby{formtovec and vectoform}&
  \frac{(-1)^p}{\epsilon p!}
  \sum\limits_\pi
  \sigma\of{\pi}
  \left(
    \delgen{\vec x} - \delgen{\vec x - \edge_{a_{\pi\of{p+1}}}}
  \right)
  \fatpartial_{a_{\pi\of{1}}}\cdots \fatpartial_{a_{\pi\of{p}}}
  \,.
\end{eqnarray}

\begin{figure}
\hspace{1.5cm}
\begin{picture}(200,200)
\includegraphics{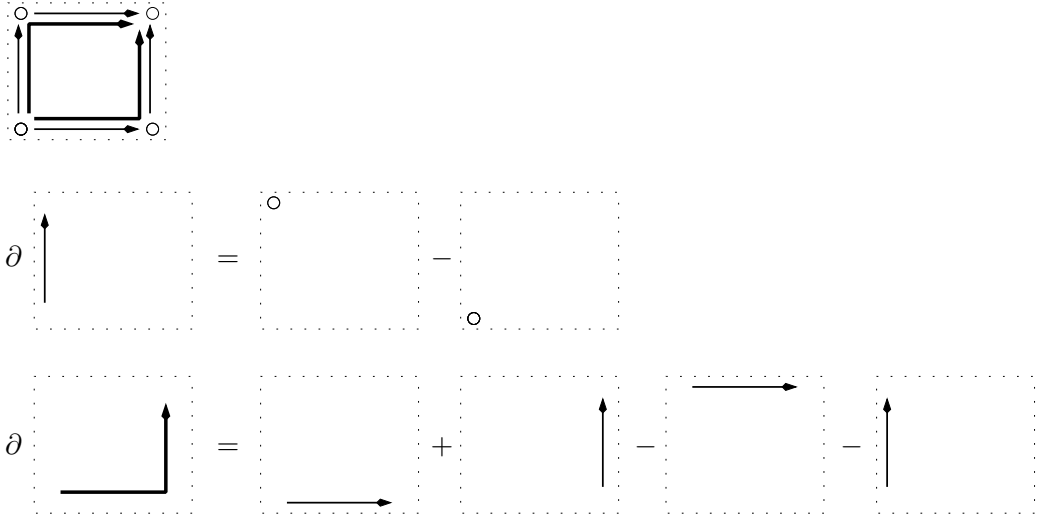}
\put(-390,94){$\partial$}
\put(-310,94){$=$}
\put(-229,94){$-$}
\put(-390,23){$\partial$}
\put(-310,23){$=$}
\put(-229,23){$+$}
\put(-152,23){$-$}
\put(-74,23){$-$}
\end{picture}
\caption{
{\it Boundaries}. 
}
The action 
\refer{explicit action of discrete boundary operator}, 
\refer{example formulas for divergence of elementary edges}
of the boundary operator
$\partial$ 
\refer{definition of discrete boundary operator} 
is illustrated. The square at the top depicts some part of
a cubic complex as in figure 
\ref{cubicgraph}. 
The second and third line indicate 
the boundary of an elementary edge and of an elementary plaquette, respectively.
\label{boundary.eps}
\end{figure}

This in particular demonstrates how the various occurences of $\gop$ and $dV$ in the
expression $\formtovec \ecoextd \vectoform$ mutually cancel. Examples of how 
\refer{explicit action of discrete boundary operator} looks like in terms of
elementary edges were given (in dual form) in 
equation \refer{example formulas for divergence of elementary edges},
\cf figure \ref{boundary.eps}.

\section{Flat topologically hypercubic complexes}
\label{Flat topologically hypercubic complexes}

The above considerations show that 
a thorough understanding of $N=2$ 
differential calculus on flat cubic graphs with
self-adjoint functions
is the basis for the study of the more general case.

So consider the situation where the structure functions
\refer{strucure functions of discrete lattice manifold} are of the trivial form 
\begin{eqnarray}
  \label{trivial structure functions of discrete calculus}
  C^{ij}{}_k
  &\shallbe&
  \delta^{ij}\delta^i{}_k
\end{eqnarray}
corresponding to a cubic graph with $e_a{}^i = \delta_a^j$ (and hence 
$\edgesCreator^i = \coordCreator^i$).
In this case we have in particular
\begin{eqnarray}
  \label{relations for flat discrete diffgeo}
  \commutator{X^i}{\coordCreator^j}
  &=&
  - \epsilon\, \delta^{ij}\,\coordCreator^i
  \nonumber\\
  \commutator{X^i}{\coordAnnihilator^j}
  &=&
  +\epsilon\,\delta^{ij}\,\coordAnnihilator^i
\end{eqnarray}
and
\begin{eqnarray}
  \label{correct anticommutator of discrete forms in flat case}
  \antiCommutator{\edgesCreator^i}{\edgesCreator^j}
  &=&
  0
  \\
  \label{creator annihilator anticommutator in flat discrete case}
  \antiCommutator{\edgesCreator^a}{\edgesAnnihilator^b}
  &\shallbe&
  \eta^{ab}
  \,.
\end{eqnarray}

\subsection{Partial derivative operator}
 \label{partial derivative operator}

Define the discrete version of the partial derivative operator $\partial_\mu$ in analogy with
\refer{partial coordinate derivative as anticommutator with coordannihilator} by
\begin{eqnarray}
  \label{definition discrete flat derivative operator}
  \partial_i 
  &\defas&
  -\antiCommutator
  {
    \extd
  }
  {
   \commutator
   {
     \coextd
   }
   { X_i}
  }
  \nonumber\\
  &=&
  \antiCommutator
  {\extd}
  {\coordAnnihilator_i}
  \,.
\end{eqnarray}
We need to find the commutation relations of $\partial_i$ with the $\coordCreator^i$
and $\coordAnnihilator^i$. First of all we have
\begin{eqnarray}
  \label{commutator discrete partial with creator}
  \commutator{\partial_i}{\coordCreator^j}
  &=&
  \commutator{\antiCommutator{\extd}{\coordAnnihilator_i}}{\coordCreator^j}
  \nonumber\\
  &=&
  \commutator{\extd}{\antiCommutator{\coordAnnihilator_i}{\coordCreator^j}}
  -
  \commutator{\antiCommutator{\extd}{\coordCreator^j}}{\coordAnnihilator_i}
  \nonumber\\
  &\equalby{creator annihilator anticommutator in flat discrete case}&
  0
  \,.
\end{eqnarray}
However the commutator with $\coordAnnihilator_i$ is ambiguous. All we know is that 
\begin{eqnarray}
  \commutator{\partial_i}{\coordAnnihilator_j}
  &=&
  \commutator{\antiCommutator{\extd}{\coordAnnihilator_i}}{\coordAnnihilator_j}
  \nonumber\\
  &=&
  -  
  \commutator{\antiCommutator{\extd}{\coordAnnihilator_j}}{\coordAnnihilator_i}
  \nonumber\\
  &=&
  -
  \commutator{\partial_j}{\coordAnnihilator_i}
  \,.
\end{eqnarray}
But this implies that the above expression can only be non-vanishing if one singles out certain
planes, i.e. pairs of coordinate indices, in the discrete graph. 
In the case of trivial flat geometry we instead expect ``discrete isotropy''
in the sense that
\begin{eqnarray}
  \commutator{\partial_i}{\coordAnnihilator_j} &\stackrel{!}{\propto}& \delta_{ij}
  \,.
\end{eqnarray} 
This implies
\begin{eqnarray}
  \label{commutator flat discrete partial with annihilator}
  \Rightarrow
  \commutator{\partial_i}{\coordAnnihilator_j} &=& 0
  \,.
\end{eqnarray}
It follows that
\begin{eqnarray}
  \label{commutator partial partial in flat discrete case}
  \commutator{\partial_i}{\partial_j}
  &=&
  0
  \,.
\end{eqnarray}
But the commutator of $\partial_i$ with the coordinate functions has a lattice correction:\footnote{
The canonical commutation relation is instead satisfied by the modified coordinate
\begin{eqnarray}
  \label{modified coordinate}
  \tilde X^i 
  &\defas&
  X^i
  \sum_{n=0}^\infty  (\epsilon \partial_i)^n
  \nonumber\\
  &=&
  X^i
  + 
  \epsilon X^i \partial_i
  +
  \epsilon X^i \partial_i \partial_i
  +
  \cdots
  \hspace{1cm}\mbox{(no sum over $i$)}
  \,.
\end{eqnarray}
Namely
\begin{eqnarray}
  \commutator
  {
    \partial_i
  }
  {
    X^j \sum_{n=0}^\infty (\epsilon \partial_j)^n
  }
  &=&
  \delta_i^j \sum_{n=0}^\infty (1-\epsilon \partial_i) (\epsilon \partial_i)^n
  \nonumber\\
  &=&
  \delta_i^j
  \,.
\end{eqnarray}
This seems to be an example of \emph{umbral calculus}, as discussed in 
\cite{LeviTempestWinternitz:2003}.
}
\begin{eqnarray}
  \label{commutator of discrete partial with discrete coordinate}
  \commutator{\partial_i}{X^j}
  &=&
  \antiCommutator
  {\extd}
  {\commutator{\edgesAnnihilator_i}{X^j}}
  +
  \antiCommutator{
    \commutator{\extd}{X^j}
  }
  {\edgesAnnihilator_i}
  \nonumber\\  
  &=&
  \antiCommutator
  {\extd}
  {- \epsilon\delta_i^j \edgesAnnihilator_i}
  +
  \antiCommutator
  {\edgesCreator^j}
  {\edgesAnnihilator_i}
  \nonumber\\
  &=&
  \delta_i^j\left( 1 - \epsilon\partial_i\right)
  \,.
\end{eqnarray}
Using this it is possible to show that $\extd$ may be represented in analogy to the continuum case by
\begin{eqnarray}
  \label{flat discrete extd explicitly}
  \extd &=& \coordCreator^i \partial_i
  \,.
\end{eqnarray}
To prove this it suffices to check that the operator on the right has the correct grading
and is nilpotent (which is, due to \refer{correct anticommutator of discrete forms in flat case},
\refer{commutator discrete partial with creator}
and
\refer{commutator partial partial in flat discrete case}, 
clearly the case) and that it has the correct commutator with
the coordinates:\footnote{
  Alternatively write
\begin{eqnarray}
  \commutator
  {\coordCreator^i \antiCommutator{\extd}{\coordAnnihilator_i}}
  {X^j}
  &\equalby{relations for flat discrete diffgeo}&
  \coordCreator^i \antiCommutator{\commutator{\extd}{X^j}}{\coordAnnihilator_i}
  \nonumber\\
  &=&
  \coordCreator^i \antiCommutator{\coordCreator^j}{\coordAnnihilator_i}
  \nonumber\\
  &\equalby{creator annihilator anticommutator in flat discrete case}&
  \coordCreator^j  
  \,.
\end{eqnarray}
}
\begin{eqnarray}
  \commutator{\coordCreator^i \partial_i}{X^j}  
  &=&
  \coordCreator^i \commutator{\partial_i}{X^j}
  +
  \commutator{\coordCreator^i}{X^j}\partial_i
  \nonumber\\
  &\stackrel{\refer{relations for flat discrete diffgeo}
   \refer{commutator of discrete partial with discrete coordinate}}{=}&
  \coordCreator^j(1 - \epsilon \partial_j)
  +
  \epsilon \coordCreator^j
  \partial_j
  \hspace{1cm}\mbox{(no sum over $j$)}
  \nonumber\\
  &=&
  \coordCreator^\nu
  \nonumber\\
  &\equalby{coordinate differentials in the discrete case}&
  \commutator{\extd}{X^j}
  \,.
\end{eqnarray}

All these relations have adjoint analogs:
\begin{eqnarray}
  \label{brackets of discrete adjoint partial}
  \commutator{\partial_i^\dag}{\coordCreator^j} &=& 0
  \nonumber\\
  \commutator{\partial_i^\dag}{\coordAnnihilator_j} &=& 0
  \nonumber\\
  \commutator{\partial^\dag_i}{X^j}
  &=&
  -\delta_i^j\left( 1 - \epsilon\partial^\dag_i\right)  
  \nonumber\\
  \coextd
  &=&
  \coordAnnihilator^i \partial_i^\dag
  \,.
\end{eqnarray}
An important point is that $\partial_i$ and $\partial_i^\dag$ are \emph{not}
proportional:
\begin{eqnarray}
  \label{discrete flat partial not minus partial adjoint}
  \partial_i^\dag
  \neq
  -\partial_i
  \,.
\end{eqnarray}
This is because
\begin{eqnarray}
  \commutator{(-\partial_i)}{X^j}
  &\equalby{commutator of discrete partial with discrete coordinate}&
  -\delta_i^j
  \left(
    1 + \epsilon(-\partial_i)
  \right)
  \,,
\end{eqnarray}
which would be in conflict with \refer{brackets of discrete adjoint partial}.
 
Furthermore we have
\begin{eqnarray}
  \partial_i \pm \partial_i^\dag
  &=&
  \antiCommutator{\extd}{\commutator{\coextd}{X_i}}
  \mp
  \antiCommutator{\coextd}{\commutator{\extd}{X_i}}
  \nonumber\\
  &=&
  \antiCommutator{\extd}{\commutator{\coextd}{X_i}}
  \pm
  \antiCommutator{\extd}{\commutator{\coextd}{X_i}}
  \mp
  \commutator{\antiCommutator{\extd}{\coextd}}{X_i}
\end{eqnarray}
The antihermitian partial derivative akin to that of the continuum limit is therefore
\begin{eqnarray}
  \label{antihermitian discrete partial derivative}
  \frac{1}{2}\left(\partial_i - \partial_i^\dag\right)
  &=&
  -
  \frac{1}{2}
  \commutator{\antiCommutator{\extd}{\coextd}}{X_i}
  \,.
\end{eqnarray}
The operator 
\begin{eqnarray}
  \label{lattice correction to discrete partial}
  \partial_i + \partial_i^\dag
  &=&
  \antiCommutator{\extd}{\commutator{\coextd}{X_i}}
  +
  \antiCommutator{\extd}{\commutator{\coextd}{X_i}}
  -
  \commutator{\antiCommutator{\extd}{\coextd}}{X_i}  
\end{eqnarray} 
vanishes in the continuum limit but is non-trivial due to 
a lattice correction in the discrete case.

In order to understand this correction consider the commutator $\commutator{\partial_\mu}{\partial_\nu^\dag}$.
According to the above commutation relations it commutes with $\coordCreator^\mu$, $\coordAnnihilator_\mu$.
If we again assume discrete isotropy in the sense that
\begin{eqnarray}
  \commutator{\partial_i}{\partial^\dag_j} &\propto& \delta_{ij}
\end{eqnarray}
then this operator also commutes with the coordinate functions, since
\begin{eqnarray}
  \commutator{
    \commutator{\partial_i}{\partial_i^\dag}
  }
  {
    X^i
  }
  &=&
    \commutator{\partial_i}{\epsilon\partial_i^\dag}  
    +
    \commutator{-\epsilon\partial_i}{\epsilon\partial_i^\dag} 
  \nonumber\\
  &=&
  0
  \,.
\end{eqnarray}
It should follow that the commutator is $\delta_{ij}$ times some constant $p$:
\begin{eqnarray}
  \commutator{\partial_i}{\partial^\dag_j} &=& p\delta_{ij}
  \,.
\end{eqnarray} 	
For the Laplace operator this implies
\begin{eqnarray}
  \antiCommutator{\extd}{\coextd}
  &=&
  \antiCommutator{\coordCreator^i \partial_i}{\coordAnnihilator^i\partial_i^\dag}
  \nonumber\\
  &=&
  \partial^\dag_i \partial^i
  +
  p\; \coordCreator^i \coordAnnihilator_i
  \,.
\end{eqnarray}

But of course, if there is a ``vacuum'' state $\ket{1}$ which is annihilated by both
$\partial_i$ and $\partial_i^\dag$ then $\commutator{\partial_i}{\partial_i^\dag}\ket{1} = 0$
implies $p=0$.

Finally we note for later use that
\begin{eqnarray}
  \commutator{
    \frac{1}{2}\left(\partial_i \pm \partial_i^\dag\right)
  }
  {
    X^j
  }
  &=&
  \delta_i^j
  \left(
    1 
    -
    \epsilon
    \frac{1}{2}
    \left(
      \partial_i \mp \partial_i^\dag
    \right) 
  \right)
  \,,
\end{eqnarray}
and that in analogy to \refer{modified coordinate} ``modified coordinates'' 
\begin{eqnarray}
  X_\pm^i
  &\defas&
  X^i
  \sum\limits_{n=0}^\infty
  (\epsilon\frac{1}{2}\left(\partial_i \mp \partial_i^\dag\right))^n
  \hspace{1cm}
  \mbox{(no sum over $i$)}
\end{eqnarray}
can be defined, which satisfy
\begin{eqnarray}
  \commutator{\frac{1}{2}\left(\partial_i \pm \partial_i\right)}
  {X^j_\pm}
  &=&
  \delta_i^j
  \sum\limits_{n=0}^\infty
  \left(1-\epsilon\frac{1}{2}\left(
    \partial_i \mp \partial_i^\dag
  \right)\right)
  (\epsilon\frac{1}{2}\left(\partial_i \mp \partial_i^\dag\right))^n  
  \nonumber\\
  &=&
  \delta_i^j
  \,.
\end{eqnarray}

\paragraph{The derivative operator and finite differences.}

We want to determine (for flat cubic complexes) 
the relation between the derivative operator
$\partial_i$ defined by \refer{definition discrete flat derivative operator} 
and the forward and backward difference
quotients defined in \refer{forward discrete derivative} and \refer{forward discrete derivative}.

Comparison of
\begin{eqnarray}
  \commutator{\extd}{f}
  &=&
  (\lpartial_i f)
  \commutator{\extd}{X^i}
  \nonumber\\
  &=&
  \commutator{\extd}{X^i}(\rpartial f)
\end{eqnarray}
with
\begin{eqnarray}
  \commutator{\extd}{f}
  &\equalby{flat discrete extd explicitly}&
  \commutator{\commutator{\extd}{X^i}\partial_i}{ f }
  \nonumber\\
  &=&
  \commutator{\extd}{X^i}
  \commutator{\partial_i}{ f }+
  \commutator{\commutator{\extd}{X^i}}{ f } \partial_i
  \nonumber\\
  &=&
  \sum\limits_i
  \left(
    \commutator{\extd}{X^i}
    \commutator{\partial_i}{ f }
    +
    \epsilon\commutator{\extd}{X^i}(\rpartial_i f)\partial_i  
  \right)
\end{eqnarray}
and
\begin{eqnarray}
  \commutator{\extd}{f}
  &=&
  \commutator{\partial_i\commutator{\extd}{X^i}}{ f }
  \nonumber\\
  &=&
  \commutator{\partial_i}{ f }
  \commutator{\extd}{X^i}
  +
  \partial_i
  \commutator{\commutator{\extd}{X^i}}{ f } 
  \nonumber\\
  &=&
  \sum\limits_i
  \left(
    \commutator{\partial_i}{ f }
    \commutator{\extd}{X^i}
    +
    \epsilon\, \partial_i \circ (\lpartial_i f)\commutator{\extd}{X^i} 
  \right)
\end{eqnarray}
shows that
\begin{eqnarray}
  \label{commutator discrete partial_i with function}
  \commutator{\partial_i}{f}
  &=&
  (\rpartial_i f)(1-\epsilon \partial_i)
  \nonumber\\
  &=&
  (1-\epsilon \partial_i)
  (\lpartial_i f)
  \,.
\end{eqnarray}
The adjoint of this is
\begin{eqnarray}
  \label{commutator discrete partial_i^dag with function}
  \commutator{\partial_i^\dag}{f}
  &=&
  -
  (\lpartial_i f)(1-\epsilon \partial^\dag_i)
  \,.
\end{eqnarray}

That this makes sense can be checked by noting that
\begin{eqnarray}
  \partial_i
  \;
  f \,g \ket{1}
  &=&
  \commutator{\partial_i}{f}\,g \ket{1} + f\, \commutator{\partial_i}{g}\ket{1}
  \nonumber\\
  &\equalby{commutator discrete partial_i with function}&
  (\rpartial_i f)\,g \ket{1} + f\, (\rpartial_i g)\ket{1} - \epsilon(\partial_i f)(\rpartial_i g)\ket{1}
  \,,
\end{eqnarray}
and similarly
\begin{eqnarray}
  -\partial^\dag_i
  \;
  f \,g \ket{1}
  &=&
  ((\lpartial_i f)g + f (\lpartial_i g) + (\lpartial_i f)(\lpartial_i g))\ket{1}
  \,,
\end{eqnarray}
which agrees with \refer{product rule for discrete almost partial derivatives}.

If
\begin{eqnarray}
  \partial_i^\edag
  &=&
  \partial_i^\dag
\end{eqnarray}
(which is the case for flat cubic complexes with cartesian preferred coordinates)
this allows to reproduce 
\refer{relation coextd and lpartial} and
\refer{relation extd and rpartial} by writing
\begin{eqnarray}
  \extd \, \commutator{\extd}{X^{i_1}}\cdots \commutator{\extd}{X^{i_p}}\,f\,\ket{1}
  &=&
  \commutator{\extd}{X^j}
  \partial_j
   \, \commutator{\extd}{X^{i_1}}\cdots \commutator{\extd}{X^{i_p}}\,f\,\ket{1}
  \nonumber\\
  &\equalby{commutator discrete partial with creator}&
  \commutator{\extd}{X^j}
   \, \commutator{\extd}{X^{i_1}}\cdots \commutator{\extd}{X^{i_p}} \commutator{\partial_j}{f}\ket{1}
  \nonumber\\
  &\equalby{commutator discrete partial_i with function}&
  \commutator{\extd}{X^j}
   \, \commutator{\extd}{X^{i_1}}\cdots \commutator{\extd}{X^{i_p}} \,(\rpartial_j f)\ket{1}    
\end{eqnarray}
and
\begin{eqnarray}
  \ecoextd \, \commutator{\extd}{X^{i_1}}\cdots \commutator{\extd}{X^{i_p}}\,f\,\ket{1}
  &=&
  (\commutator{\coextd}{X^j})^\edag
  \partial^\dag_j
   \, \commutator{\extd}{X^{i_1}}\cdots \commutator{\extd}{X^{i_p}}\,f\,\ket{1}
  \nonumber\\
  &=&
  (\commutator{\extd}{X^j})^{\edag}
   \, \commutator{\extd}{X^{i_1}}\cdots \commutator{\extd}{X^{i_p}} \commutator{\partial^\dag_j}{f}\ket{1}
  \nonumber\\
  &\equalby{commutator discrete partial_i^dag with function}&
  -
  (\commutator{\extd}{X^j})^\edag
   \, \commutator{\extd}{X^{i_1}}\cdots \commutator{\extd}{X^{i_p}} \,(\lpartial_j f)\ket{1}    
  \,.
\end{eqnarray}

Furthermore it follows that the Laplace-Beltrami operator 
\begin{eqnarray}
  \label{Laplace-Beltrami on flat cubic graphs}
  \antiCommutator{\extd}{\ecoextd}
  &=&
  g^{\mu\nu} \partial_\mu \partial_\nu^\dag
\end{eqnarray}
(on flat cubic complexes) acts componentwise as
\begin{eqnarray}
  \label{flat cubic Laplace-Beltrami operator}
  \antiCommutator{\extd}{\coextd} \ket{\extd X^{i_1}\cdots \extd X^{i_p} \;f}
  &=&
  -
  \ket{\extd X^{i_1}\cdots \extd X^{i_p} \;g^{\mu\nu}\lpartial_\mu \rpartial_\nu\,f }
  \,,
\end{eqnarray}
where $-g^{\mu\nu}\lpartial_\mu \rpartial_\nu$ is a discrete version of the 
flat wave operator.

This operator can be rewritten as
\begin{eqnarray}
  -g^{ab}\lpartial_a \rpartial_b
  &=&
  -\frac{1}{2}
  g^{ab}
  \left(
    \lpartial_a \rpartial_b
    +
    \lpartial_b \rpartial_a
  \right)
  \nonumber\\
  &\equalby{interchanging discrete derivatives by translation}&
  -\frac{1}{2}
  g^{ab}
  \left(
    T_{\edge_b-\edge_a}
    \rpartial_a \lpartial_b
    +
    \lpartial_b \rpartial_a
  \right)  
  \nonumber\\
  &=&
  -\frac{1}{2}
  g^{ab}
  \left(
    1 + 
    T_{\edge_b-\edge_a}
  \right)  
  \rpartial_a \lpartial_b
  \nonumber\\
  &=&
  -\frac{1}{2}
  g_{cd}
  \sum\limits_{a,b}
  \left(
    1 + 
    T_{\edge_b-\edge_a}
  \right)  
  g^{ca}g^{db}
  \rpartial_a \lpartial_b
  \,.
\end{eqnarray}

\subsection{Diamond complexes}
\label{section: Diamond complexes}

On every cubic complex there is a special metric operator $\gop = \gopg$
which is induced by the \emph{graph operator} 
\refer{definition graph operator} as follows:

Because of \refer{graph operator in terms of onb creators}
\begin{eqnarray}
  {\bf G} &=& \frac{1}{\epsilon}\sum\limits_a \edgesCreator^a
\end{eqnarray}
and hence
\begin{eqnarray}
  \antiCommutator{{\bf G}}{{\bf G}^\dag}
  &=&
  \frac{D}{\epsilon^2}
\end{eqnarray}
it follows that
\begin{eqnarray}
  \label{metric operator from graph operator}
  \gopg
  &\defas&
  \frac{\epsilon^2}{D}
  \left(
    {\bf G}^\dag{\bf G}
    -
    {\bf G}{\bf G}^\dag
  \right)
  \nonumber\\
  &=&
  \frac{\epsilon^2}{D}
  \commutator{{\bf G}^\dag}{{\bf G}}
  \nonumber\\
  &=&
  \frac{\epsilon^2}{D}
  \left({\bf G} + {\bf G}^\dag\right)
  \left({\bf G} - {\bf G}^\dag\right)  
\end{eqnarray}
is a valid metric operator, since it is invertible and satisfies
\begin{eqnarray}
  \gopg^\dag &=& \gopg
  \nonumber\\
  {\rm grade}\of{\gopg} &=& 0
  \,.
\end{eqnarray}
This is furthermore the \emph{unique} metric operator that can be construted from combinations
of ${\bf G}$ and ${\bf G}^\dag$ alone.\footnote{
  Due to the fact that the metric operator must be of grade 0 the building blocks need to be
${\bf G}{\bf G}^\dag$ and ${\bf G}^\dag {\bf G}$. Products of these produce nothing new, because
\begin{eqnarray}
   \left({\bf G}{\bf G}^\dag\right) \left({\bf G}^\dag {\bf G}\right) &=& 0
  \nonumber\\
   \left({\bf G}^\dag {\bf G}\right)\left({\bf G}{\bf G}^\dag\right) &=& 0
  \nonumber\\
  \left({\bf G}^\dag {\bf G}\right)\left({\bf G}^\dag {\bf G}\right) &=& 
  \frac{D}{\epsilon^2} {\bf G}^\dag {\bf G}
  \,,
\end{eqnarray}
and therefore only additive combinations remain. But the sum
${\bf G}{\bf G}^\dag + {\bf G}^\dag {\bf G}$ is equal to a constant, so that
the only non-trivial combination is the one given above:
\begin{eqnarray}
  {\bf G}{\bf G}^\dag - {\bf G}^\dag {\bf G}
  \,.
\end{eqnarray}
}
\newpage
In this sense the graph operator induces a metric on the cubic complex. Cubic complexes
equipped with this metric shall be called \emph{diamond complexes}, for reasons that 
become apparent below.

The graph operator singles out a direction on the discrete space. This can
be made manifest by introducing the coordinate
\begin{eqnarray}
  \label{time coordinate on diamonds}
  t &\defas&
  \frac{1}{\sqrt{D}} 
  \left(
    X^1 
    + 
    X^2
    +
    \cdots
    +
    X^D
  \right)
  \,.
\end{eqnarray}
The graph operator is just the differential of this coordinate:\footnote{
In the set of preferred coordinates of the cubic complex the graph operator
can obviously be written as
\begin{eqnarray}
  {\bf G}
  &=&
  \frac{1}{\epsilon}
  \left(
    \commutator{\extd}{X^1} 
    + 
    \commutator{\extd}{X^2}
    +
    \cdots
    +
    \commutator{\extd}{ X^D}
  \right)
  \nonumber\\
  {\bf G}^\dag
  &=&
  -
  \frac{1}{\epsilon}
  \left(
    \commutator{\coextd}{X^1} 
    + 
    \commutator{\coextd}{X^2}
    +
    \cdots
    +
    \commutator{\coextd}{X^D}
  \right)
  \,.
\end{eqnarray}
}
\begin{eqnarray}
  {\bf G}
  &=&
  \frac{\sqrt{D}}{\epsilon}
  \commutator{\extd}{t}
  \,.
\end{eqnarray}
and the metric operator can alternatively be written as
\begin{eqnarray}
  \gopg
  &=&
  \commutator{\extd}{t}\commutator{\coextd}{t}
  -
  \commutator{\coextd}{t}\commutator{\extd}{t}
  \nonumber\\
  &=&
  \clifford_-^t \clifford_+^t
  \,,
\end{eqnarray}
where $\clifford_\pm^t = \commutator{\extd}{t} \pm \commutator{t}{\coextd}$.

When further coordinates $r^i$ orthogonal to $t$ are introduced one finds
\begin{eqnarray}
  \gopg \ket{\extd t \extd r^{i_1}\cdots \extd r^{i_n}} &=& - \ket{\extd t \extd r^{i_1}\cdots \extd r^{i_n}}
  \nonumber\\
  \gopg \ket{\extd r^{i_1}\cdots \extd r^{i_n}} &=& + \ket{\extd r^{i_1}\cdots \extd r^{i_n}}
\end{eqnarray}
and comparison with \refer{action of gop} reveals that $\gopg$ gives a metric with nonvanishing
components
\begin{eqnarray}
  g_{tt} &=& -1
  \nonumber\\
  g_{r^i r^j} &=& \delta^{ij}
  \,,
\end{eqnarray}
i.e. a flat \emph{Lorentzian} metric with $t$ being a timelike coordinate.
In other words, the direction singled out by the graph operator is time.

Discrete models of spacetime in the form of cubic lattices with time running along the
main diagonal of the cubes are sometimes addressed as "diamond lattices" in the
literature. That's why we refer to cubic complexes with $\gopg$ being the
metric operator as diamond complexes.

\paragraph{Causal sets.}

The above shows that the graph operator singles out a causal
structure on our discrrete space. We could more generally consider \emph{scalar multiples}
of the preferred metric operator \refer{metric operator from graph operator}:
\begin{eqnarray}
  \gopg\of{V} &\defas& V\of{x} \gopg
  \,.
\end{eqnarray}
These describe geometries where each discrete lightcone is identified by $\gopg$ and
carries a spacetime volume given by $V\of{x}$. This is the data used in
\emph{causal set} theory \cite{Sorkin:2003} to describe spacetime geometry. We here
see that the formalism used here, with the preferred role that the graph operator
$\mathbf{G}$ plays in it, naturally makes contact with concepts known from
causal set theory.

\subsection{Clifford algebra on diamond complexes}
\label{Clifford algebra on diamond complexes}

Let
\begin{eqnarray}
  \clifford_\pm^a
  &=&
  \commutator{\extd}{X^a}
  \pm
  \commutator{\extd}{X^a}^\dag
  \nonumber\\
  &=&
  \commutator{\extd}{X^a}
  \pm
  \commutator{X^a}{\coextd}
  \nonumber\\
  &=&
  \edgesCreator^a \pm \edgesAnnihilator^a  
\end{eqnarray}
be the usual Clifford generators with respect to the Euclidean metric which satisfy
\begin{eqnarray}
  \antiCommutator{\clifford_\pm^a}{\clifford_\pm^b} &=& \pm 2 \delta^{ab}
  \nonumber\\
  \antiCommutator{\clifford_\pm^a}{\clifford_\mp^b} &=& 0
\end{eqnarray}
and let
\begin{eqnarray}
  \clifford_{\gop \pm}^a
  &\defas&
  \commutator{\extd}{X^a}
  \pm
  \commutator{\extd}{X^a}^{\edag}
  \nonumber\\
  &=&
  \edgesCreator^a
  \pm
  \gop \edgesAnnihilator^a \gop^{-1}
\end{eqnarray}
be the Clifford generators asscoiated with the metric induces by the metric operator $\gop$.

We can split these into the timelike generators 
\begin{eqnarray}
 \clifford_\pm^t
 &\defas&
  \frac{1}{\sqrt{D}}\left(\clifford_\pm^1 + \cdots + \clifford_\pm^D\right)
  \nonumber\\
 \clifford_{\gopg\pm}^t
 &\defas&
  \frac{1}{\sqrt{D}}\left(\clifford_{\gopg\pm}^1 + \cdots + \clifford_{\gopg\pm}^D\right)
\end{eqnarray}
and the spacelike ones
\begin{eqnarray}
  \clifford_\pm^{a-t/\sqrt{D}}
  &\defas&
  \clifford_\pm^a - \frac{1}{\sqrt{D}}\clifford_\pm^t
  \nonumber\\
  \clifford_{\gopg\pm}^{a-t/\sqrt{D}}
  &\defas&
  \clifford_{\gopg\pm}^a - \frac{1}{\sqrt{D}}\clifford_{\gopg\pm}^t
  \,,
\end{eqnarray}
so that in particular
\begin{eqnarray}
  \antiCommutator{\clifford_\pm^t}{\clifford_\pm^t} &=& \pm 2
  \nonumber\\
  \antiCommutator{\clifford_\pm^{a-t/\sqrt{D}}}{\clifford_\pm^t} &=& 0
  \,.
\end{eqnarray}

For the Lorentzian Clifford generators one finds\footnote{
The $\gopg$-adjoint of $\edgesCreator^t$ is
\begin{eqnarray}
  \gop \, \edgesAnnihilator^t\, \gop
  &=&
  \clifford_-^t \clifford_+^t\, 
    \frac{1}{2}\left(\clifford_+^t - \clifford_-^t\right)\, \clifford_-^t \clifford_+^t
  \nonumber\\
  &=&
  -\frac{1}{2}\left(\clifford_+^t - \clifford_-^t\right)
  \nonumber\\
  &=&
  - \edgesAnnihilator^t
  \nonumber
\end{eqnarray}
and that of $\edgesCreator^a - \frac{1}{\sqrt{D}}\edgesCreator^t$ is
\begin{eqnarray}
  \gop \, \left(\edgesAnnihilator^a - \frac{1}{\sqrt{D}}\edgesAnnihilator^t\right)\, \gop
  &=&
  \clifford_-^t \clifford_+^t
  \frac{1}{2} 
  \left(
    \clifford_+^a - \clifford_-^a -\frac{1}{\sqrt{D}}\clifford_+^t + \frac{1}{\sqrt{D}}\clifford_-^t
  \right)
  \clifford_-^t \clifford_+^t
  \nonumber\\
  &=&
  \frac{1}{2} 
  \left(
    -\clifford_+^t
    \clifford_+^a\clifford_+^t - 
   \clifford_-^t\clifford_-^a\clifford_-^t 
    +\frac{1}{\sqrt{D}}\clifford_+^t - \frac{1}{\sqrt{D}}\clifford_-^t
  \right)
  \nonumber\\
  &=&
  \frac{1}{2} 
  \left(
    \clifford_+^a - \clifford_-^a -\frac{1}{\sqrt{D}}\clifford_+^t + \frac{1}{\sqrt{D}}\clifford_-^t
  \right)
  \nonumber\\
  &=&
  \edgesAnnihilator^a - \frac{1}{\sqrt{D}}\edgesAnnihilator^t
  \,.
  \nonumber
\end{eqnarray}
}
\begin{eqnarray}
  \label{Lorentzian Clifford generators on hyper diamond}
  \clifford_{\gopg\pm}^t &=& \clifford^t_\mp
  \nonumber\\
  \clifford_{\gopg\pm}^{a-t/\sqrt{D}} &=& \clifford_{\pm}^{a-t/\sqrt{D}}
\end{eqnarray}
and hence they have Lorentzian signature
\begin{eqnarray}
  \antiCommutator{\clifford_{\gopg\pm}^t}{\clifford_{\gopg\pm}^t} &=& \pm 2 (-1)
  \nonumber\\
  \antiCommutator{\clifford_{\gopg\pm}^{a-t/\sqrt{D}}}{\clifford_{\gopg\pm}^t} &=& 0
\end{eqnarray}
as expected.

\subsection{Non-commutative coordinates on diamond complexes}
\label{noncommutative coordinates on diamond complex}

\subparagraph{Adjoint of the coordinates.}
With the Clifford algebra at hand it is now easy to determine the adjoint $(X^a)^\edag$ of the
preferred coordinate functions $X^a$ with respect to $\bracket{\cdot}{\cdot}_\gop$.
One finds
\begin{eqnarray}
  X^a - (X^a)^{\edag}
  &=&
  \epsilon\frac{2\sqrt{D+1}}{D}\; {\bf J}_\gopg^{a-t/\sqrt{D},t},
\end{eqnarray}
where
\begin{eqnarray}
  {\bf J}^{a-t/\sqrt{D},t}_\gopg
  &\defas&
  \frac{1}{2\sqrt{1+1/D}}
  \left(
    \clifford_{\gopg+}^{a-t/\sqrt{D}}
    \clifford_{\gopg+}^t
    -
    \clifford_{\gopg-}^{a-t/\sqrt{D}}
    \clifford_-^t
  \right)      
\end{eqnarray}
is the Clifford algebra generator of Lorentz boosts along 
the spatial direction $X^{a-t/\sqrt{D}}$.

{\it Proof:}
Using
\begin{eqnarray}
  \label{commutator Clifford with coordinates}
  \commutator{\clifford_\pm^a}{X^b}
  &=&
  \epsilon \delta^{ab} \clifford_\mp^a
  \,.
\end{eqnarray}
and
\begin{eqnarray}
  \gop
  &=&
  \frac{1}{D}
  \left(
    \clifford_-^1 + \cdots + \clifford_-^D
  \right)
  \left(
    \clifford_+^1 + \cdots + \clifford_+^D
  \right)
\end{eqnarray}
one finds the commutator
\begin{eqnarray}
  \label{commutator of gopg with preferred coordinates}
  \frac{1}{\epsilon}\commutator{\gopg}{X^a}
  &=&
  \frac{1}{D}
  \left(
  \left(
    \clifford_-^1 + \cdots + \clifford_-^D
  \right)
  \clifford_-^a
  +
  \clifford_+^a
  \left(
    \clifford_+^1 + \cdots + \clifford_+^D
  \right)
  \right)
  \nonumber\\
  &=&
  \frac{1}{D}
  \sum\limits_{b\neq a}
  \left(
    \clifford_+^a
    \clifford_+^b
    -
    \clifford_-^a 
    \clifford_-^b
  \right)
  \nonumber\\
  &=&
  \frac{1}{2D}
  \sum\limits_{b}
  \left(
    \commutator{
    \clifford_+^a}{
    \clifford_+^b}
    -
    \commutator{
    \clifford_-^a}{ 
    \clifford_-^b}
  \right)
  \nonumber\\
  &=&
  \frac{1}{2\sqrt{D}}
  \left(
    \commutator{
    \clifford_+^a}{
    \clifford_+^t}
    -
    \commutator{
    \clifford_-^a}{ 
    \clifford_-^t}
  \right)
  \nonumber\\
  &=&
  \frac{1}{2\sqrt{D}}
  \left(
    \commutator{
    \clifford_+^a-\frac{1}{\sqrt{D}}\clifford_+^t}{
    \clifford_+^t}
    -
    \commutator{
    \clifford_-^a-\frac{1}{\sqrt{D}}\clifford_-^t}{ 
    \clifford_-^t}
  \right)    
  \nonumber\\
  &=&
  \frac{1}{\sqrt{D}}
  \left(
    \left(
    \clifford_+^a-\frac{1}{\sqrt{D}}\clifford_+^t\right)
    \clifford_+^t
    -
    \left(
      \clifford_-^a-\frac{1}{\sqrt{D}}\clifford_-^t
    \right) 
    \clifford_-^t
  \right)      
\end{eqnarray}
and hence
\begin{eqnarray}
  (X^a)^\edag
  &=&
  \gopg\, X^a\, \gopg
  \nonumber\\
  &=&
  X^a + \commutator{\gopg}{X^a}\gopg
  \nonumber\\
  &\equalby{commutator of gopg with preferred coordinates}&
  X^a
  +
  \epsilon
  \frac{1}{\sqrt{D}}
  \left(
    \left(
    \clifford_+^a-\frac{1}{\sqrt{D}}\clifford_+^t\right)
    \clifford_+^t
    -
    \left(
      \clifford_-^a-\frac{1}{\sqrt{D}}\clifford_-^t
    \right) 
    \clifford_-^t
  \right)      
  \clifford_-^t \clifford_+^t
  \nonumber\\
  \label{intermediate result in calculation of coord commutators on diamond complex}
  &=&
  X^a
  -
  \epsilon
  \frac{1}{\sqrt{D}}
  \left(
    \left(
    \clifford_+^a-\frac{1}{\sqrt{D}}\clifford_+^t\right)
    \clifford_-^t
    -
    \left(
      \clifford_-^a-\frac{1}{\sqrt{D}}\clifford_-^t
    \right) 
    \clifford_+^t
  \right)      
  \\
  &\equalby{Lorentzian Clifford generators on hyper diamond}&
  X^a
  -
  \epsilon
  \frac{1}{\sqrt{D}}
  \left(
    \clifford_{\gopg+}^{a-t/\sqrt{D}}
    \clifford_{\gopg+}^t
    -
    \clifford_{\gopg-}^{a-t/\sqrt{D}}
    \clifford_{\gopg-}^t
  \right)
  \,.        
\end{eqnarray}
\endofproof

\subparagraph{(Anti)-self adjoint part of the coordinates.}

The $\gopg$-hermitian and $\gopg$-anti-hermitian part of the coordinates
is
\begin{eqnarray}
  {}_{\rm s} X^a &\defas& \left(X^+ + (X^a)^\edag\right)/2
  \nonumber\\
  {}_{\rm a} X^a &\defas& \left(X^+ - (X^a)^\edag\right)/2
  \,.
\end{eqnarray}
These have the following commutators:
\begin{eqnarray}
  \label{commutators of gopg modified discrete coordinates}
  \commutator{ {}_{\rm s}X^a }{ {}_{\rm s}X^b }
  &=&
  -\frac{\epsilon^2}{2D}\left(\clifford_+^a\clifford_+^b - \clifford_-^a \clifford_-^b\right)
  \nonumber\\
  \commutator{ {}_{\rm a}X^a }{ {}_{\rm a}X^b }
  &=&
  \frac{\epsilon^2}{2D}\left(\clifford_+^a\clifford_+^b - \clifford_-^a \clifford_-^b\right)   
  \nonumber\\
  \commutator{ {}_{\rm a}X^a }{ {}_{\rm s}X^b }  
  &=&
  -\epsilon^2\frac{1}{\sqrt{D}}
  (\delta^{ab}-\frac{2}{D})
  \left(
    \left(\clifford_+^a + \clifford_+^b\right) \clifford_{+}^t
    -
    \left(\clifford_-^a+ \clifford_-^b\right) \clifford_{-}^t
  \right)  
  +
  \epsilon^2\frac{8}{D^2}
  \,.
\end{eqnarray}

{\it Proof:}
Because of $\commutator{X^a}{X^b} = 0$ and $\commutator{(X^a)^\edag}{(X^b)^\edag} = 0$
the only non-trivial part of these commutators is
\begin{eqnarray}
  \commutator{\gopg X^a \gopg}{X^b}
  &\equalby{intermediate result in calculation of coord commutators on diamond complex}&
  -\epsilon^2\frac{1}{\sqrt{D}}
  \left(
    (\delta^{ab}-\frac{1}{D})\clifford_-^a \clifford_{-}^t
    +
    \frac{1}{\sqrt{D}}
    \clifford_{+}^{a-t/\sqrt{D}} \clifford_+^b
  \right)
  \nonumber\\
  &&
  +\epsilon^2\frac{1}{\sqrt{D}}
  \left(
    (\delta^{ab}-\frac{1}{D})\clifford_+^a \clifford_{+}^t
    +
    \frac{1}{\sqrt{D}}
    \clifford_{-}^{a-t/\sqrt{D}} \clifford_-^b
  \right)
  \nonumber\\
  &=&
  -\epsilon^2\frac{1}{\sqrt{D}}
  \left(
    (\delta^{ab}-\frac{1}{D})\clifford_-^a \clifford_{-}^t
    +
    \frac{1}{\sqrt{D}}
    \clifford_{+}^{a} \clifford_+^b
    -\frac{1}{D}\clifford_+^t\clifford_+^b
  \right)
  \nonumber\\
  &&
  +\epsilon^2\frac{1}{\sqrt{D}}
  \left(
    (\delta^{ab}-\frac{1}{D})\clifford_+^a \clifford_{+}^t
    +
    \frac{1}{\sqrt{D}}
    \clifford_{-}^{a} \clifford_-^b
    -
    \frac{1}{D}\clifford_-^t\clifford_-^b
  \right)
  \,.
\end{eqnarray}
Antisymmetrizing yields
\begin{eqnarray}
  \commutator{\gopg X^a \gopg}{X^b} - \commutator{\gopg X^b \gopg}{X^a}
  &=&
  -\epsilon^2\frac{1}{\sqrt{D}}
  \left(
    -\frac{1}{D}\clifford_-^a \clifford_{-}^t
    +
    \frac{1}{\sqrt{D}}
    \clifford_{+}^{a} \clifford_+^b
    -\frac{1}{D}\clifford_+^t\clifford_+^b
  \right)
  \nonumber\\
  &&
  +\epsilon^2\frac{1}{\sqrt{D}}
  \left(
    -\frac{1}{D}\clifford_+^a \clifford_{+}^t
    +
    \frac{1}{\sqrt{D}}
    \clifford_{-}^{a} \clifford_-^b
    -
    \frac{1}{D}\clifford_-^t\clifford_-^b
  \right)
  \nonumber\\
  &&
  +\epsilon^2\frac{1}{\sqrt{D}}
  \left(
    -\frac{1}{D}\clifford_-^b \clifford_{-}^t
    +
    \frac{1}{\sqrt{D}}
    \clifford_{+}^{b} \clifford_+^a
    -\frac{1}{D}\clifford_+^t\clifford_+^a
  \right)
  \nonumber\\
  &&
  -\epsilon^2\frac{1}{\sqrt{D}}
  \left(
    -\frac{1}{D}\clifford_+^b \clifford_{+}^t
    +
    \frac{1}{\sqrt{D}}
    \clifford_{-}^{b} \clifford_-^a
    -
    \frac{1}{D}\clifford_-^t\clifford_-^a
  \right)  
  \nonumber\\
  &=&
  -\epsilon^2\frac{1}{\sqrt{D}}
  \left(
    (-\frac{1}{D})\clifford_-^a \clifford_{-}^t
    +
    \frac{1}{\sqrt{D}}
    \clifford_{+}^{a} \clifford_+^b
    +\frac{1}{D}\clifford_+^b\clifford_+^t
  \right)
  \nonumber\\
  &&
  +\epsilon^2\frac{1}{\sqrt{D}}
  \left(
    (-\frac{1}{D})\clifford_+^a \clifford_{+}^t
    +
    \frac{1}{\sqrt{D}}
    \clifford_{-}^{a} \clifford_-^b
    +
    \frac{1}{D}\clifford_-^b\clifford_-^t
  \right)
  \nonumber\\
  &&
  +\epsilon^2\frac{1}{\sqrt{D}}
  \left(
    (-\frac{1}{D})\clifford_-^b \clifford_{-}^t
    -
    \frac{1}{\sqrt{D}}
    \clifford_{+}^{a} \clifford_+^b
    +\frac{1}{D}\clifford_+^a\clifford_+^t
  \right)
  \nonumber\\
  &&
  -\epsilon^2\frac{1}{\sqrt{D}}
  \left(
    (-\frac{1}{D})\clifford_+^b \clifford_{+}^t
    -
    \frac{1}{\sqrt{D}}
    \clifford_{-}^{a} \clifford_-^b
    +
    \frac{1}{D}\clifford_-^a\clifford_-^t
  \right)
  \nonumber\\
  &=&
  -\epsilon^2 \frac{2}{D}
  \left(
    \clifford_+^a\clifford_+^b - \clifford_-^a\clifford_-^b
  \right)  
  \nonumber 
\end{eqnarray}
and symmetrizing gives
\begin{eqnarray}
  \commutator{\gopg X^a \gopg}{X^b} + \commutator{\gopg X^a \gopg}{X^b}
  &=&
  -\epsilon^2\frac{1}{\sqrt{D}}
  \left(
    (\delta^{ab}-\frac{1}{D})\clifford_-^a \clifford_{-}^t
    -\frac{1}{D}\clifford_+^t\clifford_+^b
  \right)
  \nonumber\\
  &&
  +\epsilon^2\frac{1}{\sqrt{D}}
  \left(
    (\delta^{ab}-\frac{1}{D})\clifford_+^a \clifford_{+}^t
    -
    \frac{1}{D}\clifford_-^t\clifford_-^b
  \right)
  \nonumber\\
  &&
  -\epsilon^2\frac{1}{\sqrt{D}}
  \left(
    (\delta^{ab}-\frac{1}{D})\clifford_-^b \clifford_{-}^t
    -\frac{1}{D}\clifford_+^t\clifford_+^a
  \right)
  \nonumber\\
  &&
  +\epsilon^2\frac{1}{\sqrt{D}}
  \left(
    (\delta^{ab}-\frac{1}{D})\clifford_+^b \clifford_{+}^t
    -
    \frac{1}{D}\clifford_-^t\clifford_-^a
  \right)  
  \nonumber\\
  &=&
  -\epsilon^2\frac{1}{\sqrt{D}}
  \left(
    (\delta^{ab}-\frac{1}{D})\clifford_-^a \clifford_{-}^t
    +\frac{1}{D}\clifford_+^b\clifford_+^t
  \right)
  \nonumber\\
  &&
  +\epsilon^2\frac{1}{\sqrt{D}}
  \left(
    (\delta^{ab}-\frac{1}{D})\clifford_+^a \clifford_{+}^t
    +
    \frac{1}{D}\clifford_-^b\clifford_-^t
  \right)
  \nonumber\\
  &&
  -\epsilon^2\frac{1}{\sqrt{D}}
  \left(
    (\delta^{ab}-\frac{1}{D})\clifford_-^b \clifford_{-}^t
    +\frac{1}{D}\clifford_+^a\clifford_+^t
  \right)
  \nonumber\\
  &&
  +\epsilon^2\frac{1}{\sqrt{D}}
  \left(
    (\delta^{ab}-\frac{1}{D})\clifford_+^b \clifford_{+}^t
    +
    \frac{1}{D}\clifford_-^a\clifford_-^t
  \right)  
  \nonumber\\
  &&
  +
  \epsilon^2\frac{8}{D^2}
  \nonumber\\
  &=&
  -\epsilon^2\frac{1}{\sqrt{D}}
  \left(
    (\delta^{ab}-\frac{2}{D})\clifford_-^a \clifford_{-}^t
  \right)
  \nonumber\\
  &&
  +\epsilon^2\frac{1}{\sqrt{D}}
  \left(
    (\delta^{ab}-\frac{2}{D})\clifford_+^a \clifford_{+}^t
  \right)
  \nonumber\\
  &&
  -\epsilon^2\frac{1}{\sqrt{D}}
  \left(
    (\delta^{ab}-\frac{2}{D})\clifford_-^b \clifford_{-}^t
  \right)
  \nonumber\\
  &&
  +\epsilon^2\frac{1}{\sqrt{D}}
  \left(
    (\delta^{ab}-\frac{2}{D})\clifford_+^b \clifford_{+}^t
  \right)  
  \nonumber\\
  &&
  +
  \epsilon^2\frac{8}{D^2}
  \nonumber\\
  &=&
  -\epsilon^2\frac{1}{\sqrt{D}}
  (\delta^{ab}-\frac{2}{D})
  \left(
    -
    \clifford_-^a \clifford_{-}^t
    +
    \clifford_+^a \clifford_{+}^t
    -
    \clifford_-^b \clifford_{-}^t
    +
    \clifford_+^b \clifford_{+}^t
  \right)  
  +
  \epsilon^2\frac{8}{D^2}
  \nonumber
  \,.
\end{eqnarray}

This has the following important implication: We started with the differential
calculus of a \emph{commutative} algebra $\cal A$ of functions on an enumerable set. 
The introduction of a Lorentzian metric required us to use an inner product
$\bracket{\cdot}{\cdot}_\gopg$ on (the vector space of) this algebra with
respect to which elements of {\cal A} are \emph{not self-adjoint}. We may
switch to the self-adjoint component of these coordinates and regard this as
a new set of modified coordinates. As shown by \refer{commutators of gopg modified discrete coordinates} 
the algebra of these
modified coordinates is however no longer commutative.

\subsection{The $1+1$ dimensional diamond complex.}

The non-relativistic limit of the 1+1 dimensional diamond complex
has already been discussed in \S\fullref{Example: Hyper diamond model and stochastic calculus.}.

The volume form on every diamond complex can be set to
\begin{eqnarray}
  \ket{\evol} &=& 
  \ket{\extd X^1 \extd X^2 \cdots \extd X^D}
\end{eqnarray}
by choosing $c=-1$ in \refer{relation dV to det(eta) in discrete geometry}.

In the $1+1$ dimensional case it is convenient and conventional to use
indices in $\set{+,-}$, so we identify
\begin{eqnarray}
  X^+ &\defas& X^1
  \nonumber\\
  X^- &\defas& X^2
  \,.
\end{eqnarray}

A general element of $\Omega$ is 
\begin{eqnarray}
  \ket{\fatalpha}
  &\defas&
  \ket{
    \alpha^{(0)}
  +
   \extd X^\mu\, \alpha^{(1)}_\mu
  +
  \extd X^1 \extd X^2\, \alpha^{(2)}
  } 
  \,.
\end{eqnarray}

We now list the action of various operators on such states. First of all the metric operator
acts as
\begin{eqnarray}
  \gopg \ket{\alpha}
  &=&
  \ket{\alpha}
  \nonumber\\
  \gopg \ket{\extd X^1\,\alpha}
  &=&
  -  \ket{\extd X^2\,\alpha}
  \nonumber\\
  \gopg \ket{\extd X^2\,\alpha}
  &=&
  -  \ket{\extd X^1\,\alpha}
  \nonumber\\
  \gopg
  \ket{\extd X^1 \extd X^2\,\alpha}
  &=&
  -\ket{\extd X^1 \extd X^2\,\alpha}  
  \,.
\end{eqnarray}

The Hodge star operator associated with the unmodified inner product $\bracket{\cdot}{\cdot}$
gives
\begin{eqnarray}
  \hodge \ket{\alpha} &=& \ket{\alpha \,\extd X^1 \extd X^2}
  \nonumber\\
  \hodge \ket{\extd X^1\, \alpha} &=& \ket{\alpha\, \extd X^2}
  \nonumber\\
  \hodge \ket{\extd X^2\, \alpha} &=& -\ket{\alpha\, \extd X^1}
  \nonumber\\
  \hodge \ket{\extd X^1 \extd X^2\,\alpha} &=& \ket{\alpha}    
\end{eqnarray}
or equivalently
\begin{eqnarray}
  \hodge \ket{\alpha} &=& \ket{\extd X^1 \extd X^2\,T_{\vec e_1 + \vec e_2}[\alpha] }
  \nonumber\\
  \hodge \ket{\alpha\,\extd X^1} &=& \ket{\extd X^2\,T_{\vec e_1 + \vec e_2}[\alpha]}
  \nonumber\\
  \hodge \ket{\alpha\, \extd X^2} &=& -\ket{\extd X^1\, T_{\vec e_1 + \vec e_2}[\alpha]}
  \nonumber\\
  \hodge \ket{\alpha\,\extd X^1 \extd X^2} &=& \ket{T_{\vec e_1 +\vec e_2}[\alpha]}    
  \,.
\end{eqnarray}
Inverting these relations gives the action of $\invhodge$:
\begin{eqnarray}
  \invhodge \ket{\alpha} &=& \ket{T_{-\vec e_1-\vec e_2}[\alpha]\,\extd X^1 \extd X^2}
  \nonumber\\
  \invhodge \ket{\extd X^1 \, \alpha} &=& - \ket{T_{-\vec e_1 - \vec e_2}[\alpha]\,\extd X^2}
  \nonumber\\
  \invhodge \ket{\extd X^2 \, \alpha} &=&  \ket{T_{-\vec e_1 - \vec e_2}[\alpha]\,\extd X^1}
  \nonumber\\
  \invhodge \ket{\extd X^1 \extd X^2\,\alpha} &=& \ket{T_{-\vec e_1 - \vec e_2}[\alpha]}
  \,.
\end{eqnarray}

Acting on this with $\extd$
\begin{eqnarray}
  \extd \invhodge \ket{\alpha} &=& 0
  \nonumber\\
  \extd \invhodge \ket{\extd X^1 \, \alpha} &=& 
    - \ket{T_{-\vec e_1 - \vec e_2}[\lpartial_1\alpha]\,\extd X^1 \extd X^2}
  \nonumber\\
  \extd \invhodge \ket{\extd X^2 \, \alpha} &=&  
    -\ket{T_{-\vec e_1 - \vec e_2}[\lpartial_2\alpha]\,\extd X^1\extd X^2}
  \nonumber\\
  \extd\invhodge \ket{\extd X^1 \extd X^2 \, \alpha} &=& 
   \ket{T_{-\vec e_1 - \vec e_2}[\lpartial_1\alpha]\,\extd X^1}
   +
  \ket{T_{-\vec e_1 - \vec e_2}[\lpartial_2\alpha]\,\extd X^2}
\end{eqnarray}
and then again with $\hodge$
\begin{eqnarray}
  \hodge\extd \hodge^{-1} \ket{\alpha} &=& 0
  \nonumber\\
  \hodge\extd \hodge^{-1} \ket{\extd X^1 \, \alpha} &=& 
    - \ket{\lpartial_1\alpha}
  \nonumber\\
  \hodge\extd \hodge^{-1} \ket{\extd X^2 \, \alpha} &=&  
    -\ket{\lpartial_2\alpha}
  \nonumber\\
  \hodge\extd\hodge^{-1} \ket{\extd X^1 \extd X^2\,\alpha} &=& 
   \ket{\extd X^2\, \lpartial_1\alpha}
   -
  \ket{\extd X^1\, \lpartial_2\alpha}
\end{eqnarray}
gives, up to a sign, the ``matrix elements'' of the Euclidean $\coextd$:
\begin{eqnarray}
  \coextd \ket{\alpha} &=& 0
  \nonumber\\
  \coextd \ket{\extd X^1 \, \alpha} &=& 
    - \ket{\lpartial_1\alpha}
  \nonumber\\
  \coextd \ket{\extd X^2 \, \alpha} &=&  
    -\ket{\lpartial_2\alpha}
  \nonumber\\
  \coextd \ket{\extd X^1 \extd X^2\,\alpha} &=& 
  \ket{\extd X^1\, \lpartial_2\alpha}
   -
   \ket{\extd X^2\, \lpartial_1\alpha}
  \,.
\end{eqnarray}

Those of $\extd$ are of course
\begin{eqnarray}
  \extd\ket{\alpha} &=& \ket{\extd X^1\, (\rpartial_1 \alpha)} + \ket{\extd X^2\, (\rpartial_2)}
  \nonumber\\
  \extd\ket{\extd X^1\, \alpha} &=& -\ket{\extd X^1 \extd X^2\,\rpartial_2\alpha}
  \nonumber\\
  \extd\ket{\extd X^2\, \alpha} &=& \ket{\extd X^1 \extd X^2\,\rpartial_1\alpha}
  \nonumber\\
  \extd \ket{\extd X^1 \extd X^2\,\alpha} &=& 0
  \,.
\end{eqnarray}

This implies
\begin{eqnarray}
  \coextd\extd \ket{\fatalpha}
  &=&
  \ket{
  -(\lpartial_1 \rpartial_1  + \lpartial_2 \rpartial_2) \alpha^{(0)}
  +
    \extd X^1\,
    \lpartial_2
    (\rpartial_1 \alpha^{(1)}_2 - \rpartial_2 \alpha^{(1)}_1)
  -
    \extd X^2\,
    \lpartial_1
    (\rpartial_1 \alpha^{(1)}_2 - \rpartial_2 \alpha^{(1)}_1)
  }
  \nonumber\\
\end{eqnarray}
and
\begin{eqnarray}
  \extd\coextd\ket{
    \fatalpha
  }
  &=&
  \ket{
  -
  \extd X^1\, (\rpartial_1\lpartial_1 \alpha^{(1)}_1 + \rpartial_1\lpartial_2 \alpha^{(1)}_2)
  -
  \extd X^2\, (\rpartial_2\lpartial_1 \alpha^{(1)}_1 + \rpartial_2\lpartial_2 \alpha^{(1)}_2)
  -
    \extd X^1 \extd X^2\,
    (\rpartial_1\lpartial_1 + \rpartial_2 \lpartial_2 )\alpha^{(2)}
  }  
  \,.
  \nonumber\\
\end{eqnarray}

Therefore the action of the Euclidean Laplace-Beltrami operator is

\hspace{-2cm}\parbox{0cm}{
\begin{eqnarray}
  &&\antiCommutator{\extd}{\coextd}\ket{\fatalpha}
  \nonumber\\
  &=&
  -
  \ket{
  (\lpartial_1 \rpartial_1  + \lpartial_2 \rpartial_2) \alpha^{(0)}
  +
    \extd X^2 \, (\lpartial_1 \rpartial_1 + \rpartial_2 \lpartial_2)\alpha^{(1)}_2  
  +
    \extd X^1 \, (\rpartial_1 \lpartial_1 + \lpartial_2 \rpartial_2)\alpha^{(1)}_1  
  +
    \extd X^1 \extd X^2\,
    (\rpartial_1\lpartial_1 \alpha^{(2)} + \rpartial_2 \lpartial_2 \alpha^{(2)})
  }    
  \nonumber\\
  &&
  +
  \underbrace{
  \ket{
    \extd X^1 (\lpartial_2\rpartial_1 - \rpartial_1\lpartial_2)\alpha^{(1)}_2
  +
    \extd X^2 (\lpartial_1\rpartial_2 - \rpartial_2\lpartial_1)\alpha^{(1)}_2
  }
  }_{\equalby{discrete cubic forward and backward differences commute} 0}
  \,,
\end{eqnarray}
}

i.e. every component of $\fatalpha$ is acted on with the discrete Euclidean
wave operator $(\lpartial_1 \rpartial_1 + \lpartial_2 \rpartial_2)$.

$\,$\\

Now turn to the Lorentzian objects:

The action of the true Hodge star operator is determined by
$\ehodge = \gopg \circ \hodge$ \refer{relation hodge to ehodeg}:
\begin{eqnarray}
  \ehodge \ket{\alpha} &=& -\ket{\extd X^1 \extd X^2\, T_{\vec e_1 + \vec e_2}[\alpha]}
  \nonumber\\
  \ehodge \ket{\alpha\,\extd X^1} &=& -\ket{\extd X^1\, T_{\vec e_1 + \vec e_2}[\alpha]}
  \nonumber\\
  \ehodge \ket{\alpha\, \extd X^2} &=& \ket{\extd X^2	\, T_{\vec e_1 + \vec e_2}[\alpha]}
  \nonumber\\
  \ehodge \ket{\alpha\,\extd X^1 \extd X^2} &=& \ket{T_{\vec e_1 + \vec e_2}[\alpha]}    
  \,.
\end{eqnarray}
Its inverse therefore is
\begin{eqnarray}
  \invehodge \ket{\alpha} &=& \ket{T_{-\vec e_1 - \vec e_2}[\alpha]\, \extd X^1 \extd X^2}
  \nonumber\\
  \invehodge \ket{\extd X^1\, \alpha}
  &=&
  -\ket{T_{-\vec e_1 - \vec e_2}[\alpha]\, \extd X^1}
  \nonumber\\
  \invehodge \ket{\extd X^2\, \alpha}
  &=&
  \ket{T_{-\vec e_1 - \vec e_2}[\alpha]\, \extd X^2}
  \nonumber\\
  \invehodge
  \ket{\extd X^1 \extd X^2\, \alpha}
  &=&
  -\ket{T_{-\vec e_1 - \vec e_2}[\alpha]}
  \,.
\end{eqnarray}
Acting again with $\extd$
\begin{eqnarray}
  \extd\invehodge \ket{\alpha} 
   &=& 0
  \nonumber\\
  \extd\invehodge \ket{\extd X^1\, \alpha}
  &=&
  \ket{T_{-\vec e_1 - \vec e_2}[\lpartial_2 \alpha]\, \extd X^1\extd X^2}
  \nonumber\\
  \extd\invehodge \ket{\extd X^2\, \alpha}
  &=&
  \ket{T_{-\vec e_1 - \vec e_2}[\lpartial_1\alpha]\, \extd X^1\extd X^2}
  \nonumber\\
  \extd\invehodge
  \ket{\extd X^1 \extd X^1\, \alpha}
  &=&
  -\ket{T_{-\vec e_1 - \vec e_2}[\lpartial_1\alpha]\, \extd X^1}
  -\ket{T_{-\vec e_1 - \vec e_2}[\lpartial_2\alpha]\, \extd X^2}
\end{eqnarray}
and then with $\ehodge$
\begin{eqnarray}
  \ehodge\extd\invehodge \ket{\alpha} 
   &=& 0
  \nonumber\\
  \ehodge\extd\invehodge \ket{\extd X^1\, \alpha}
  &=&
  \ket{\lpartial_2 \alpha}
  \nonumber\\
  \ehodge\extd\invehodge \ket{\extd X^2\, \alpha}
  &=&
  \ket{\lpartial_1\alpha}
  \nonumber\\
  \ehodge\extd\invehodge
  \ket{\extd X^1 \extd X^2\, \alpha}
  &=&
  \ket{\extd X^1\, \lpartial_1\alpha}
  -
  \ket{\extd X^2 \lpartial_2\alpha}
\end{eqnarray}
yields the Lorentzian $\ecoextd$:
\begin{eqnarray}
  \ecoextd \ket{\alpha} 
   &=& 0
  \nonumber\\
  \ecoextd \ket{\extd X^1\, \alpha}
  &=&
  \ket{\lpartial_2 \alpha}
  \nonumber\\
  \ecoextd \ket{\extd X^2\, \alpha}
  &=&
  \ket{\lpartial_1\alpha}
  \nonumber\\
  \ecoextd
  \ket{\extd X^1 \extd X^2\, \alpha}
  &=&
  \ket{\extd X^2 \lpartial_2\alpha}
  -
  \ket{\extd X^1\, \lpartial_1\alpha}
  \,.
\end{eqnarray}

This implies
\begin{eqnarray}
  \ecoextd \extd \ket{\fatalpha}
  &=&
  \ket{
  (\lpartial_2 \rpartial_1 + \lpartial_1 \rpartial_2)\alpha^{(0)}
  +
    \extd X^2 \, \lpartial_2(\rpartial_1 \alpha^{(1)}_2 - \rpartial_2 \alpha^{(1)}_1)
  -
    \extd X^1 \, \lpartial_1(\rpartial_1 \alpha^{(1)}_2 - \rpartial_2 \alpha^{(1)}_1)
  }
\end{eqnarray}
and
\begin{eqnarray}
  \extd \ecoextd \ket{\fatalpha}
  &=&
  \ket{
    \extd X^1 \; \rpartial_1 (\lpartial_2 \alpha^{(1)}_1 + \lpartial_1 \alpha^{(1)}_2)
    +
    \extd X^2 \; \rpartial_2 (\lpartial_2 \alpha^{(1)}_1 + \lpartial_1 \alpha^{(1)}_2)
  +
    \extd X^1 \extd X^2 \, 
    \left(
       \rpartial_1 \lpartial_2 \alpha^{(2)}
       +\rpartial_2 \lpartial_1 \alpha^{(2)}
    \right)
  }
  \,.
  \nonumber\\
\end{eqnarray}

Therefore the Lorentzian Laplace-Beltrami operator atcs as
\begin{eqnarray}
  &&
  \antiCommutator{\extd}{\ecoextd}\ket{\fatalpha}
  \nonumber\\
  &=&
  \ket{
    (\lpartial_2 \rpartial_1 + \lpartial_1 \rpartial_2)\alpha^{(0)}
  +
  \extd X^1 \,(\lpartial_1\rpartial_2 + \rpartial_1\lpartial_2)\alpha^{(1)}_1
  +
  \extd X^2 \,(\lpartial_2\rpartial_1 + \rpartial_2\lpartial_1)\alpha^{(1)}_2
  +
    \extd X^1 \extd X^2 \, \left(
       \rpartial_1 \lpartial_2 
       +\rpartial_2 \lpartial_1 
    \right)\alpha^{(2)}
  }
  \,,
  \nonumber\\
\end{eqnarray}
i.e. by multiplying every component by the Lorentzian wave operator
$\lpartial_2 \rpartial_1 + \lpartial_1 \rpartial_2$. Due
to \refer{discrete cubic forward and backward differences commute}
all the $\lpartial_a$ and $\rpartial_b$ mutually commute and hence the general solution
to the Lorentzian wave equation is just as in the
continuum a superposition of ``left-moving'' and ``right moving''
waves:
\begin{eqnarray}
  (\lpartial_2 \rpartial_1 + \lpartial_1 \rpartial_2)f = 0
  \;&\Leftrightarrow&\;
  f\of{\vec x} = f_+\of{X^+} + f_-\of{X^-}
  \,.
\end{eqnarray}

\newpage
\section{Lattice Yang-Mills and fermions}
\label{noncom function algebras}

\subsection{Lattice Yang-Mills theory}
\label{lattice Yang-Mills theory}

All the previous results generalize straightforwardly to the case where the underlying
function algebra is the product of that generated by $\set{\delgen{x}}_x$ with 
any other (possibly noncommutative) algebra $G$, where the algebra product is
$(\delgen{x}\otimes g_x)(\delgen{y}\otimes g_y) = (\delgen{x}\delgen{y}\otimes g_x g_y)$,
 if we decree that elements of 
$G$ commute with $\extd$: For instance a typical elementary edge then looks like
\begin{eqnarray}
  (\delgen{x}\otimes g_x) \commutator{\extd}{(\delgen{y}\otimes g_y)}
  &=&
  (\delgen{x}\otimes g_x g_y) \commutator{\extd}{(\delgen{y}\otimes 1)}
  \nonumber\\
  &\defas&
  (g_x g_y)
  \delgen{x} \commutator{\extd}{\delgen{y}}
  \nonumber\\
  &\defas&  
  \delgen{x} \commutator{\extd}{\delgen{y}}
  (g_x g_y)
  \,.
\end{eqnarray}
Here the first line is a condition on $\extd$ while the last two lines are
just definition of notation.

In order to adapt the integral to this new algebra we need to have some
map 
\begin{eqnarray}
  \lgtrace : G \to \C  
  \,.
\end{eqnarray}
Using this map the integral \refer{definition discrete integral} is generalized to
\begin{eqnarray}
  \int A \; dV
  &\defas&
  \lgtrace
  A\of{\vec x}\; dV\of{\vec x}
  \,,
\end{eqnarray}
where the notation should be self-explanatory.

One needs furthermore to define the adjoint of group-valued 0-forms. With the simple
$U\of{1}$ example in mind and using hindsight it is clear that we need
\begin{eqnarray}
  g^{\dag} &=& g^{-1}
  \,.
\end{eqnarray}

As the notation suggest, an important special case is that where $G$ is a group algebra.
In order to see how the discrete version of covariant derivative operators with respect to some
group bundle should look like, consider for a moment $G$ to be the algebra of
$N\times N$ matrices for some $N$. In the continuum the covariant derivative would be
$\extd + A = \extd + A_\mu dx^\mu$, where $A_\mu$ is a matrix-valued function. But
$A_\mu$ should take values in the tangent space of $G$, not in $G$ itself, whereas
by the above any discrete 1-form is necessarily $G$-valued. This is resolved as follows:
The covariant derivative of any $p$-form $B$ is
$\superCommutator{\extd + A}{B}$. For well behaved graphs we may rewrite this using the
$G$-generalized graph operator $\bf G$ which is simply 
obtained from that in \S\fullref{set of edges and graph operator}
by setting all the $G$ coefficients to unity:
\begin{eqnarray}
   \superCommutator{\extd + A}{B}
   &\equalby{extd action in terms of G with no intermediate edges}&
   \superCommutator{{\bf G} + A}{B}
   \,.
\end{eqnarray}
But now 
\begin{eqnarray}
  {\bf G} + A
  &\equalby{graph operator in terms of onb creators}&
  \sum\limits_\mu
  \left(
    \frac{1}{\epsilon}
    \edgesCreator^\mu
    +
    A_\mu \edgesCreator^\mu
  \right)
  \nonumber\\
  &=&
  \frac{1}{\epsilon}
  \sum\limits_\mu
  (1 + \epsilon A_\mu)\edgesCreator^\mu
\end{eqnarray}
and we see that ${\bf G} + A$ is just the infinitesimal holonomy of $A$ along $\edge^\mu$.
But since the holonomy is $G$-valued this suggests that the correct discrete version of  
the covariant derivative is
\begin{eqnarray}
  {\rm d}_A B &\defas& \commutator{{\bf H}}{B}
  \,,
\end{eqnarray}
where ${\bf H}$ is the discrete $G$-valued 1-form which assigns the $G$-holonomy
$H_\mu\of{\vec x}$
in question to every edge $\delgen{\vec x,\vec x+\edge^\mu}$.

In particular the Yang-Mills 
\emph{field strength} ${\bf F}$ is now just the square of ${\bf H}$:
\begin{eqnarray}
  {\rm d}_A A &\to&
  \frac{1}{2}
  \superCommutator{{\bf H}}{\bf H}
  \nonumber\\
  &=&
  {\bf H}^2
  \,.
\end{eqnarray}
It should be noted that even though ${\bf H}$ is a 1-form, its square does in general 
\emph{not} vanish, as it does in the continuum (\cf \refer{anticommutator of 1-form}).
The non-commutativity precisely produces the desired field strength,
by the above reasoning.
With 
\begin{eqnarray}
  {\bf H} = \sum\limits_{\vec x} H_\mu\of{\vec x}\delgen{\vec x, \vec x + \edge^\mu}
\end{eqnarray}
we find (on cubic graphs) that
\begin{eqnarray}
  {\bf H}^2
  &=&
  \left(
  H_\mu\of{\vec x}H_\nu\of{\vec x + \edge^\mu}
  -
H_\nu\of{\vec x}H_\mu\of{\vec x + \edge^\nu}
  \right)
  \delgen{\vec x, \vec x + \edge^\mu, \vec x + \edge^\mu + \edge^\nu}
  \,,
\end{eqnarray}
which is, just as expected, the two form that assigns to every
plaquette the difference of holonomies around that plaquette.

It is now easy to write down the general action $S_{\rm YM}$ for $G$-Yang-Mills theory on the
discrete space:
\begin{eqnarray}
  \label{H square version of YM lattice action}
  S_{\rm YM}
  &=&
  \bracket{{\bf H}^2}{{\bf H}^2}
  \nonumber\\
  &=&
  \int 
    \lgtrace
    \left({\bf H^{\dag 2}}{\bf H}^2\right)_{\cal S}
  \,.
\end{eqnarray}
Note that this action is defined for arbitrary background metrics.

\begin{figure}[h]
\begin{center}
\begin{picture}(100,100)
\includegraphics{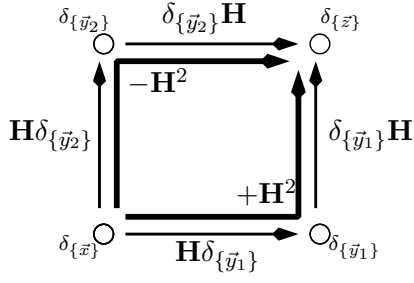}
\put(-37,16){$+{\bf H}^2$}
\put(-78,58){$-{\bf H}^2$}
\put(-60,-6){${\bf H}\delgen{\vec y_1}$}
\put(-65,84){$\delgen{\vec y_2}{\bf H}$}
\put(-2,40){$\delgen{\vec y_1}{\bf H}$}
\put(-123,40){${\bf H}\delgen{\vec y_2}$}
\put(-104,0){${}_{\delgen{\vec x}}$}
\put(-2,0){${}_{\delgen{\vec y_1}}$}
\put(-104,86){${}_{\delgen{\vec y_2}}$}
\put(-5,86){${}_{\delgen{\vec z}}$}
\end{picture}
\end{center}
\caption{Discrete gauge field strength on a plaquette.}
\label{plaquette.eps}
\end{figure}

It is maybe remarkable that the non-commutativity of 0-forms and 1-forms on the
lattice drastically simplifies the notion of gauge covariant derivative to 
a simple algebraic porduct of the gauge holonomy 1-form with itself. The
discrete gauge theory is in this sense conceptually actually simpler than
the continuum theory. In order to illustrate the relevant mechanism in more
detail let us restrict attention to a single plaquette as illustrated
in figure \ref{plaquette.eps} and work out the value of ${\bf H}^2$ on that plaquette
in full detail.

For notational simplicity we assume that the holonomy along any of the 4-edges
is obtained from a gauge connection which is constant along that edge. Then the
holonomy 1-form reads
\begin{eqnarray}
  {\bf H} &=& 
    H_1\of{\vec x}\delgen{\vec x,\vec y_1}
    +
    H_1\of{\vec y_2}\delgen{\vec y_2,\vec z}
    +
    H_2\of{\vec x}\delgen{\vec x,\vec y_2}
    +
    H_2\of{\vec y_1}\delgen{\vec y_1,\vec z}
  \nonumber\\
  &=&
    \exp\of{\epsilon A_1\of{\vec x}}\delgen{\vec x,\vec y_1}
    +
    \exp\of{\epsilon A_1\of{\vec y_2}}\delgen{\vec y_2,\vec z}
    +
    \exp\of{\epsilon A_2\of{\vec x}}\delgen{\vec x,\vec y_2}
    +
    \exp\of{\epsilon A_2\of{\vec y_1}}\delgen{\vec y_1,\vec z}
  \nonumber\\
  &=&
    (1+ \epsilon A_1\of{\vec x})\delgen{\vec x,\vec y_1}
    +
    (1+ \epsilon A_1\of{\vec y_2})\delgen{\vec y_2,\vec z}
    +
    (1+ \epsilon A_2\of{\vec x})\delgen{\vec x,\vec y_2}
    +
    (1+ \epsilon A_2\of{\vec y_1})\delgen{\vec y_1,\vec z}
  +
  \order{\epsilon}
  \,.
  \nonumber\\
\end{eqnarray}
Its square is
\begin{eqnarray}
   {\bf H}^2
   &=&
  \delgen{\vec x, \vec y_1,\vec z}
  \left(
    (1+ \epsilon A_1\of{\vec x})
    (1+ \epsilon A_2\of{\vec y_1})
    -
    (1+ \epsilon A_2\of{\vec x})
    (1+ \epsilon A_1\of{\vec y_2})
  \right)
  + \order{\epsilon}
  \nonumber\\
  &=&
  \epsilon^2
  \delgen{\vec x, \vec y_1,\vec z}
  \left(
     \frac{
     A_2\of{\vec y_1}
     -
     A_2\of{\vec x}
     }{\epsilon}
     -
     \frac{
     A_1\of{\vec y_2}
     -
     A_1\of{\vec x}     
     }{\epsilon}
     +
     A_1\of{\vec x}A_2\of{\vec y_1}
     -
     A_2\of{\vec x}A_1\of{\vec y_2}
  \right)
  + \order{\epsilon^3}
  \nonumber\\
  &=&
  \epsilon^2
  \delgen{\vec x, \vec y_1,\vec z}
  \left(
     \lpartial_1 A_2\of{\vec x}
     -
     \lpartial_2 A_1\of{\vec x}
     +
     \commutator{A_1}{A_2}\of{\vec x}
  \right)
    +
    \order{\epsilon^3}
  \,.	
\end{eqnarray}
Therefore this expression does indeed reduce to the correct continuum quantity for 
$\epsilon \to 0$:
\begin{eqnarray}
  {\bf F} &=& \frac{1}{\epsilon^2}{\bf H}^2 + \order{\epsilon}
  \,.
\end{eqnarray}

\paragraph{Wilson lines.}

The action \refer{H square version of YM lattice action} is indeed equal to the well-known Wilson-line action 
\cite{Wilson:1974} for
lattice gauge theory (see chapter 1 of \cite{Necco:2003} for a brief review):
If we write $R,L \in G$ for the holonomy along $(\vec x \to \vec y_1 \to \vec z)$
and $(\vec x \to \vec y_2 \to \vec z)$ respectively we have
\begin{eqnarray}
  {\bf H}^2 &=&
  \frac{1}{\epsilon^2}
  \delgen{\vec x, \vec y_1,\vec z}
  \left(
    R - L
  \right)
  \,.
\end{eqnarray}
This gives
\begin{eqnarray}
  \bracket{{\bf H}^2}{{\bf H}^2}
  &=&
  \bra{1}
  {\bf H}^{2\dag}{\bf H}^2 \ket{1}
  \nonumber\\
  &=&
  {\rm tr}
  \left(R^{-1}-L^{-1}\right)
  \left(
    R - L
  \right)
  \nonumber\\
  &=&
  2
  \rm tr
  \left(
    1 
    -\frac{1}{2}( R^{-1}L
    + L^{-1}R
    )
  \right)
  \nonumber\\
  &=&
  2N
  \left(
   1
   -
   \frac{1}{N}{\rm Re}W
  \right) 
  \,, 
\end{eqnarray}
(\cf eq. (1.5) of \cite{Necco:2003})
where $W$ is the \emph{Wilson line}
\begin{eqnarray}
  W &=& {\rm tr}L^{-1}R
  \,.
\end{eqnarray}

\paragraph{Equations of motion.}

The form \refer{H square version of YM lattice action} of the lattice YM action  makes
it easy to derive the classical equations of motion:
Varying the action yields
\begin{eqnarray}
  \superCommutator{{\bf H}}{\cdot}^\dag
  {\bf H}^2 \ket{1} &=& 0
  \,.
\end{eqnarray}
The \emph{Bianchi identity} hold trivially:
\begin{eqnarray}
  \superCommutator{{\bf H}}{{\bf H}^2} &=& 0
  \,.
\end{eqnarray}

\paragraph{Gauge-covariant derivative.}

With the notation
\begin{eqnarray}
  \hextd &\defas& \superCommutator{{\bf H}}{\cdot}
  \nonumber\\
  \hcoextd &\defas& (\extd_A)^\dag
\end{eqnarray}
the above equations of motion take the more familiar form
\begin{eqnarray}
  \label{lattice YM equations of motion a}
  \hextd {\bf H}^2 &=& \hextd (\frac{1}{2}\hextd H) \;=\; 0
  \nonumber\\
  \hcoextd {\bf H}^2 &=& \hcoextd (\frac{1}{2}\hextd H) \;=\; 0
  \,.
\end{eqnarray}
We purposefully write $\hextd$ instead of $\extd_A$. Note that even
for $G= {\rm U}\of{1}$ where $\extd_A = \extd$ in the continuum
we have $\hextd \neq \extd$ on the lattice. 

Of course $\hextd$ is nilpotent:
\begin{eqnarray}
  \hextd^2 &=& 0
  \,.
\end{eqnarray}

\newpage
\subsection{Fermions}

On cubic graphs with $\ell = 1$ the operators $\edgesCreator^a$ and $\edgesAnnihilator^a$
generate the canonical creation/annihilation algebra 
\refer{anticommutation relations creators/creators on cubic graphs},
\refer{condition for discrete canonical anticommutators}:
\begin{eqnarray}
  \antiCommutator{\edgesCreator^a}{\edgesCreator^b} &=& 0
  \nonumber\\ 
  \antiCommutator{\edgesAnnihilator^a}{\edgesAnnihilator^b} &=& 0
  \nonumber\\
  \antiCommutator{\edgesCreator^a}{\edgesAnnihilator^b}
  &=&
  \delta^{a,b}
  \,,
\end{eqnarray}
as familiar from the continuum, which is isomorphic to the two mutually anticommuting Clifford
algebras
\begin{eqnarray}
  \label{equation in section discrete spinors}
  \antiCommutator{\clifford_\pm^a}{\clifford_\pm^b} &=& \pm 2 \delta^{ab}
  \nonumber\\
  \antiCommutator{\clifford_\pm^a}{\clifford_\mp^b} &=& 0  
\end{eqnarray}
via
\begin{eqnarray}
  \clifford_\pm^a &\defas& \edgesCreator^a \pm \edgesAnnihilator^a
  \,.
\end{eqnarray}

We now consider the case that, in the presence of a non-trivial metric tensor $\gop$,
Clifford generators $\clifford_{\gop \pm}^a : \Omega \to \Omega$ may be constructed, which
satisfy
\begin{eqnarray}
  \label{some discrete Clifford algebra}
  \antiCommutator{\clifford_{\gop \pm}^a}{\clifford_{\gop \pm}^b}
  &=&
  \pm 2 \eta^{ab}
  \nonumber\\
  \antiCommutator{\clifford_{\gop \pm}^a}{\clifford_{\gop \mp}^b}
  &=& 0
  \,,
\end{eqnarray} 
where $\eta^{ab}$ is the Minkowski metric,
as well as
\begin{eqnarray}
  (\clifford_{\gop\pm}^a)^\edag
  &=&
  \pm
  \clifford_{\gop\pm}^a
  \,.
\end{eqnarray}

The possibility to satisfy this simplifying assumption for non-flat background metrics needs yet to be
studied, as do the possibilities of sensibly modifying it in such a situation. In the following 
we concentrate
on an analysis of the case of diamond graphs \S\fullref{section: Diamond complexes}, 
where a construction \refer{equation in section discrete spinors} is
well possible, as discussed in \S\fullref{Clifford algebra on diamond complexes}.

With the Clifford algebra \refer{some discrete Clifford algebra} at hand the treatment of
spinors very closely parallels that of the continuum theory 
(\cf \S\fullref{Dirac, Laplace-Beltrami, and spinors}): Spinors live in minimal left (right)
ideals of the Clifford algebra as represented on $\Omega\of{\cal A}$
and are transformed under the Lorentz group by the 
action of ``rotors'', which in the continuum have the form \refer{a rotor}. In order for
spinor bilinears to be Lorentz invariant these operators have to be unitary, which requires
adapting \refer{a rotor} in a particular way by replacing
\begin{eqnarray}
  R_\pm \;=\; 
  \exp\of{\rho_{[ab]}\clifford_{\gop\pm}^a\clifford_{\gop \pm}^b}
  &\longrightarrow&
  \exp\of{\sum\limits_{[ab]} 
   \sqrt{\rho_{ab}}^\prime\clifford_{\gop\pm}^a\clifford_{\gop \pm}^b \sqrt{\rho_{[ab]}}^\prime} 
  \,, 
\end{eqnarray}
where the prime indicates projection on the hermitian part of the 0-form:
\begin{eqnarray}
  \sqrt{\rho_{ab}}^\prime
  &\defas&
  \frac{1}{2}
  \left(
    \sqrt{\rho_{ab}} + (\sqrt{\rho_{ab}})^\edag
  \right)
  \,.
\end{eqnarray}
This construction enforces that $\rho$ is written in terms of the self-adjoint but non-commutative
coordinates of the diamond complex, which were discussed in \S\fullref{noncommutative coordinates on diamond complex}.

This way the rotors $R_\pm$ are manifestly unitary with respect to
$\bracket{\cdot}{\cdot}_{\gop}$, 
\begin{eqnarray}
  R_{\pm}(R_\pm)^\edag &=& 1
  \,,
\end{eqnarray}
and hence the inner product is invariant under the
``discrete Lorentz transformations'' generated by $R_\pm$:
\begin{eqnarray}
  \bracket{R_\pm \psi}{R_\pm \phi}_\gop
  &=&
  \bracket{\psi}{\phi}_\gop
  \,.
\end{eqnarray}

(Note that the metric operator $\gop$ plays the role of the ``$\clifford^0$-matrix'' used
to construct Lorentz invariant bilinears of spinors when using their matrix representation.)

It is now a triviality to write down action functionals for lattice spinor fields:
Let $\Dirac$ be any Dirac operator on $\Omega\of{\cal A}$ which in particular is
$\gop$-hermitian
\begin{eqnarray}
  \Dirac^\edag
  &=&
  \Dirac
\end{eqnarray}
then
\begin{eqnarray}
  \label{simple lattice Dirac action}
  S[\psi] &=& \bracket{\psi}{\Dirac \, \psi}_\gop
\end{eqnarray}
defines a lattice version of the Dirac action for the spinor field $\psi$. 

In particular, on flat complexes one could consider 
the ``K{\"ahler}-Dirac operator'' $\Dirac = \extd + \ecoextd$, which
however, as a lattice effect, mixes the left-ideals in which the spinors live:
According to \S\fullref{partial derivative operator} we have
\begin{eqnarray}
  \label{Dirac-Kaehler operator on flat diamond}
  \extd \pm \ecoextd
  &=&
  \clifford_\mp^a \frac{1}{2}\left(\partial_a - \partial^\dag_a \right)
  +
  \clifford_\pm^a \frac{1}{2}\left(\partial_a + \partial^\dag_a\right)
\end{eqnarray}
so that
\begin{eqnarray}
  (\extd \pm \ecoextd)
  \ket{(\clifford_\mp^{a_1}\cdots \clifford_\mp^{a_p}) \,\alpha}
  &=&
  \ket{\clifford_\mp^a\;(\clifford_\mp^{a_1}\cdots \clifford_\mp^{a_p}) \lrpartial_a\alpha}  
  +
  (-1)^{p}
  \frac{\epsilon}{2}
  \ket{(\clifford_\mp^{a_1}\cdots \clifford_\mp^{a_p})\;\, \osmotial_a\alpha\, \clifford_\mp^a}  
  \,.
\end{eqnarray}

This can be avoided, if desired, by instead using $\Dirac = \tilde \Dirac$
where
\begin{eqnarray}
  \tilde \Dirac_\pm &\defas&
  \tilde \extd \pm \tilde \coextd
  \nonumber\\
  &=&
  \clifford_\mp^\mu
  \frac{1}{2}
  \left(
    \partial_\mu - \partial_\mu^\dag
  \right)
  \,.
\end{eqnarray}

All this is closely related to the Wilson-Dirac operator.
(See eq (3.3) of \cite{Luescher:1998}).

\paragraph{Relation to the approach by Becher and Joos.}

The issues considered here concerning the Dirac-K{\"a}hler operator on \emph{flat}
lattices are of course not new. The formalism presented here, with its emphasis on 
methods known from non-commutative geometry, provides a transparent and systematic
instrument to generalize all these constructions to arbitrary background metrics, but
for the flat case we should expect overlap with other formulations. Indeed, already long
ago the Dirac-K{\"a}hler operator on flat cubic lattices has been considered by
Becher and Joos in \cite{BecherJoos:1982}. Comparison with their equations 
(3.21) and (3.22) shows that there is an interesting way to rewrite our
\refer{Dirac-Kaehler operator on flat diamond}:

From \refer{commutator discrete partial_i with function} and 
\refer{interchanging discrete derivatives by translation} it follows that
\begin{eqnarray}
  T_{-\edge_a} \circ \partial_a \;=\; \partial_a \circ T_{-\edge_a} &=& -\partial^\dag_a
  \nonumber\\
  T_{\edge_a} \circ -\partial^\dag_a \;=\; -\partial^\dag_a \circ T_{\edge_a} &=& \partial_a   
\end{eqnarray}
and it is clear that
\begin{eqnarray}
  \commutator{T_{\pm\edge_a}}{\edgesCreator} 
  \;=\;0
  \;=\;
  \commutator{T_{\pm\edge_a}}{\edgesAnnihilator} 
  \,.
\end{eqnarray}
Using this one can rewrite \refer{Dirac-Kaehler operator on flat diamond} as
\begin{eqnarray}
  \extd \pm \ecoextd
  &=&
  \edgesCreator^a \partial_a \mp \edgesAnnihilator^a \partial_a^\dag
  \nonumber\\
  &=&
  \left(
    \edgesCreator^a \pm T_{-\edge_a}\circ \edgesAnnihilator^a
  \right)
  \partial_a
  \,.
\end{eqnarray}
This suggests to define \emph{pseudo-Clifford} generators
\begin{eqnarray}
  \label{pseudo-Clifford generators}
  \pseudoCliff^a_\pm
  &\defas&
  \edgesCreator^a \pm T_{-\edge_a}\circ \edgesAnnihilator^a
  \,,
\end{eqnarray}
which satisfy the \emph{pseudo-Clifford algebra}\footnote{
  The way fermions anticommute to translations in this algebra  
is reminiscent of supersymmetry. However, the translation operators on the
right are not \emph{generators} of translations.
}
\begin{eqnarray}
  \antiCommutator{\pseudoCliff^a_\pm}{\pseudoCliff^b_\pm}
  &=&
  \pm 2 \eta^{ab} T_{-\edge_a}
  \nonumber\\
  \antiCommutator{\pseudoCliff^a_\pm}{\pseudoCliff^b_\mp}
  &=& 0
  \,.
\end{eqnarray}
(With respect to (3.15) of \cite{BecherJoos:1982}
note that, using the ordinary operator product, this still extends to a fully associative
Clifford product.)

In terms of this pseudo-Clifford algebra \refer{Dirac-Kaehler operator on flat diamond}
is equivalently rewritten as
\begin{eqnarray}
  \extd \pm \ecoextd
  &=&
  q^a_\pm \partial_a
\end{eqnarray}
with no second (pseudo-)Clifford algebra appearing.

Projectors on eigenspaces of $q^i q^j$ are of the form
\begin{eqnarray}
  P &=& \left(T_{-\edge_i - \edge_j} + s q^i q^j \right)
\end{eqnarray}
(with $s\in \set{\pm 1,\pm i}$) and can be used to build minimal left ideals
of the $q_\pm$-algebra. But, these projectors should act on a non-constant function
instead of on the ``vacuum'' $\ket{1}$. 

This is however of little use, because
0-forms and partial derivative (difference) operators
\emph{do not respect} these ideals:
While for the ordinary $\clifford^a_\pm$ one has
\refer{commutator Clifford with coordinates}
$\commutator{\clifford^a_\pm}{X^b} = \epsilon \delta^{ab}\clifford_\mp^a$
the pseudo-Clifford generators satisfy
\begin{eqnarray}
  q^a_\pm X^a &=& (X^a+\epsilon) \edgesCreator^a \pm X^a T_{-\edge_a}\edgesAnnihilator^a
\end{eqnarray}
so that
\begin{eqnarray}
  \commutator{q^a_\pm}{X^b}
  &=&
  \epsilon
  \delta^{ab}
  \edgesCreator^a
  \nonumber\\
  &=&
  \epsilon \delta^{ab}
  \frac{1}{2}
  \left(
    q^a_+ + q^a_-
  \right)
  \,.
\end{eqnarray}

\paragraph{Chiral symmetry.}
But we don't necessarily need ideals to model spinors on the exterior bundle. Using the 
Dirac-Hestenes representation we can identify spinors with Clifford algebra elements
(e.g. even graded elements in 3+1 dimensions). By the symbol map these can be 
mapped to inhomogeneous forms $\ket{\psi}$ and \refer{simple lattice Dirac action}
defines the Dirac action\footnote{
Really the Dirac-K{\"a}hler action. But for flat metrices both are equivalent.
} of $\psi$. This action has several interesting properties:
It has
\begin{enumerate}
  \item no \emph{fermion doubling}
  \item chiral symmetry
  \item a clear geometrical origin and interpretation.
\end{enumerate}
The absence of fermion doubling is discussed in \S\fullref{fermion doubling}.
The presence of chiral symmetry can be seen as follows:

In even dimensions the pseudoscalar $\bar q$ of the pseudo-Clifford algebra \refer{pseudo-Clifford generators}
satisfies
\begin{eqnarray}
  &&\bar q \defas q_-^1 q_-^2 \cdots q_-^p
  \nonumber\\
  &&\antiCommutator{\Dirac}{\bar q} = 0
  \nonumber\\
  &&\bar q^2 = (-1)^{D(D-1)2 + s}T_{-\sqrt{D}\edge^t}
  \,,
\end{eqnarray}
where $\edge^t = \frac{1}{\sqrt{D}}\left(\edge^1 + \cdots + \edge^D\right)$ is the
timelike direction on the diamond defined in \refer{time coordinate on diamonds}.
Since all translation operators $T_{\cdot}$ commute with $\Dirac$ we can form
the modified pseudoscalar
\begin{eqnarray}
  \tilde q &\defas& 
  i^{D(D-1)/2+s}\left(T_{-\sqrt{D}\edge^t}\right)^{-1/2}\; \bar q_+
  \,,
\end{eqnarray}
which is constructed so as to satisfy
\begin{eqnarray}
  \tilde q^2 &=& 1
  \nonumber\\
  \antiCommutator{\Dirac}{\tilde q} &=& 0
  \,.
\end{eqnarray}
Therefore this operator indeed serves as a substitute for the usual chirality operator.
(This particular construction however relies on the translation invariance of $\Dirac$.
A generalization to interacting fermions needs further investigation, 
\cf \S\fullref{Dirac ops from holonomy forms}.)

\paragraph{Dirac operators from holonomy 1-forms}
\label{Dirac ops from holonomy forms}

We have seen in \S\fullref{lattice Yang-Mills theory} 
that the correct translation of the gauge covariant
derivative (of Lie algebra valued forms) to the lattice is the supercommutator with the 
holonomy 1-form ${\bf H}$.
In order to couple Lie algebra-valued fermions (gauginos) to the gauge field we therefore need to express
the Dirac operator in terms of this supercommutator.  To this end, rewrite the
Dirac-K{\"a}hler action as follows:
\begin{eqnarray}
  \bracket{\psi}{(\extd + \coextd)\psi}
  &=&
  \bracket{\psi}{\extd \psi} + \bracket{\extd \psi }{\psi}
  \nonumber\\
   \;=\;
  |\bracket{\psi}{\extd \psi}|^2
  \,.
\end{eqnarray}

On graphs without opposite and intermediate edges this may be expressed in terms
of the supercommutator with the graph operator:
\begin{eqnarray}
  |\bracket{\psi}{\extd \psi}|^2
  &=&
    |\bracket{\psi}{\superCommutator{{\bf G}}{\psi}}|^2
  \,.
\end{eqnarray}
The graph operator is nothing but the holonomy 1-form for vanishing gauge connection and the
generalization of the above to non-vanishing gauge connection coupling is
\begin{eqnarray}
  \label{version of gauge coupled Dirac action}
  S &=& |\bracket{\psi}{\superCommutator{{\bf H}}{\psi}}|^2
  \,.
\end{eqnarray}

This should be a viable lattice version of the continuum expression 
$\bracket{\psi}{(\gamma^\mu(\partial_\mu + A_\mu))\psi}$. It is not
however expressed in terms of a nice Dirac operator.\footnote{
The following might
be a way to remedy this:

Assume for the moment that for a given holonomy 1-form ${\bf H}$ there
exists a $(p<D)$-form $\ket{V}$ such that
\begin{eqnarray}
  \label{vacuum of holonomy 1-form}
  {\bf H}\ket{V} &=& 0
  \,.
\end{eqnarray}
(For instance for vanishing gauge connection we can choose $V = {\bf H} = {\bf G}$.)
Next consider instead of
\refer{version of gauge coupled Dirac action}.
the 
action $|\bracket{\psi V}{\superCommutator{{\bf H}}{\psi} V}|^2
 = |\bra{V}\psi^\dag \superCommutator{{\bf H}}{\psi}\ket{V}|^2$
whose variation with respect to $\psi$ still reproduces the
lattice version of the minimally coupled massless Dirac equation.
Due to \refer{vacuum of holonomy 1-form} it is equal to
\begin{eqnarray}
  \bra{V}\psi^\dag \superCommutator{{\bf H}}{\psi}\ket{V}
  &=&
  \bra{V}\psi^\dag \left({\bf H} + {\bf H}^\dag\right) \psi\ket{V}
  \,,
\end{eqnarray}
and one sees that now 
\begin{eqnarray}
  {\bf H} + {\bf H}^\dag
  &\equalby{pseudo-Clifford generators}&
  H_a q_+^a
  \,.
\end{eqnarray}
plays the role of the Dirac operator in the presence of the gauge field.
}

In the same spirit the Dirac action of group-valued fermions on the lattice should be
\begin{eqnarray}
  \left|\bracket{\psi}{\epsilon{\bf H}\psi + (-1)^\numberOperator \psi {\bf G}}\right|^2
\end{eqnarray}
because $\epsilon{\bf H} = {\bf G} + \epsilon{\bf A} + \order{\epsilon^2}$
so that
\begin{eqnarray}
  \epsilon{\bf H}\psi + (-1)^\numberOperator \psi {\bf G}
  &=&
  \left(\extd + \epsilon{\bf A}\right)\psi
  +
  \order{\epsilon^2}
  \,.
\end{eqnarray}

\newpage

\section{Summary and Outlook}

It was shown how the abstract differential geometric formalism which has been developed in
great detail by Dimakis and M{\"u}ller-Hoissen nicely describes abstract \emph{metric} geometry when 
augmented by an \emph{inner product} which generalizes the Hodge inner product. 
This way a hybrid formalism of noncommutative spectral geometry in the sense
of Connes, with its emphasis on inner product spaces and Dirac operators, and D\&MH-like formulations
of differential geometry is obtained. There is no restriction to compact and/or Riemannian spaces and
in fact for the discrete hypercubic examples worked out in detail here (which introduce noncommutativity
from 1-forms on upwards) the familiar continuum formalism
largely translates in a \emph{mimetic} fashion to the noncommutative case, allowing to deal with 
issues of Lorentzian signature in the usual way (e.g. by restricting physical inner products 
to spatial hyperslices). 

Apart from the introduction of the inner product itself, we found that the study of its 
\emph{deformations} is very fruitful. It was shown explicitly how geometry is encoded in
\emph{metric operators} $\hat g$ which transform the inner product as $\bracket{\cdot}{\cdot}
\mapsto \bracket{\cdot}{\hat g^{-1}\, \cdot}$ and that this allows for instance the construction of 
the Hodge star operator or of lattice field theory actions for \emph{curved} generalized
(and in particular discrete) spaces.

In fact, the formalism singles out, by means of the \emph{graph operator} $\mathbf{G}$, which encodes the
connectedness of the ``manifold'', a special class of metric operators of the form
$\hat g = V\of{x} (\mathbf{G} \mathbf{G}^\dagger - \mathbf{G}^\dagger \mathbf{G})$ which encode
\emph{pseudo-Riemannian} geometry in a way familiar from the ``causal sets''-approach
\cite{Sorkin:2003}, in that
$(\mathbf{G} \mathbf{G}^\dagger - \mathbf{G}^\dagger \mathbf{G})$ determines the lightcone structure
and $V\of{x}$ the volume element.

The deformation formalism used here seems to be deeper than one might suspect at first sight. 
We mentioned that when superstrings are formulated in loop space the background fields 
(at least massless fields from the NS and NS-NS sector) as well as background gauge and duality 
transformation are
associated with deformations of the Hodge inner product on loop space. From this point of view
the formalism studied here may be regarded as the super\emph{particle} limit of 
loop space formalism in superstring theory \cite{Schreiber:2004,Schreiber:2004f}.

This suggests that the work presented here could and should be extended and deepened in several 
directions. In future work we will more clearly work out the relation of
the approach used here to, for instance, the theory of Whitney forms, causal sets, and
other concepts.

\newpage
\appendix 
\section{Fermion doubling}
\label{fermion doubling}

An old problem of lattice field theory is known as \emph{fermion doubling}
(see for instance section 4.2 of \cite{Langfeld:2002}).

The problem may be stated as follows:

A lattice Laplace operator $\Delta$ constructed from finite difference operators
of the form
\begin{eqnarray}
  D_\mu &\propto& T_{-a\edge_\mu} - T_{(s-a)\edge_\mu}
\end{eqnarray}
with $2a,s\in \N$ by
\begin{eqnarray}
  \Delta &=& D^\mu D_\mu
\end{eqnarray}
has eigenvalues whose absolute value is
\begin{eqnarray} 
  -4\sum\limits_\mu \eta^{\mu\mu} \sin^2\of{\frac{s}{2}\epsilon k_\mu}
  \,.
\end{eqnarray}
For small $k_\mu$ this is close to the $k^\mu k_\mu$-behaviour of the
continuum theory, but for larger $k_\mu$ the difference becomes noticable.
In particular when $\frac{s}{2}\epsilon k_\mu > \pi/2$ new massless states
appear, the so-called ``doublers''. Since $k_{\rm max} = \pi/\epsilon$ is
the largest wavenumber supported on the lattice this happens whenever
\begin{eqnarray}
  s &>& 1
  \,.
\end{eqnarray}
In order to avoid the doubling one hence has to choose $s=1$. But since furthermore
the operator $iD_\mu$ has real eigenvalues only for $a = (s-a)$ hermiticity of
$iD_\mu$ 
restricts $a$ to $a = 1/2$. This is however acceptable only for bosons, since
$D^\mu D_\mu$ with $s=1,\, a=1/2$ is well defined on the lattice, while
$\gamma^\mu D_\mu$ is not, because it involves translations by half a lattice spacing.
The choice $a=1,\, s=2$ would give a well defined hermitian $iD_\mu$ for fermions, but
it will have doublers due to $s>1$.

Therefore, in order to avoid the fermion doubling, the lattice Dirac operator $\gamma^\mu D_\mu$
has to be modified in some sense. 
There exist many different approaches, one of the
more prominent ones being the use of ``Wilson fermions'' \cite{Wilson:1974}. 
A theorem called the Nielson-Ninomiya theorem 
\cite{NielsonNinomiya:1980a,NielsonNinomiya:1980b}
states that a lattice action with 
translational invariance, locality and hermiticity can only avoid fermion
doubling by beaking chiral symmetry.

Note that the Dirac-K{\"a}hler operator \refer{Dirac-Kaehler operator on flat diamond} 
corresponds to 
$s=1$ and hence does not produce any fermion doubling, even though it
is manifestly self-adjoint. This is possible due to the
non-locality which comes from the non-commutativity of 0-forms with 1-forms.

\newpage
\acknowledgments
We would like to thank Ricardo Mosna for drawing our attention to the work by
Becher and Joos and Klaus Fredenhagen for making us clarify the assumptions in 
\S\ref{length, volume and divergence}.

\newpage
\bibliography{std}

\end{document}